%% file: main.tex
\documentclass[journal,transmag]{IEEEtran}

\input{math}
\usepackage{graphicx}
\usepackage{graphics} 
\usepackage{epsfig} 
\usepackage{amssymb}  
\usepackage{amsthm}
\usepackage{subcaption}
\usepackage{amsmath}
\usepackage{nomencl}
\usepackage{longtable}
\usepackage{pifont}
\usepackage{diagbox}
\usepackage{multirow}
\usepackage{amsmath}
\usepackage{xcolor}
\usepackage{enumitem}

\newtheorem{theorem}{Theorem}[section]
\newtheorem{lemma}{Lemma}[theorem]

\ifCLASSINFOpdf
\else
\fi

\begin{document}
\title{Mixed $H_2/H_\infty$ Data-Driven Control Design for Hard Disk Drives}

\author{\IEEEauthorblockN{Omid Bagherieh\IEEEauthorrefmark{1}, and
Roberto Horowitz\IEEEauthorrefmark{2}}
\IEEEauthorblockA{\IEEEauthorrefmark{1,2}Department of Mechanical Engineering,
University of California, Berkeley, CA 94720 USA}
\thanks{Manuscript received March 1, 2018; revised August 26, 2015. 
Corresponding author: O. Bagherieh (email: omidba@berkeley.edu).}}

\IEEEtitleabstractindextext{%
\input{0-Abstract}
}
\maketitle
\IEEEdisplaynontitleabstractindextext
\IEEEpeerreviewmaketitle

\section{Introduction}\label{sec:Intro}
\input{1-Introduction}

\section{Preliminaries}\label{sec:Preliminaries}
\input{2-Preliminaries}

\section{Control Objectives}\label{sec:ConrolObj}
\input{3-ControlObjectives}

\section{Control Algorithms}\label{sec:Conrolalg}
\input{4-ControlAlgorithm}

\section{Design Results}\label{sec:Results}
\input{5-Example}

\section{Conclusion}
\input{6-Conclusion}

\section*{Appendix A}
\input{7-Appendix/7-Appendix_Proof.tex}

\section*{Appendix B}
\input{7-Appendix/7-Appendix_Hinf.tex}

\section*{Acknowledgment}
This research was supported by the Western Digital Inc, and the Computer Mechanics Laboratory (CML) sponsors.
\ifCLASSOPTIONcaptionsoff
  \newpage
\fi

\bibliographystyle{amsplain}
\bibliography{ref}

\begin{IEEEbiography}[{\includegraphics[width=1in,height=1.25in,clip,keepaspectratio]{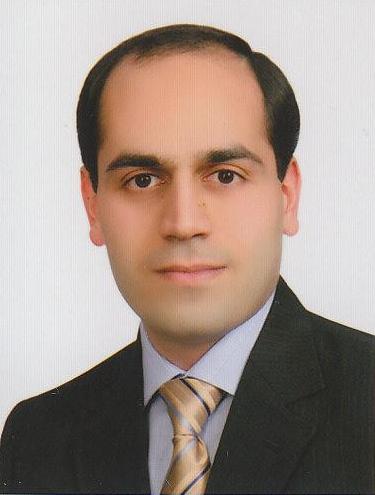}}]
{Omid Bagherieh}
was born in Isfahan, Iran, in 1989. He received the B.Sc. degree in mechanical engineering from the Sharif University of Technology, Tehran, Iran, in 2011, and M.Sc. degree in mechanical engineering from the University of British Columbia, Vancouver, Canada, in 2013.
Since 2013, he has been with the Department of Mechanical Engineering, University of California, Berkeley, as a Ph.D. student. His current research interests include control applications to servo systems and wind turbines.

\end{IEEEbiography}


\begin{IEEEbiography}[{\includegraphics[width=1in,height=1.25in,clip,keepaspectratio]{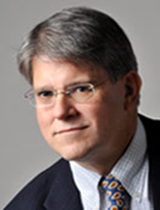}}]
{Roberto Horowitz}
received the B.S. degree with
highest honors in mechanical engineering and the
Ph.D. degree from the University of California,
Berkeley, in 1978 and 1983, respectively.
He is currently a Professor and the chair in the Department of Mechanical
Engineering, University of California, Berkeley. He
is teaching and conducting research in the areas of
adaptive, learning, nonlinear and optimal control
and mechatronics, with applications to disk file
systems, robotics, microelectromechanical systems
(MEMS’s), and intelligent vehicle and highway systems (IVHS’s).
Dr. Horowitz was the recipient of a 1984 IBM Young Faculty Development
Award and a 1987 National Science Foundation Presidential Young Investigator
Award.
\end{IEEEbiography}

\end{document}

%% file: math.tex
\makeatletter
\def\dbackdots{\mathinner{\mkern1mu\raise\p@
    \vbox{\kern7\p@\hbox{.}}\mkern2mu
    \raise4\p@\hbox{.}\mkern2mu\raise7\p@\hbox{.}\mkern1mu}}
\makeatother

\def\matrixx{{\rm MATRIX \kern-.5em\lower.7ex\hbox{X}\kern-.25em \ }}

\newcommand{\abs}[1]{\left| #1 \right|}
\newcommand{\norm}[1]{\left\| #1 \right\|}

\newcommand{\eform}[1]{\if #11{\times 10}\else{\times 10^{#1}} \fi}

\newcommand{\matrixone}[1]{ \left[
        \begin{array}{c}
                 #1
            \end{array}\right]
}

\newcommand{\matrixonetwo}[2]{ \left[
        \begin{array}{cc}
                #1 & #2
        \end{array} \right]
}

\newcommand{\matrixthreeone}[3]{ \left[
        \begin{array}{c}
                #1 \\ #2 \\ #3
        \end{array} \right]
}

\newcommand{\matrixfourone}[4]{\left[ \begin{array}{c}
                                #1 \\ #2 \\ #3 \\ #4
                        \end{array} \right]}
\newcommand{\matrixonefour}[4]{\left[ \begin{array}{cccc}
                                #1 & #2 & #3 & #4
                        \end{array} \right]}



%% file: 0-Abstract.tex
\begin{abstract}
A frequency based data-driven control design considering mixed $H_2$/$H_\infty$ control objectives is developed for multiple input-single output systems. The main advantage of the data-driven control over the model-based control is its ability to use the frequency response measurements of the controlled plant directly without the need to identify a model for the plant. In the proposed methodology, multiple sets of measurements can be considered in the design process to accommodate variations in the system dynamics. The controller is obtained by translating the mixed $H_2$/$H_\infty$ control objectives into a convex optimization problem. The $H_\infty$ norm is used to shape closed loop transfer functions and guarantee closed loop stability, while the $H_2$ norm is used to constrain and/or minimize the variance of signals in the time domain.

The proposed data-driven design methodology is used to design a track following controller for a dual-stage HDD. The  sensitivity decoupling structure\cite{SD_2007} is considered as the controller structure. The compensators inside this controller structure are designed and compared by decoupling the system into two single input-single-output systems as well as solving for a single input-double output controller.
\end{abstract}

%% file: 1-Introduction.tex
\IEEEPARstart{R}{apid} development of Internet and computer technologies continue to demand ever increasing digital data storage. Hard disk and solid state drives are the dominant players in the digital storage game. Hard Disk Drives (HDDs) are primarily used in data centers, while Solid State Drives (SSDs) are mainly used in personal computers and portable devices~\cite{HDDvsSSD2011}. It is predicted that the amount of data stored in data centers will increase by a factor of six from 2015 to 2020~\cite{StatsDataCenters}. As a consequence, data centers need to be equipped with high storage density devices, with continuous improvement of their reliability and performance.

The feedback controllers in HDDs are designed primarily for stabilizing the actuators as well as achieving the desired performance specifications required for reliable and precise positioning of the read/write head mounted on the edge of the servo assembly~\cite{hernandez1999dual,yi1999two}. The most common configuration for the servo assembly is to use the \textit{dual-stage actuation}, which uses two actuators to control the head position~\cite{al2003dual}. The controllers for the dual-stage actuation can be designed using either the model of actuators or their frequency response measurement data sets. The former is called the model-based control design~\cite{abramovitch2002brief,hernandez1999dual}, while the later one is called the data-driven control design~\cite{galdos2010h,karimi2016H2,karimi2016HinfSISO}.


The robust control theory has been developed to consider dynamics uncertainties in the design process~\cite{herrmann2004hdd, robustbook1996}. In the model-based robust control, the nominal model as well as uncertainties are modeled based on the available measurements~\cite{ljung1999system,richardson1982parameter}. However, this methodology has two major drawbacks. The first drawback is that the accurate modeling requires high order dynamics for both the nominal model and the uncertainties, which will lead to a high order controller~\cite{robustbook1996}. The second drawback is that the modeled uncertainties may not be representing the actual variations in the system dynamics~\cite{goodwin1992quantifying}. In the data-driven control design, the measurements represent the real dynamics of the system. Therefore, if the number of measurements are adequate enough to represent uncertainties and modes of the system, the stability and performance level achieved in the design step are guaranteed to be achieved when implementing the controller on the real system~\cite{datadriven_book_H2}. Moreover, the data-driven control methodologies can be useful in designing a common controller for a set of plants produced in a production line \cite{hou2013model}. The dynamics variations among all these plants are represented by their frequency response measurements.

In order to design the controller using the data-driven control methodology in the frequency domain, the state of the art is to convert the problem into an optimization problem where $H_2$ and/or $H_\infty$ norms of the closed loop transfer functions can be considered as the objective and/or constraints. The data-driven $H_\infty$ control problem for Single Input-Single Output (SISO) systems was addressed in \cite{karimi2016HinfSISO}, where a necessary and sufficient convex condition for the $H_\infty$ norm constraint was obtained. This $H_\infty$ control methodology is extended to systems with Multiple Input-Single Output (MISO) in section~\ref{sec:Conrolalg}. A sufficient convex condition for the $H_\infty$ norm of Multi Input-Multi Output (MIMO) systems was also developed in~\cite{karimi2016H2}. The data-driven $H_2$ control problem for SISO systems with pre-specified control structures such as Finite Impulse Response (FIR) filters was given in \cite{datadriven_book_H2}. The data-driven $H_2$ control for MIMO systems with general control structures was developed in \cite{karimi2016H2}, where a sufficient convex condition for the upper-bound of the $H_2$ norm of closed loop transfer functions was derived.

The $H_\infty$ and $H_2$ norm criteria for the closed loop transfer functions can be considered as a mixed $H_2/H_\infty$ control problem~\cite{karimi2016H2,khargonekar1991mixed}. This problem is addressed in section~\ref{sec:Conrolalg} for MISO systems. The proposed algorithm combines the necessary and sufficient convex conditions for the $H_\infty$ norm and the sufficient convex conditions for the $H_2$ norm~\cite{karimi2016H2}. The mixed $H_2/H_\infty$ developed in~\cite{karimi2016H2} used a sufficient convex condition for the $H_\infty$ norm, whereas a necessary and sufficient convex condition for the $H_\infty$ norm is considered here. Moreover, the mixed $H_2/H_\infty$ in~\cite{karimi2016H2} designs the controller based on one set of frequency response measurement of the system. However, one set of measurement cannot represent the uncertainties in the plant dynamics, and multiple measurements are required to capture these uncertainties~\cite{pintelon2012system}. Therefore, multiple measurements are considered in the proposed algorithm.

The proposed data-driven mixed $H_2/H_\infty$ methodology is applied to a dual-stage HDD in section~\ref{sec:Results}. The dual-stage HDD utilizes two actuators for the precision positioning of the read/write head~\cite{aggarwal1997micro,al2006hard,ding2000design}. These actuators have several resonance modes~\cite{atsumi2003vibration,huang2001active}. Including each of these resonance modes in the actuator model directly increases the controller order~\cite{robustbook1996}. However by using the data-driven control methodology, all these modes are already included in the system frequency response measurements and are considered in the controller design step without any direct effect on the controller order~\cite{datadriven_book_H2}. In the proposed methodology, the controller order is fixed and is a function of the control objectives rather than a direct function of the model complexity.

The dual-stage HDD accepts two control inputs, while having only one measurement output. Therefore, the controller for this system will be a single input-double output controller. The conventional design methodology in the HDD industry is to decouple the system into two SISO systems using a well-known methodology called the sensitivity decoupling approach~\cite{abramovitch2002brief, kobayashi2001track,SD_2007}. Therefore, the individual controllers can be obtained in two sequential steps by considering the mixed $H_2/H_\infty$ design process for SISO systems. Section~\ref{sec:Results} proposes to use the mixed $H_2/H_\infty$ methodology developed for MISO systems, where the complete SIMO control block will be obtained in one step. The controller synthesized using the MISO design strategy will be compared with the controller synthesized using the sequential SISO design strategy. The main advantage of the MISO design is its ability to design both controllers in one step, rather than having one controller fixed in the design of the other controller.

%% file: 2-Preliminaries.tex
In this section, the notations and preliminary concepts used for the data driven controller design are presented, first. Then, the stable factorizations of the plant and controller will be reviewed. Subsequently, the closed loop feedback structure and the resulting closed loop transfer functions are introduced.

\subsection{Notations and Preliminaries}
The notations used throughout this paper is introduced in this section. The set of real and complex matrices with $m$ rows and $n$ columns are denoted as $\mathbb{R}^{m\times n}$ and $\mathbb{C}^{m\times n}$, respectively. The set of real and complex scalars use the same notation without the superscript. Considering a complex scalar $a\in \mathbb{C}$, the real and imaginary parts of the scalar $a$ are denoted as $Re(a)$ and $Im(a)$. The magnitude of this scalar is also denoted as $\abs{a}$. The dimensions of a complex matrix are shown as a subscript, where $M_{m\times n}\in \mathbb{C}^{m\times n}$ represents a complex matrix with $m$ rows and $n$ columns. $I_n$ and $0_n$ respectively denote the identity and zero matrices of size $n$. The dimension of a matrix, shown as a subscript, may be eliminated to simplify the notation. The trace of a matrix $M$ is denoted as $Tr(M)$ and its minimum and maximum singular values are denoted as $\underline{\sigma}(M)$ and $\bar{\sigma}(M)$, respectively. $M^*$ represents the complex conjugate transpose of $M$. The matrix inequalities are shown with $\succ$ and $\prec$.

The notations $G_{m\times n}(z)$ and $G_{m\times n}(e^{j\omega})$ represent the transfer function $G_{m\times n}$ in the z-domain and the frequency domain, respectively. Moreover, the subscript $m\times n$ indicates that the transfer function has $n$ inputs and $m$ outputs. The notations for the transfer function dimensions, z-domain and frequency domain may be eliminated to simplify the notations. $\mathbb{R}_{p}^{m\times n}$ denotes the class of all rational causal transfer functions with $m$ outputs and $n$ inputs. $\mathbb{R}_{p,q}^{m \times n}$ denotes the same class of transfer functions as $\mathbb{R}_{p}^{m\times n}$ with order $q$. $\mathbb{RH}_{\infty}^{m\times n}$ denotes the class of all asymptotically stable rational proper transfer functions with $m$ outputs and $n$ inputs.

The transfer function norms are defined considering a transfer function $H_{m\times n}(e^{j\omega})$ with $m$ outputs and $n$ inputs in the frequency domain, where $\omega \in \Omega$ and $\Omega = (-\pi, \pi]$. Here, the infinity norm and two norm for the transfer function $H_{m\times n}$ are defined.
\begin{itemize}
\item \textit{Infinity norm also known as $H_\infty$ norm}: The supremum of the largest singular value of the transfer function $H_{m\times n}(e^{j\omega})$ across the entire frequency region.
\begin{equation}\label{eq:intro:MISO:Hinf_def}
\norm{H_{m\times n}}_{\infty} \overset{\Delta}{=} \mathop{sup}_{\omega \in \Omega} \bar{\sigma} (H_{m\times n}(e^{j\omega}))
\end{equation}
where $\bar{\sigma}(.)$ denotes the largest singular value of the matrix $H_{m\times n}(e^{j\omega})$. According to Eq. (\ref{eq:intro:MISO:Hinf_def}), the $H_\infty$ norm of a system will be helpful in the control design process to constrain the maximum singular value of the system. 

\item \textit{Two norm also known as $H_2$ norm}: The $H_2$ norm is a representative for the system energy and is defined as
\begin{equation}\label{eq:intro:MISO:H2_def}
\norm{H_{m\times n}}^2_{2} \overset{\Delta}{=} \frac{1}{2\pi} \int_{\Omega} { \text{Tr}[H_{m\times n}^{\ast}(e^{j\omega})H_{m\times n}(e^{j\omega})] ~d\omega}
\end{equation}
According to the Parseval's relation~\cite{robustbook1996}, the $H_2$ norm of a transfer function is equal to the square root of the variance of the transfer function output in the time domain, if the input to the transfer function is zero mean white noise with a unit variance. Therefore, the $H_2$ norm is a useful criteria for constraining the variance of stochastic signals in the time domain.
\end{itemize}

\subsection{Plant and Controller Factorizations}

\begin{figure}
	\centering
	\includegraphics[width=8cm]{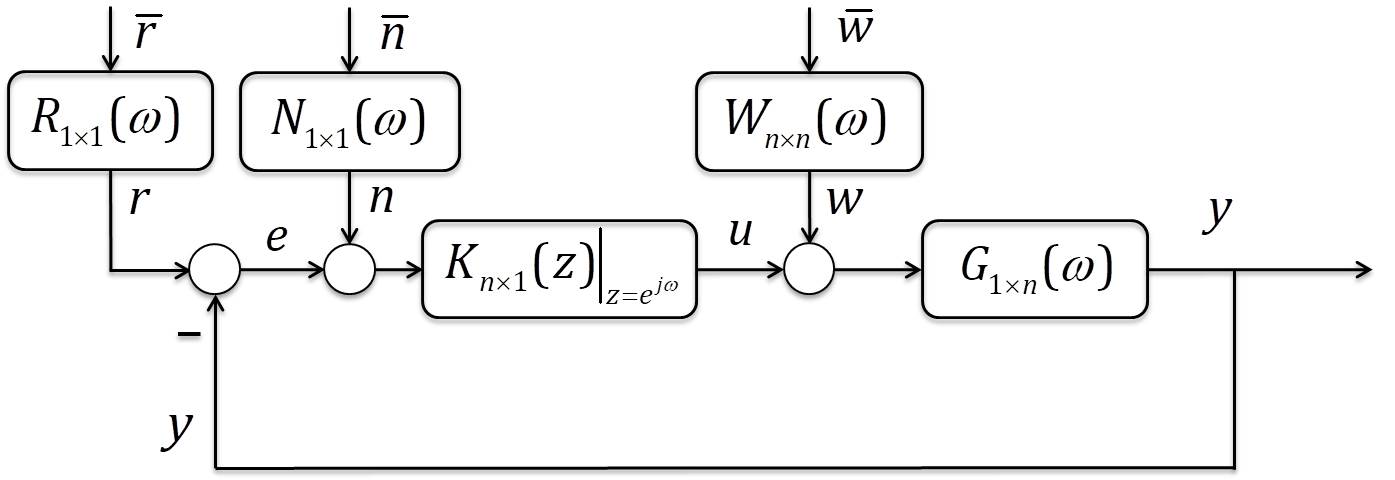}
	\caption{Control block diagram. $G_{1\times n}(\omega)$ represents the frequency response data of the plant. The disturbances to the system are colored by the stable weighting functions $R(\omega) \in \mathbb{RH}_\infty$, $N(\omega) \in \mathbb{RH}_\infty$ and $W(\omega) \in \mathbb{RH}_\infty^{n \times n}$.}
	\label{fig:MISO:MISOblk}
\end{figure}

The well-known stable factorization results~\cite{tay1989indirect} are used to factorize the plant and controller. Consider the feedback system in Fig.~\ref{fig:MISO:MISOblk}, and assume that a negative feedback controller $K_{n\times 1}\in \mathbb{R}_{p}^{n\times 1}$ is designed, which stabilizes the plant $G_{1\times n}\in \mathbb{R}_{p}^{1\times n}$. The plant and controller can respectively be represented by the factorizations
\begin{equation}
G_{1\times n}(\omega)=\tilde{M}_{1\times 1}^{-1}(\omega)\tilde{N}_{1\times n}(\omega),\label{eq:MISO:G_parameterization}
\end{equation}
\begin{equation}
K_{n\times 1}(z)=X_{n\times 1}(z)Y_{1\times 1}^{-1}(z),\label{eq:MISO:K_parameterization}
\end{equation}
where $\tilde{M}_{1\times 1}\in \mathbb{RH}_{\infty}^{1\times 1}$, $\tilde{N}_{1\times n}\in \mathbb{RH}_{\infty}^{1\times n}$, $X_{n\times 1}\in \mathbb{RH}_{\infty}^{n\times 1}$ and $Y_{1\times 1}\in \mathbb{RH}_{\infty}^{1\times 1}$. It is worth mentioning that the stable factorizations can be obtained in both the frequency and $z$ domains. Here, the plant $G_{1\times n}(\omega)$ in Eq. (\ref{eq:MISO:G_parameterization}) is factorized in the frequency domain, while the controller $K_{n\times 1}(z)$ in Eq. (\ref{eq:MISO:K_parameterization}) is factorized in the $z$ domain. The stable factorizations in the $z$ domain can be easily converted to the frequency domain using $z=e^{j\omega}$.

Deriving the stable factorizations for the plant $G_{1\times n}(\omega)$ in Eq. (\ref{eq:MISO:G_parameterization}) can be challenging, since inspecting the stability of frequency response data is not straightforward. Here, three different scenarios for obtaining stable factorizations for the plant $G_{1\times n}(\omega)$ in the frequency domain are considered.

\begin{itemize}
\item \textit{ Stable $G_{1\times n}(\omega)$}: The most straightforward stable factorizations for a stable plant $G_{1\times n}(\omega)$ is
\begin{eqnarray}
\tilde{N}_{1\times n}(\omega) = G_{1\times n}(\omega),  \label{eq:MISO:Factorization:Gstable:N}\\[3pt]
\tilde{M}_{1\times n}(\omega) = 1.\label{eq:MISO:Factorization:Gstable:M}
\end{eqnarray}

\item \textit{ Unstable $G_{1\times n}(\omega)$}: A stabilizing controller $K^0_{n \times 1}(z)$ will be used to obtain the stable factorizations.
\begin{eqnarray}
\tilde{N}_{1\times n}(\omega) = \frac{G_{1\times n}(\omega)}{1 + G_{1\times n}(\omega)K^0_{n \times 1}(e^{j\omega})}, \label{eq:MISO:Factorization:Guns:N} \\[3pt]
\tilde{M}_{1\times 1}(\omega) = \frac{1}{1 + G_{1\times n}(\omega)K^0_{n \times 1}(e^{j\omega})}.\label{eq:MISO:Factorization:Guns:M}
\end{eqnarray}
The factorizations given in Eqs (\ref{eq:MISO:Factorization:Guns:N}) and (\ref{eq:MISO:Factorization:Guns:M}) are stable, since both of these factorizations represent closed loop transfer functions of stable loops. It should be noted that $\tilde{N}_{1\times n}(\omega)$ and $\tilde{M}_{1\times 1}(\omega)$ in Eqs.~(\ref{eq:MISO:Factorization:Guns:N}) and (\ref{eq:MISO:Factorization:Guns:M}) are themselves frequency response data, which are generated by computing the right side of the equation point-wise at the frequency point $\omega \in \Omega$. Also, $K^0_{n \times 1}(e^{j\omega}) \in \mathbb{C}^{n \times 1}$ is the data generated by computing the frequency response of the compensator $K^0_{n \times 1}(z)\big|_{z=e^{j\omega}}$ at the frequency point $\omega \in \Omega$, where $\Omega = (-\pi, \pi]$.

\item \textit{ Known marginally stable poles times a stable $G_{1\times n}^0(\omega)$}: 
In the case of known marginally stable poles, these poles can be considered as zeros in the factorization $\tilde{M}_{1\times 1}(\omega)$. As an example, the stable factorizations of the following plant 
\begin{equation}
G_{1\times n}(\omega)=(\frac{e^{j\omega}}{e^{j\omega}-1})^2G^0_{1\times n}(\omega)
\end{equation}
can be derived as
\begin{eqnarray}
\tilde{N}_{1\times n}(\omega) = G^0_{1\times n}(\omega),\\[3pt]
\tilde{M}_{1\times 1}(\omega) = (\frac{e^{j\omega}-1}{e^{j\omega}})^2,
\end{eqnarray}
where  $\tilde{N}_{1\times n}(\omega)$ and $\tilde{M}_{1\times 1}(\omega)$ will be stable factorizations of the plant $G_{1\times n}(\omega)$.
\end{itemize}

The controller $K(z)$ is implemented in the $z$ domain. Subsequently, the factorizations of this controller given in Eq. (\ref{eq:MISO:K_parameterization}) are obtained in the $z$ domain as well. These factorizations are stable if and only if all their poles are chosen to be inside the unit circle. Considering the following structure for the SIMO controller with $m$ outputs in Fig.~\ref{fig:MISO:MISOblk}
\begin{multline}
K_{n\times 1}(z)=\frac{1}{y_n z^n+..+y_1 z+y_0}\\
\cdot \matrixthreeone{x_{1,n} z^n+..+x_{1,1} z+x_{1,0}}{..}{x_{m,n} z^n+..+x_{m,1} z+x_{m,0}},
\end{multline}
a simple set of stable factorizations will be the finite impulse response, $FIR$, filter with all the poles located at the origin.
\begin{equation}\label{eq:MISO:fact:X}
X_{n\times 1}(z)=\frac{1}{z^n}\matrixthreeone{x_{1,n} z^n+..+x_{1,1} z+x_{1,0}}{..}{x_{m,n} z^n+..+x_{m,1} z+x_{m,0}},
\end{equation}
\begin{equation}\label{eq:MISO:fact:Y}
Y_{1\times 1}(z)=\frac{y_n z^n+..+y_1 z+y_0}{z^n}
\end{equation}
where $X(z)$ and $Y(z)$ represent stable factorizations of the controller in Eq. (\ref{eq:MISO:K_parameterization}). The stable factorizations can be written as the product of the controller coefficients and filter terms
\begin{equation}\label{eq:MISO:roFx}
X_{n\times 1}(z) = \rho_xF_x(z),
\end{equation}
\begin{equation}\label{eq:MISO:roFy}
Y_{1\times 1}(z) = \rho_yF_y(z),
\end{equation}
where $\rho_x$ and $\rho _y$ represent the controller coefficients
\begin{equation}
\rho_x = \matrixthreeone{x_{1,n},~~ ..,~~ x_{1,1},~~ x_{1,0}}{..}{x_{m,n},~~ ..,~~ x_{m,1},~~ x_{m,0}},
\end{equation}
\begin{equation}
\rho_y = \matrixone{y_{n},~~ ..,~~ y_{1},~~ y_{0}},
\end{equation}
and the filter terms $F_x(z)$ and $F_y(z)$ are written as follows
\begin{eqnarray}
F_x(z) = \frac{1}{z^n}\matrixonefour{z^n}{..}{z^{1}}{z^{0}}^{T},\label{eq:MISO:Fx}\\
F_y(z) = \frac{1}{z^n}\matrixonefour{z^n}{..}{z^{1}}{z^{0}}^{T}.\label{eq:MISO:Fy}
\end{eqnarray}

Here for simplicity, the poles of stable factorizations are chosen to be at the origin. However, these poles can be chosen to be anywhere inside the unit circle. Any change in the location of these poles will directly affect the denominator of filter terms given in Eqs. (\ref{eq:MISO:Fx}) and (\ref{eq:MISO:Fy}).

A fixed structure can be considered inside the controller, while deriving the stable factorizations for the controller. Assuming that the first element of the controller should include an integrator of the form $\frac{z}{z-1}$. The stable factorizations of the controller given in Eqs. (\ref{eq:MISO:fact:X}) and (\ref{eq:MISO:fact:Y}) are modified to include this integrator in the structure.
\begin{multline}\label{eq:MISO:fact:Xi}
X_{n\times 1}(z)=\frac{1}{(z-\alpha)z^{n-1}}\\
.\matrixfourone{z ~ (x_{1,n-1} z^{n-1}+..+x_{1,1} z+x_{1,0})}{(z-1) ~ (x_{2,n-1} z^{n-1}+..+x_{2,1} z+x_{2,0})}{..}{(z-1) ~ (x_{m,n-1} z^{n-1}+..+x_{m,1} z+x_{m,0})},
\end{multline}
\begin{equation}\label{eq:MISO:fact:Yi}
Y_{1\times 1}(z)=\frac{z-1}{z-\alpha}\frac{y_{n-1} z^{n-1}+..+y_1 z+y_0}{z^{n-1}},
\end{equation}
where $\abs{\alpha}<1$ is a pole inside the unit circle. The stable factorizations in Eqs. (\ref{eq:MISO:fact:Xi}) and (\ref{eq:MISO:fact:Yi}) can also be written in terms of controller coefficients using Eqs. (\ref{eq:MISO:roFx}) and (\ref{eq:MISO:roFy}).

\subsection{Feedback structure}
The feedback structure given in Fig.~\ref{fig:MISO:MISOblk} is used to design the controller $K(z)$. In this block diagram, $r$ represents the reference trajectory which is desired to be followed. In addition, signals $n$, $w$ represent measurement and control input noises, respectively. Signals $\bar{r}, \bar{n}, \bar{w}$ are white noises with unit variances, and the filter blocks $R \in \mathbb{RH}_\infty$, $N \in \mathbb{RH}_\infty$, $W \in \mathbb{RH}_\infty^{n\times n}$ are used to color these white noises.

In Fig.~\ref{fig:MISO:MISOblk}, the lower-case letters are used to represent signals, where the upper-case letters are used to represent open loop transfer functions. The closed loop transfer functions are also denoted using upper-case letters, where their subscripts represent the input to output causality. For example, $E_{r \rightarrow e}$ denotes the closed loop transfer function from input $r$ to output $e$. Since the plant $G(\omega)$ in Fig.~\ref{fig:MISO:MISOblk} is represented in the frequency domain, all the closed loop transfer functions are derived in the frequency domain. To make the notations simple, the frequency domain arguments $(e^{j\omega})$ and $(\omega)$ will be eliminated from transfer function notations.

Eqs. (\ref{eq:MISO:G_parameterization}) and (\ref{eq:MISO:K_parameterization}) can be used to derive the closed loop transfer functions from external signals in Fig.~\ref{fig:MISO:MISOblk} to tracking error $e$, control input $u$ and the measurement output $y$ in terms of the stable factorizations of the plant and controller
\begin{multline}\label{eq:MISO:cl:MN}
\begin{bmatrix}
E_{r \rightarrow e}            & E_{n \rightarrow e}               & E_{w \rightarrow e}      \\
U_{r \rightarrow u}     		 & U_{n \rightarrow u} 		       & U_{w \rightarrow u}      \\
Y_{r \rightarrow y}     		 & Y_{n \rightarrow y} 		       & Y_{w \rightarrow y}
\end{bmatrix}
=\\
\frac{1}{X\tilde{N}+Y\tilde{M}}
\begin{bmatrix}
\tilde{M}Y      & -X\tilde{N}        & -\tilde{N}Y \\
\tilde{M}X      &  \tilde{M}X   	   & -\tilde{N}X \\
\tilde{N}X		 &  \tilde{N}X		   &  \tilde{N}Y
\end{bmatrix}.
\end{multline}
It is worth mentioning that using the stable factorizations given in Eqs.~(\ref{eq:MISO:G_parameterization}) and (\ref{eq:MISO:K_parameterization}) will create the common scalar denominator, $X\tilde{N}+Y\tilde{M}$, for all the closed loop transfer functions. Moreover, both the numerators and the denominator are all linear functions of both plant and controller stable factorizations, where these factorizations are also linear functions of controller coefficients in Eqs.~(\ref{eq:MISO:roFx}) and (\ref{eq:MISO:roFy}).

%% file: 3-ControlObjectives.tex
The very first step to design the controller $K_{n\times 1}(z)$ in Fig.~\ref{fig:MISO:MISOblk} is to define the desired control objectives. The $H_\infty$ and $H_2$ norms of the closed loop transfer functions will be used to define the control objectives, since these norms are representing both the average performance of the closed loop systems across the frequency region as well as the worst performance at a single frequency.

Assume that there are $l$ frequency response measurements available for the plant $G$ in Fig.~\ref{fig:MISO:MISOblk}. Each of these measurements is denoted by $G_i$, where $i$ represents the $i$'s measurement. As a result for each specific closed loop path in Fig.~\ref{fig:MISO:MISOblk}, there are $l$ closed-loop transfer functions corresponding to each individual plant measurement. These closed loop transfer functions are denoted as $H_i$.

The $H_\infty$ norm of a system was defined in Eq. (\ref{eq:intro:MISO:Hinf_def}). This norm is helpful for constraining the maximum singular value of the system across the entire frequency region, and can be used to shape the closed-loop transfer functions. The $H_\infty$ control objective for multiple plant measurements is defined based on the worst case scenario. Consider the bounded weighting function $W_H \in \mathbb{C}^{o\times p}$ in the frequency domain, the $H_\infty$ norm can be constrained as
\begin{equation}\label{eq:MISO:multiple:Hinf:cst}
\forall i\in 1,.,l : \norm{W_H H_i}_{\infty}<\gamma.
\end{equation}
This inequality constrains the $H_\infty$ norm of the closed loop transfer function for each individual measurements. The $H_\infty$ norm can also be considered as an objective
\begin{equation}\label{eq:MISO:multiple:Hinf:obj}
\min_{\omega \in \Omega,~K \in \mathbb{R}_{p,g}^{n \times 1}} \max_{i} \norm{W_H H_i}_{\infty},
\end{equation}
where the goal is to find the controller $K(z)$ of order $g$, which minimizes the largest $H_\infty$ norm among all the measurements.

The $H_2$ control objective for multiple plant measurements can be defined as the average of or the worst $H_2$ norm square of all the plant measurements\cite{OmidPhDThesis}. Here, the average $H_2$ norm is considered in the case of multiple plant measurements. This $H_2$ norm can be constrained by a scalar $\eta$
\begin{equation}\label{eq:MISO:multiple:H2:avg:cst}
\frac{1}{l}\sum_{i=1}^{l} \norm{H_i}_2^2 \leq \eta
\end{equation}
or be minimized 
\begin{equation}\label{eq:MISO:multiple:H2:avg:obj}
\min_{K \in \mathbb{R}_{p,g}^{n \times 1}} \sum_{i=1}^{l} \frac{1}{l} \norm{H_i}_2^2
\end{equation}
where the goal is to find the controller $K(z)$ of order $g$.

%% file: 4-ControlAlgorithm.tex
The algorithm used to design the controller $K(z)$ in Fig.~\ref{fig:MISO:MISOblk} will be discussed in this section. This algorithm does not rely on the use of a plant model. Instead, the algorithm synthesizes the controller by directly utilizing the available frequency response data from the plants.


In this section, the individual convex conditions of $H_{\infty}$ norm control objectives given in Eqs. (\ref{eq:MISO:multiple:Hinf:cst})-(\ref{eq:MISO:multiple:Hinf:obj}) as well as $H_2$ norm control objectives given in Eqs. (\ref{eq:MISO:multiple:H2:avg:cst})-(\ref{eq:MISO:multiple:H2:avg:obj}) are first described. Subsequently, the mixed $H_2/H_\infty$ control problem is formulated by combining the convex conditions for both the $H_\infty$ and $H_2$ norm control objectives.

\subsection{Data-driven $H_\infty$ control design}\label{sec:Conrolalg:Hinf}
The data-driven $H_\infty$ control design method is developed to design a stabilizing controller by minimizing or constraining the $H_\infty$ norm of selected sets of closed loop transfer functions. In~\cite{karimi2016HinfSISO}, \textit{Karimi et al.} proposed a data-driven $H_\infty$ control design methodology for SISO systems. In this section, this algorithm is extended to MISO systems. The developed control algorithm obtains a necessary and sufficient convex condition for the $H_\infty$ norm, which also guarantees the closed loop stability. Since the $H_\infty$ norm control problem is translated to a convex condition, the controller can be obtained by solving a convex optimization problem.

The SIMO control block diagram for a MISO system is shown in Fig.~\ref{fig:MISO:MISOblk}, where the goal is to obtain the controller $K(z)$ satisfying the control objectives in section~\ref{sec:ConrolObj}. Eq. (\ref{eq:MISO:cl:MN}) represents the closed loop transfer functions in terms of stable factorizations of the plant and controller given in Eqs. (\ref{eq:MISO:G_parameterization}) and (\ref{eq:MISO:K_parameterization}). As one can notice, all the closed loop transfer functions have a scalar denominator. However, the numerators can be scalar, vector or matrix transfer functions.

As an example, the convex condition for the $H_\infty$ norm of the closed loop transfer function from control input disturbance $w$ to control input $u$ will be presented. The procedure for obtaining the convex condition of all other closed loop transfer functions in Eq. (\ref{eq:MISO:cl:MN}) is similar.
\begin{equation}\label{eq:MISO:UwDefinition}
U_{w \rightarrow u,i}=\frac{-\tilde{N}_i X}{X\tilde{N}_i +Y\tilde{M}_i }
\end{equation}
where $U_{w \rightarrow u,i} \in \mathbb{R}_p^{n \times n}$, since control input disturbance $w \in \mathbb{R}^n$ and control input signal $u \in \mathbb{R}^n$. The weighted $H_\infty$ norm for this transfer function can be defined as follows.
\begin{equation}\label{eq:MISO:Hinf_WUw}
H_{\infty} = \max_{i} \norm{W_{U_{w \rightarrow u}}U_{w \rightarrow u,~i}}_{\infty},
\end{equation}
where the weighting function, $W_{U_{w \rightarrow u}} \in \mathbb{C}^{n \times n}$, can be any numerically shaped bounded function of frequency. As mentioned in section~\ref{sec:ConrolObj}, this weighting function will shape the largest maximum singular values of the closed loop transfer functions $U_{w \rightarrow u,i}$ across the entire frequency region~\cite{OmidPhDThesis}.

The following theorem proposes a methodology to convert the data-driven $H_\infty$ control problem into a convex optimization problem for the given $H_\infty$ norm defined in Eq.~(\ref{eq:MISO:Hinf_WUw}). The theorem was first developed for SISO systems in \cite{karimi2016HinfSISO}. Theorem~\ref{THEOREMMISOHINF} extends the results to MISO systems.

\begin{theorem}\label{THEOREMMISOHINF}
Assume that the frequency response data for the i's measurement of the plant $G_{1 \times n,i}(\omega)$ with $n$ inputs and one output is given over the frequency region $\Omega$, and is factorized according to Eq.~(\ref{eq:MISO:G_parameterization}). Given a positive scalar $\gamma$, the following two statements are equivalent.
\begin{itemize}
\item $I$)  Controller $K_{n \times 1}(z)$ stabilizes the plant $G_{1 \times n,i}(\omega)$ and 
\begin{equation}\label{eq:MISO:theorem:Hinf}
\forall i \in 1,..,l : \norm{W_{U_{w \rightarrow u}}U_{(w \rightarrow u)i}}_{\infty} < \gamma
\end{equation}

\item $II$) There exists controller stable factorizations $X(z)\big|_{z=e^{j\omega}}, Y(z)\big|_{z=e^{j\omega}}$ according to Eq. (\ref{eq:MISO:K_parameterization}), such that the following convex inequality holds,
\begin{multline}\label{eq:MISO:theorem:II}
\forall i \in 1,..,l,~\forall \omega \in \Omega : \\
\gamma^{-1} \bar{\sigma} (\ W_{U_{w \rightarrow u}}(\omega)X(e^{j\omega})\tilde{N}_i(\omega)\ ) < \\
Re(\ \tilde{N}_i(\omega)X(e^{j\omega}) + \tilde{M}_i(\omega)Y(e^{j\omega})\ )
\end{multline}
where $\bar{\sigma}(M)$ and $Re(\: r \:)$ functions represent the maximum singular value of the matrix $M$ and the real part of the complex number $r$, respectively. 
\end{itemize}
\end{theorem}
Proof: See appendix A.

This theorem defines a necessary and sufficient convex condition for the $H_\infty$ norm criterion given in Eq. (\ref{eq:MISO:theorem:Hinf}). This $H_\infty$ criterion can be used to either constraint or minimize the $H_\infty$ norm in Eq. (\ref{eq:MISO:Hinf_WUw}). It is worth mentioning that the frequency set $\Omega$ represents the entire frequency region. However, it is not practical to consider the condition in Eq.~(\ref{eq:MISO:theorem:II}) for the entire frequency region. Therefore, a linear frequency grid is utilized to estimate the set $\Omega$. This frequency grid will be considered throughout this paper for the $H_\infty$, $H_2$ and mixed $H_2/H_\infty$ control design problems.

If it is desired to constraint the $H_\infty$ norm given in Eq. (\ref{eq:MISO:Hinf_WUw}), the goal is to find a controller such that Eq. (\ref{eq:MISO:theorem:II}) holds for a given value of $\gamma$. Using the stable factorizations' coefficients $\rho_x$ and $\rho_y$ mentioned in Eqs. (\ref{eq:MISO:roFx}) and (\ref{eq:MISO:roFy}) to factorize the controller $K_{n \times 1}(z)$, Eq.~(\ref{eq:MISO:theorem:II}) becomes a convex function of the coefficients $\rho_x$ and $\rho_y$.

However, if the objective is to minimize the $H_\infty$ norm, the value of $\gamma$ which is the upper-bound of the $H_\infty$ norm in Eq.~(\ref{eq:MISO:theorem:Hinf}) should be minimized. In this case, $\gamma,~\rho_x,~\rho_y$ are all optimization variables and Eq.~(\ref{eq:MISO:theorem:II}) will become nonlinear in terms of these variables. Therefore, the following iterative bisection algorithm\cite{karimi2016HinfSISO} is used to solve this problem.
\begin{enumerate}
\item Pick a value for $\gamma$, which can obtain a feasible solution to Eq.~(\ref{eq:MISO:theorem:II}).
\item Given $\gamma$, find $\rho_x,~\rho_y$ such that Eq.~(\ref{eq:MISO:theorem:II}) is satisfied.
\item Given $\rho_x,~\rho_y$, find the minimum value for $\gamma$ such that Eq.~(\ref{eq:MISO:theorem:II}) is satisfied.
\item Go back to step 2 until the difference between the value of $\gamma$ in step 2 and 3 is smaller than a desired threshold.
\end{enumerate}
The iterative bisection algorithm will not necessarily converge to the global optimal solution. Therefore, the controller obtained using this algorithm may only locally minimize the $H_\infty$ norm given in Eq. (\ref{eq:MISO:Hinf_WUw}).

The convex condition proposed in theorem~\ref{THEOREMMISOHINF} is a necessary and sufficient condition for the $H_\infty$ norm criterion given in Eq. (\ref{eq:MISO:theorem:Hinf}). In~\cite{karimi2016H2}, a sufficient convex condition has been proposed for this criterion. The controller design results based on theorem~\ref{THEOREMMISOHINF} and the sufficient condition mentioned in~\cite{karimi2016H2} are compared in section~\ref{sec:Results:LMIvsIneq}.

\subsection{Data-driven $H_2$ control design}

The data-driven $H_2$ control design methodology used here was developed in~\cite{karimi2016H2} for MIMO systems, where a convex upper-bound for the $H_2$ norm is obtained using an affine approximation of quadratic terms~\cite{karimi2016H2,convexconcave2003}. The $H_2$ constraint or minimization control objectives defined in Eqs.~(\ref{eq:MISO:multiple:H2:avg:cst}) and (\ref{eq:MISO:multiple:H2:avg:obj}) are imposed by constraining or minimizing this upper-bound. Consider the following $H_2$ norm for the MISO system given in Fig.~\ref{fig:MISO:MISOblk}

\begin{equation}\label{eq:MISO:H2upperbound_multiple_avg_define}
H_2^{\text{average}} = \frac{1}{l} \sum_{i=1}^{l} Q_{E_{r \rightarrow e}}\norm{E_{r \rightarrow e,i}}^2_2 + Q_{U_{r \rightarrow u}}\norm{U_{r \rightarrow u,i}}^2_2,
\end{equation}
where $E_{r \rightarrow e,i}$ is a SISO and $U_{r \rightarrow u,i}$ is a SIMO transfer function. $Q_{E_{r \rightarrow e}}$ and $Q_{U_{r \rightarrow u}}$ are scalar weighting functions. The upper-bound for the $H_2$ norm defined in Eq.~(\ref{eq:MISO:H2upperbound_multiple_avg_define}) can be written as
\begin{multline}
H_2^{\text{average}} \leqslant \\
\frac{1}{l} \sum_{i=1}^{l} \int_\Omega {[Q_{E_{r \rightarrow e}} Tr(\Gamma_{E_{r \rightarrow e}}^i) + Q_{U_{r \rightarrow u}} Tr(\Gamma_{U_{r \rightarrow u}}^i)]~d\omega}
\label{eq:MISO:H2upperbound_multiple_avg0}
\end{multline}
where $\Gamma_{E_{r \rightarrow e}}^i(\omega) \in \mathbb{R}$ and $\Gamma_{U_{r \rightarrow u}}^i(\omega) \in \mathbb{R}^{n \times n}$ are the positive definite frequency based variables for each individual measurement $i$. Theorem~\ref{theorem:miso:H2:Karimi} was developed in~\cite{karimi2016H2} to impose constraint on this upper-bound of the $H_2$ norm. Please note that the upper-bound for any other closed loop transfer functions can be formulated in a similar fashion.

\begin{theorem}\label{theorem:miso:H2:Karimi}
Assume that the frequency response data for the i's plant measurement $G_{1 \times n,i}(e^{j\omega})$ with $n$ inputs and one output is given over the frequency region $\Omega$, and is factorized according to Eq.~(\ref{eq:MISO:G_parameterization}). Given a positive scalar $\eta$ and an initial stabilizing controller $K_{k-1}(z)=X_{k-1}(z)Y_{k-1}(z)^{-1}$, the $H_2$ norm defined in Eq.~(\ref{eq:MISO:H2upperbound_multiple_avg_define}) and its upper-bound in Eq.~(\ref{eq:MISO:H2upperbound_multiple_avg0}) are constrained by $\eta$,
\begin{multline}
H_2^{\text{average}} \leqslant \\
\frac{1}{l} \sum_{i=1}^{l} \int_\Omega {[Q_{E_{r \rightarrow e}} Tr(\Gamma_{E_{r \rightarrow e}}^i) + Q_{U_{r \rightarrow u}} Tr(\Gamma_{U_{r \rightarrow u}}^i)]~d\omega} \leqslant \eta
\label{eq:MISO:H2upperbound_multiple_avg}
\end{multline}
if Eq.~(\ref{eq:MISO:H2upperbound_multiple_avg}) is satisfied and there exists controller stable factorizations $X_k(z), Y_k(z)$ according to Eq. (\ref{eq:MISO:K_parameterization}), such that the following LMIs are satisfied
\begin{multline}
\forall i \in 1,..,l, \ \forall \omega \in \Omega:\\
\begin{bmatrix}
\Gamma_{E_{r \rightarrow e}}^i 				& 	Y_{k} \tilde{M}_i\\
\tilde{M}_i^\star Y_{k}^\star  	& 	P_{k, i}^\star P_{k-1} + P_{k-1, i}^\star P_{k, i} - P_{k-1, i}^\star P_{k-1, i}
\end{bmatrix}
\succ
0,
\label{eq:MISO:H2LMI_Sr_multiple_avg}
\end{multline}
\begin{multline}
\forall i \in 1,..,l, \ \forall \omega \in \Omega:\\
\begin{bmatrix}
\Gamma_{U_{r \rightarrow e}}^i 				& 	X_{k} \tilde{M}_i\\
\tilde{M}_i^\star X_{k}^\star  	& 	P_{k, i}^\star P_{k-1, i} + P_{k-1, i}^\star P_{k, i} - P_{k-1, i}^\star P_{k-1, i}
\end{bmatrix}
\succ
0,
\label{eq:MISO:H2LMI_Ur_multiple_avg}
\end{multline}
where the subscript $k$ is representing the iteration number and the parameter $P_{k,i}$ is defined as follows
\begin{equation}
P_{k, i} = \tilde{N}_i X_k + \tilde{M}_i Y_k.
\label{eq:MISO:P}
\end{equation}
\end{theorem}
Proof: See~\cite{karimi2016H2}.

Theorem~\ref{theorem:miso:H2:Karimi} imposes a sufficient convex condition on the upper-bound of the $H_2$ norm defined in Eq.~(\ref{eq:MISO:H2upperbound_multiple_avg}) using the variable $\eta$. This variable can be utilized to either constraint or minimize the $H_2$ norm upper-bound according to Eqs.~(\ref{eq:MISO:multiple:H2:avg:cst}) and (\ref{eq:MISO:multiple:H2:avg:obj}).


In this iterative approach, the controller stable factorizations $X_{k-1}$ and $Y_{k-1}$ from the previous iteration $k-1$ are used to find the stable factorizations $X_{k}$ and $Y_{k}$ for the current iteration $k$. Therefore, the controller is obtained by iterating over its stable factorizations. According to~\cite{karimi2016H2,convexconcave2003}, the following two properties can be proved for this iterative approach.

\begin{enumerate}
\item If the algorithm converges, this upper-bound converges to the real value of the $H_2$ norm defined in Eq. (\ref{eq:MISO:H2upperbound_multiple_avg_define}).
\item Considering the minimization of the $H_2$ norm upper-bound defined in Eq.~(\ref{eq:MISO:H2upperbound_multiple_avg}), the value of the upper-bound is guaranteed to decrease at each iteration.
\end{enumerate}

Theorem~\ref{theorem:miso:H2:Karimi} can be used to obtain a controller, which imposes $H_2$ constraints or minimization objectives on the upper-bound of the $H_2$ norm in terms of convex conditions. However, the obtained controller does not necessarily stabilize the closed loop system. Therefore, the following theorem has been proposed in~\cite{karimi2016H2} to guarantee closed loop stability.

\begin{theorem}\label{theorem:miso:H2Stability}
Given a set of strictly proper plants $G_{1 \times n,i}(e^{j\omega})$, an initial stabilizing controller $K_{k-1}=X_{k-1}Y_{k-1}^{-1}$ and a feasible solution to the following LMI,
\begin{equation}
\forall i \in 1,..,l, \ \forall \omega \in \Omega: P^{*}_{k,i}P_{k-1,i}+P^{*}_{k-1,i}P_{k,i} \succ 0,
\end{equation}
where $P_{k,i}$ is defined in Eq.~(\ref{eq:MISO:P}), the controller $K_k=X_{k}Y_{k}^{-1}$ stabilizes the closed loop system, if
\begin{itemize}
\item The initial controller and the controller $K$ share the same poles on the stability boundary.
\item The following inequality holds.
\begin{equation}
\forall \omega \in \Omega: Y_{k}^\star Y_{k-1} + Y_{k-1}^\star Y_{k} - Y_{k-1}^\star Y_{k-1} \succ 0
\end{equation}
\end{itemize}
\end{theorem}
Proof: See~\cite{karimi2016H2}.

Another approach to guarantee the closed loop stability is to consider the mixed $H_2/H_\infty$ control problem, where the convex condition for the $H_\infty$ norm criterion developed in theorem~\ref{THEOREMMISOHINF} guarantees the closed loop stability. The mixed $H_2/H_\infty$ control design will be illustrated in section~\ref{sec:Conrolalg:mixed}.

The $H_2$ norm can be either minimized or constrained by imposing minimization or constraint criteria on the upper-bound of the $H_2$ norm given in Eq.~(\ref{eq:MISO:H2upperbound_multiple_avg}). This inequality is linear in terms of unknown variables which are $\Gamma^i_{E_{r \rightarrow e}}$, $\Gamma^i_{U_{r \rightarrow u}}$, $\rho_x$ and $\rho_y$ and $\eta$. Moreover, the inequalities given in Eqs.~(\ref{eq:MISO:H2LMI_Sr_multiple_avg}) and (\ref{eq:MISO:H2LMI_Ur_multiple_avg}) are all linear matrix inequalities ($LMIs$). Therefore, the $H_2$ control problem can be formulated as a convex optimization problem. The convex optimization solvers are used to solve this problem. However, the solver should be capable of handling the LMIs as the constraints.



\subsection{Data-driven mixed $H_2/H_\infty$ control design}\label{sec:Conrolalg:mixed}
The convex conditions for either $H_2$ and $H_\infty$ control design criteria are illustrated above. These $H_2$ and $H_\infty$ criteria can be combined to form a mixed $H_2/H_\infty$ control design problem. The main advantage of mixing the $H_2$ and $H_\infty$ norm criteria is the capability of considering both these criteria in the controller design step.

In order to formulate the mixed $H_2/H_\infty$ control design problem, the sufficient convex conditions for the $H_2$ norm criteria in theorem~\ref{theorem:miso:H2:Karimi} are used to impose the $H_2$ norm constraints or minimization objectives. The conditions for the $H_\infty$ norm criteria are imposed by using the necessary and sufficient convex conditions in theorem~\ref{THEOREMMISOHINF}, which guarantees the closed loop stability. Here as an example, the following mixed $H_2/H_\infty$ control design problem is considered.
\begin{equation}\label{eq:opt:miso:general}
\begin{aligned}
& \underset{\substack{X_k \in \mathbb{RH}_\infty^{n \times 1},\\ Y_k \in \mathbb{RH}_\infty}}{\text{min}}
& & H_{2,~\text{objective}}~~~~\text{defined in Eq.~(\ref{eq:MISO:multiple:H2:avg:obj})} \\
& \text{subject to}
& & H_{\infty,~\text{constraints}}~~\text{defined in Eq.~(\ref{eq:MISO:multiple:Hinf:cst}}),\\
&&& H_{2,~\text{constraints}}~~~\text{defined in Eq.~(\ref{eq:MISO:multiple:H2:avg:cst})}.
\end{aligned}
\end{equation}
where the objective is the minimization of $H_{2,~\text{objective}}$ norm, subject to constraints imposed in terms of $H_\infty$ and $H_2$ norms.



\begin{figure}
	\centering
	\includegraphics[width=8cm]{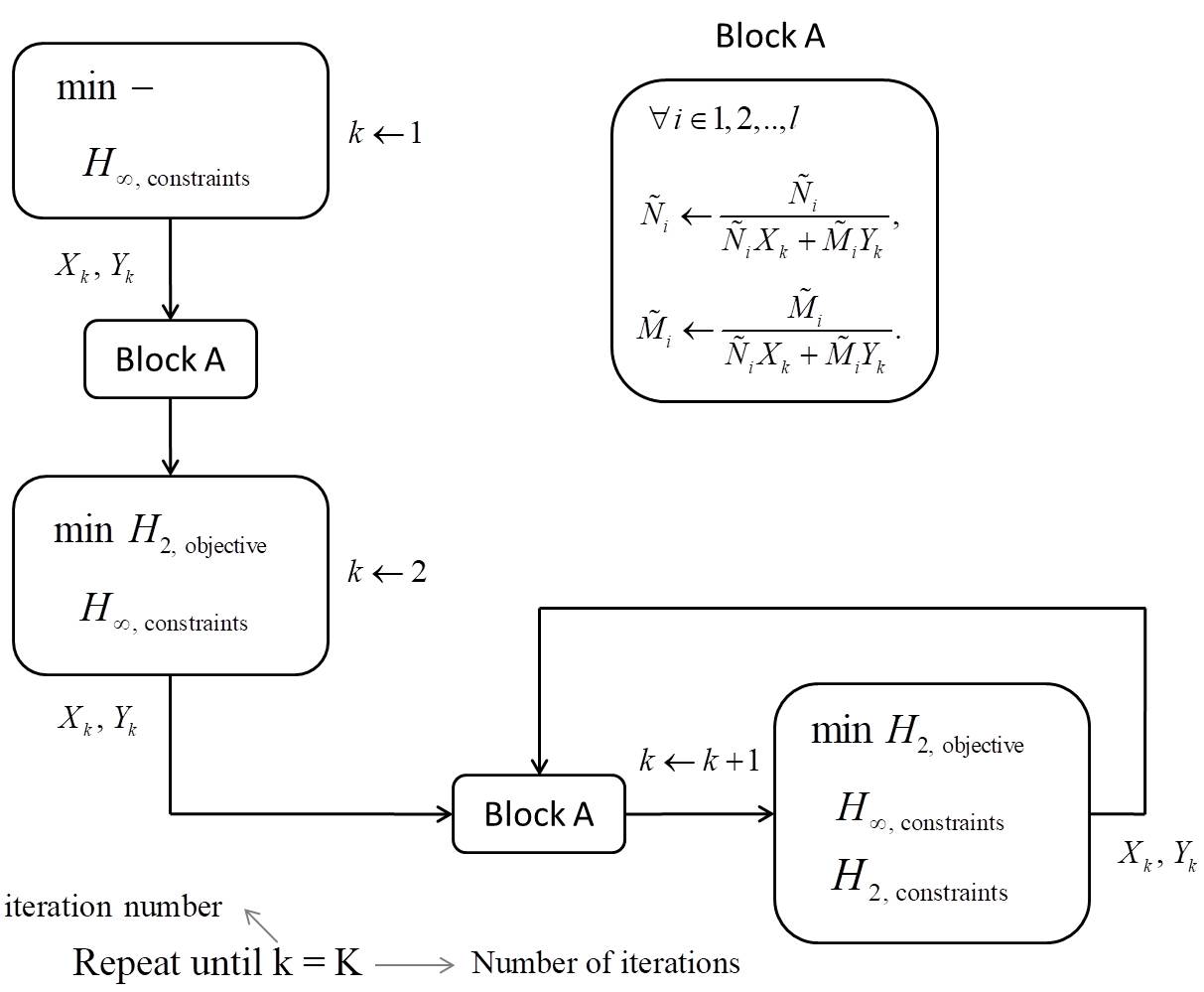}
	\caption{Mixed $H_2/H_\infty$ design algorithm flowchart.}
	\label{fig:SISO:MixedH2Hinf}
\end{figure}

The mixed $H_2/H_\infty$ control design scenario uses the design algorithm flowchart shown in Fig. \ref{fig:SISO:MixedH2Hinf}. In this flowchart, $k$ represents the iteration number and $K$ is the total number of iterations used for obtaining the final iteration of the controller. $X_k(e^{j\omega}) \in \mathbb{C}^{n \times 1}$ and $Y_k(e^{j\omega}) \in \mathbb{C}$ are the frequency responses of the controller factorizations, which are transfer functions or polynomial in the z domain, evaluated at $z=e^{j\omega}$. On the other hand, $\tilde{N}_i(\omega) \in \mathbb{C}^{1 \times n}$, $\tilde{M}_i(\omega) \in \mathbb{C}$ are frequency response data points of stable factorizations of the i's plant measurement.

The very first step to design a mixed $H2/H_\infty$ controller is to design a controller which satisfies the $H_{\infty,~\text{constraints}}$. The $H_\infty$ control design algorithm proposed in theorem~\ref{THEOREMMISOHINF} stabilizes the closed loop system. Therefore, the roots of the common denominator for all the closed loop transfer functions, $\tilde{N}_i(\omega)X_k(e^{j\omega}) + \tilde{M}_i(\omega)Y_k(e^{j\omega})$, will be inside the unit circle. This common denominator obtained at iteration $k$ is used to normalize the plant factorizations at iteration $k+1$ as follows
\begin{equation}
\begin{split}
& \forall i \in 1,..,l,~\forall \omega \in \Omega:\\
& \tilde{N}_i(\omega) \leftarrow \frac{\tilde{N}_i(\omega)}{\tilde{N}_i(\omega) X_k(e^{j\omega}) + \tilde{M}_i(\omega) Y_k(e^{j\omega})},\\
& \tilde{M}_i(\omega) \leftarrow \frac{\tilde{M}_i(\omega)}{\tilde{N}_i(\omega) X_k(e^{j\omega}) + \tilde{M}_i(\omega) Y_k(e^{j\omega})},\label{eq:MN:normalize}
\end{split}
\end{equation}
where $\tilde{N}_i(\omega)$ and $\tilde{M}_i(\omega)$ on the left side of the arrows represent the plant factorizations used at iteration $k+1$. The next step is to design a controller which minimizes the $H_{2,~\text{objective}}$ subject to $H_{\infty,~\text{constraints}}$. Finally, the $H_{2,~\text{constraints}}$ will be included in the optimization problem and according to the flowchart, this optimization problem will be iterated over for the remaining iterations. The main reason for the step by step inclusion of the $H_{2,~\text{objective}}$ and $H_{2,~\text{constraints}}$ is that the convex condition of the $H_2$ norm is a sufficient and not a necessary solution, and depends on the plant factorizations $\tilde{M}_i(\omega),~\tilde{N}_i(\omega)$. Therefore, the normalizations of these parameters in Eq. (\ref{eq:MN:normalize}) can be helpful in avoiding locally optimal solutions.


%% file: 5-Example.tex

In this section, the proposed data-driven mixed $H_2/H_\infty$ algorithm is used to design a track-following controller for a dual-stage HDD. The dual-stage HDD actuators have several resonance modes~\cite{atsumi2003vibration,huang2001active}. In most model-based robust control design methodologies, including each of these resonance modes in the actuator models directly increases the controller order~\cite{robustbook1996}. However by using the data-driven control methodology, all these modes are already included in the system frequency response measurements and will be considered in the controller design step without any direct effect on the controller order~\cite{datadriven_book_H2}.



\subsection{Feedback Structure}\label{sec:AppDD:Structure}
\input{5-Example/1-ApplicationDD-ControlStructure}

\subsection{Control Objectives}\label{sec:AppDD:Objective}
\input{5-Example/2-ApplicationDD-Obj}

\subsection{Control Algorithms}\label{sec:AppDD:Algorithm}
\input{5-Example/3-ApplicationDD-DesignAlgorithm}

\subsection{Control Design Settings}\label{sec:AppDD:Setting}
\input{5-Example/4-ApplicationDD-DesignSetting}

\subsection{Control Design Results}\label{sec:AppDD:Result}
\input{5-Example/5-ApplicationDD-DesignResult}

%% file: 5-Example/1-ApplicationDD-ControlStructure.tex
\begin{figure}
	\centering
	\includegraphics[width=8cm]{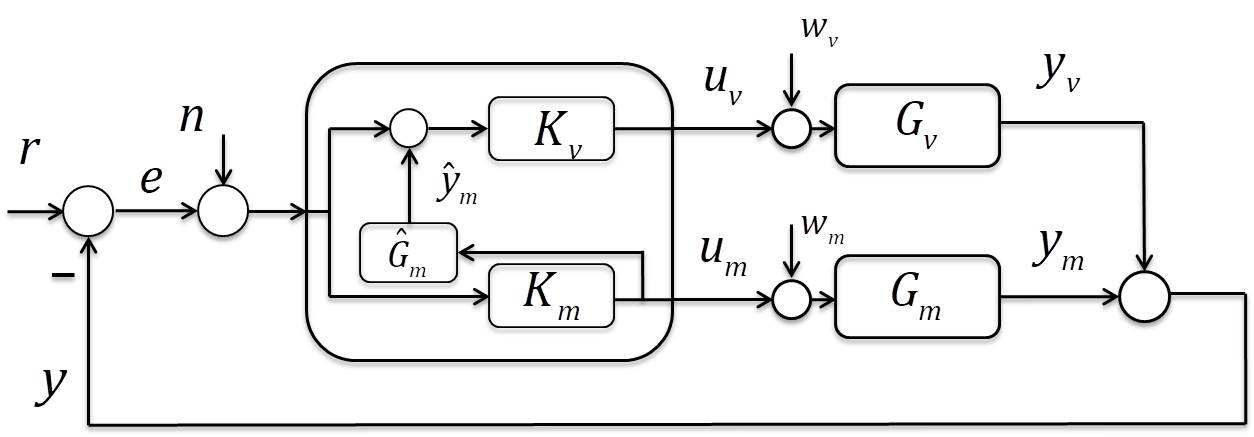}
	\caption{Sensitivity decoupling control Structure for a dual-stage HDD. $\hat{G}_m(z)$ represents an estimated Mili-Actuator (MA) transfer function. In order to switch to the single-stage HDD with VCM as the actuator also known as the single-stage VCM loop, signals $u_m$ and $\hat{y}_m$ are set to zero.}
	\label{fig:AppDD:DualSD}
\end{figure}

The Voice Coil Motor (VCM) and Mili-Actuator (MA) are the two actuators used for nano-positioning the head on the data tracks. These two actuators are respectively denoted as $G_v$ and $G_m$ in the dual-stage feedback structure shown in Fig.~\ref{fig:AppDD:DualSD}. In this block diagram, there are four sources of external noises and disturbances considered to be applied on the HDD. $n$ is contaminating the position error, $e$, and is called the measurement noise. $w_v$ and $w_m$ are the control input noises and $r$ is the track run-out. The major sources for these noises and disturbances are the windage caused by the rotation of the magnetic disk, external vibrations and also the measurement noises.

The HDD actuators have uncertain dynamics, primarily consisting of resonance modes at high frequency regions, which make the control design problem challenging. The data-driven control design introduced in section~\ref{sec:Conrolalg} uses several actuator measurements as the representative of plant uncertainties and a controller is synthesized that stabilizes the closed loop system and achieves the desired performance characteristics for all the measurements. These performance characteristics are considered to be in terms of $H_\infty$ and $H_2$ norms of closed loop transfer functions, where these norms can be constrained or minimized.

The common control structure used in the dual-stage HDD is the Sensitivity decoupling structure illustrated in Fig.~\ref{fig:AppDD:DualSD}.
The main advantage of using the sensitivity decoupling structure is that the sensitivity transfer function for the dual-stage structure, $E_{r \rightarrow e}$, can be written as the product of VCM and MA sensitivity transfer functions. Moreover, in the case of a MA failure, the signals $u_m$ and $\hat{y}_m$ in Fig.~\ref{fig:AppDD:DualSD} will be disconnected, and the HDD will perform as a single-stage HDD with VCM as the actuator. The effect of the MA control input noise $w_m$ on the VCM single-stage closed loop signals will not be considered in the controller design, since it is assumed that the effect of control input noises are lumped into the track run-out and measurement noise spectrum. This is a standard practice for designing a track following controller in the hard disk drive industry. However, if the noise spectrum for the MA control input noise is known, the transfer functions from the MA control input noise to different closed loop signals can be included in the design process.

The sensitivity decoupling control structure depicted in Fig.~\ref{fig:AppDD:DualSD} should be designed such that both the dual-stage and single-stage loops are stable and achieve the desired control objectives presented in section~\ref{sec:AppDD:Objective}. The following two control design strategies are used to obtain the controllers in the sensitivity decoupling structure.


\begin{itemize}
\item \textit{Sequential SISO design strategy}: This is a two step design process. In the first step, the VCM SISO compensator $K_v$ is designed for the single-stage VCM loop. Then in the next step, the compensator $K_v$ is fixed and the MA SISO compensator $K_m$ for the dual-stage loop is designed.

\item \textit{SIMO design strategy}: In this design strategy, the VCM compensator $K_v$ and the MA compensator $K_m$ are designed simultaneously. The controller block diagram in the sensitivity decoupling structure given in Fig.~\ref{fig:AppDD:DualSD} can be expressed as follows
\begin{equation}\label{eq:AppDD:Kbar_def}
	\bar{K} = 
    \begin{bmatrix}
    \bar{K}_1\\
    \bar{K}_2
	\end{bmatrix}
    = 
    \begin{bmatrix}
    K_v(1+K_m\hat{G}_m)\\
    K_m
	\end{bmatrix}.
\end{equation}
As a result, the sensitivity decoupling compensators $K_v$ and $K_m$ will be obtained in terms of $\bar{K}$
\begin{eqnarray}
K_v = \frac{\bar{K}_1}{1+\bar{K}_2\hat{G}_m},\label{eq:miso:KvKbar}\\
K_m = \bar{K}_2.\label{eq:miso:KmKbar}
\end{eqnarray}
In SIMO design strategy, the SIMO controller $\bar{K}$ is designed such that it stabilizes and achieves the desired control objectives for both the single-stage and dual-stage loops. Subsequently, the compensators $K_v$ and $K_m$, are obtained as a function of $\bar{K}$ utilizing Eqs. (\ref{eq:miso:KvKbar}) and (\ref{eq:miso:KmKbar}).
\end{itemize}



%% file: 5-Example/2-ApplicationDD-Obj.tex



The control objectives are considered in terms of $H_\infty$ and $H_2$ norms of closed loop transfer functions in the frequency domain. The $H_\infty$  norms will shape the frequency responses of the closed loop transfer functions and also guarantee closed loop stability. The $H_2$ norms of closed loop transfer functions will be used to constraint and/or minimize the variances of the corresponding signals in the time domain.

\subsubsection{$H_\infty$ norm}
The $H_\infty$ norm of closed loop transfer functions for both the dual-stage and the single-stage VCM loops will be constrained. According to theorem~\ref{THEOREMMISOHINF}, these constraints will guarantee closed loop stability for both of these loops.

The following $H_\infty$ norm constraints for the single-stage HDD are imposed.
\begin{multline}\label{eq:Hinf:single}
\forall i \in 1,..,l :\\
\norm{W_{E_{r \rightarrow e}^{s}}E_{r \rightarrow e,~i}^{s}}_{\infty}<1, ~\norm{W_{U_{r \rightarrow u_v}^{s}}U_{r \rightarrow u_v,~i}^{s}}_{\infty}<1,\\
\norm{W_{E_{w_v \rightarrow e}^{s}}E_{w_v \rightarrow e,~i}^{s}}_{\infty}<1, ~\norm{W_{U_{w_v \rightarrow u_v}^{s}}U_{w_v \rightarrow u_v}^{s},~i}_{\infty}<1.
\end{multline}
where $W_{H}(\omega) \in \mathbb{C}^{o \times p}$ is a bounded weighting function in the frequency domain for the closed loop transfer function $H_{p \times q}$ and the frequency argument $(\omega)$ is eliminated from the notation for simplicity. The superscript $H^{s}$ represents the closed loop transfer functions for the single-stage HDD with VCM as the actuator. As mentioned in section~\ref{sec:Conrolalg}, these weighting functions can be any bounded numerically shaped weighting functions.

The $H_\infty$ norm constraints for the dual-stage HDD in Fig.~\ref{fig:AppDD:DualSD} can be considered in a similar fashion to Eq. (\ref{eq:Hinf:single}).
\begin{multline}\label{eq:Hinf:dual:overall}
\forall i \in 1,..,l :\\
\norm{W_{E_{r \rightarrow e}}E_{r \rightarrow e,~i}}_{\infty}<1, ~\norm{W_{U_{r \rightarrow u}}U_{r \rightarrow u,~i}}_{\infty}<1,\\
\norm{W_{E_{w \rightarrow e}}E_{w \rightarrow e,~i}}_{\infty}<1, ~\norm{W_{U_{w \rightarrow u}}U_{w \rightarrow u,~i}}_{\infty}<1,
\end{multline}
where
\begin{equation}
u=
	\begin{bmatrix}
    	u_v & u_m
	\end{bmatrix}^T,~
w=
	\begin{bmatrix}
    	w_v & w_m
	\end{bmatrix}^T.
\end{equation}
Since the control input as well as control input disturbance have two components, the closed-loop transfer functions in Eq.~(\ref{eq:Hinf:dual:overall}) are not necessarily SISO transfer functions and the $H_\infty$ norm constraints will shape their maximum singular values across the frequency regions.

If it is desired to shape the magnitude of individual SISO closed loop transfer functions in the dual-stage settings, the $H_\infty$ norm for that specific SISO transfer function can be considered. The individual closed loop transfer functions for the tracking error are
\begin{multline}\label{eq:Hinf:dual:individual:E}
\forall i \in 1,..,l :\\
\norm{W_{E_{w_v \rightarrow e}}E_{w_v \rightarrow e,~i}}_{\infty}<1,~
\norm{W_{E_{w_m \rightarrow e}}E_{w_m \rightarrow e,~i}}_{\infty}<1,
\end{multline}
and for the control inputs are
\begin{multline}\label{eq:Hinf:dual:individual:U}
\forall i \in 1,..,l :\\
\norm{W_{U_{r \rightarrow u_v}}U_{r \rightarrow u_v,~i}}_{\infty}<1,\\
\norm{W_{U_{r \rightarrow u_m}}U_{r \rightarrow u_m,~i}}_{\infty}<1,\\
\norm{W_{U_{w_v \rightarrow u_v}}U_{w_v \rightarrow u_v,~i}}_{\infty}<1,\\
\norm{W_{U_{w_v \rightarrow u_m}}U_{w_v \rightarrow u_m,~i}}_{\infty}<1,\\
\norm{W_{U_{w_m \rightarrow u_v}}U_{w_m \rightarrow u_v,~i}}_{\infty}<1,\\
\norm{W_{U_{w_m \rightarrow u_m}}U_{w_m \rightarrow u_m,~i}}_{\infty}<1.\\
\end{multline}

\subsubsection{$H_2$ norm}
The primary control objective in the HDD is to minimize the variance of the position tracking error, $E[e^2(k)]=\norm{e}_2^2$, despite the existence of all the disturbances in the system. We will assume that the disturbances $r,~n,~w_v$ and $w_m$ shown in Fig.~\ref{fig:AppDD:DualSD} can be accurately described by filtering uncorrelated unit variance white noises through the following transfer functions $R(e^{j\omega})$, $N(e^{j\omega})$, $W_v(e^{j\omega})$ and $W_m(e^{j\omega})$, respectively. Therefore, the average variance of the tracking error for the dual-stage loop will be expressed using the $H_2$ norms of closed loop transfer functions in the frequency domain
\begin{multline}\label{eq:H2:E:dual}
\frac{1}{l}\sum_{i=1}^{l}\norm{e_i}^2_2 = \sum_{i=1}^{l}(~ \norm{E_{r \rightarrow e,~i} R}^2_2 + \norm{ E_{n \rightarrow e,~i} N}^2_2 \\
+ \norm{ E_{w_v \rightarrow e,~i} W_v}^2_2 + \norm{ E_{w_m \rightarrow e,~i} W_m}^2_2~).
\end{multline}
where $i$ denotes the $i^{th}$ frequency response data set. The average variance of the tracking error for the single-stage VCM loop will be expressed in a similar fashion
\begin{multline}\label{eq:H2:E:single}
\frac{1}{l}\sum_{i=1}^{l}\norm{e^{s}_i}^2_2 = \sum_{i=1}^{l}(~ \norm{ E^{s}_{r \rightarrow e,~i} R}^2_2 + \norm{E^{s}_{n \rightarrow e,~i} N }^2_2 \\
+ \norm{ E^{s}_{w_v \rightarrow e,~i} W_v}^2_2 ~).
\end{multline}

It is also necessary to constrain the average variances of several signals to take into account actuator limitations. There is a limitation on the amplitude of the VCM control input signal in the time domain and its average variance has to be constrained. The following equations constrain the average variance of the VCM control input in terms of $H_2$ norms in the frequency domain for the single-stage VCM feedback loop
\begin{multline}\label{eq:H2:vU:single}
\frac{1}{l}\sum_{i=1}^{l}\norm{u_{vi}^{s}}^2_2 = \sum_{i=1}^{l}(~ \norm{U^{s}_{r \rightarrow u_v,~i} R}^2_2 + \norm{ U^{s}_{n \rightarrow u_v,~i} N}^2_2 \\
+ \norm{ U^{s}_{w_v \rightarrow u_v,~i} W_v}^2_2~) < \eta_{u_v^s},
\end{multline}
and for the dual-stage feedback loop
\begin{multline}\label{eq:H2:vU:dual}
\frac{1}{l}\sum_{i=1}^{l}\norm{u_{vi}}^2_2 = \sum_{i=1}^{l}(~ \norm{U_{r \rightarrow u_v,~i} R}^2_2 + \norm{U_{n \rightarrow u_v,~i} N}^2_2 \\
+ \norm{U_{w_v \rightarrow u_v,~i} W_v}^2_2 + \norm{U_{w_m \rightarrow u_v,~i} W_m}^2_2~)~<~\eta_{u_v}.
\end{multline}
Assume that the maximum allowable amplitude of a zero mean Gaussian random signal in the time domain is $X$. The standard practice in the magnetic recording industry is  to constrain the 3$\sigma$ value of the signal to be within $X$, in order to assure with a $99.7\%$ probability that the random signal will remain within X, assuming that the signal is Gaussian. The maximum allowable VCM control input is usually $5~Volts$, therefore its standard deviation and variance should be limited to be smaller than $5/3~Volts$ and $(5/3)^2~Volts^2$, respectively. Since the actuator models and noise spectrums given in Figs.~\ref{fig:SISO:G} and \ref{fig:SISO:noises} are only an approximation of the real system, therefore the upper-limits for the average variances of the VCM control input both in the VCM single-stage and dual-stage loops are rounded up to be
\begin{equation}\label{eq:dual:H2:vUnumbers}
\eta_{u_v}=2.0^2~Volts^2,~\eta_{u_v^s}=2.0^2~Volts^2.
\end{equation}
In the case of the MA, there are limitations on the amplitude of the MA control input as well as the MA output signals in the time domain. Therefore, the average variances of these two signals have to be constrained. Here for simplicity, only the average variance of the MA output will be constrained. However if necessary, the average variance of the MA input can also be considered as an additional constraint. The MA is only used in the dual-stage feedback loop, and the average variance of its output can be constrained in terms of $H_2$ norms in the frequency domain as follows
\begin{multline}\label{eq:H2:mY:dual}
\frac{1}{l}\sum_{i=1}^{l}\norm{y_{mi}}^2_2 = \sum_{i=1}^{l}(~ \norm{Y_{r \rightarrow y_m,~i} R}^2_2 + \norm{Y_{n \rightarrow y_m,~i} N}^2_2 \\
+ \norm{Y_{w_v \rightarrow y_m,~i} W_v}^2_2 + \norm{Y_{w_m \rightarrow y_m,~i} W_m}^2_2~)~<~\eta_{y_m} .
\end{multline}
The MAs used in HDDs have limited output strokes. A MA with a smaller output stroke is generally cheaper to fabricate, more reliable and has higher resonance mode frequencies than MAs with larger strokes. On the other hand, decreasing the MA stroke can affect the performance of the servo system. In order to study the effect of the MA output stroke on the overall HDD track following servo performance, four different values for the average variance upper-limit of the MA output stroke are considered here
\begin{equation}\label{eq:dual:H2:mYnumbers}
\eta_{y_m}=44^2,~42^2,~40^2,~38^2~nm^2.
\end{equation}


%% file: 5-Example/3-ApplicationDD-DesignAlgorithm.tex
In the previous section, the mixed $H_2/H_\infty$ control objectives were defined by the general optimization problem given in Eq. (\ref{eq:opt:miso:general}). In this section, this general optimization problem will be specified in terms of the dual-stage feedback structure given in Fig.~\ref{fig:AppDD:DualSD}. The optimization problem will then be solved using both the sequential SISO and SIMO control design strategies described in section~\ref{sec:AppDD:Structure}.

These optimization problems are formulated as convex optimization problems. The convex conditions for the control objectives defined in terms of $H_\infty$ and $H_2$ norms were derived in section~\ref{sec:Conrolalg}. The $H_\infty$ norms control objectives are converted into their necessary and sufficient convex conditions, and the $H_2$ norms control objectives are approximated by their sufficient convex conditions.

In the sequential SISO design strategy, first the VCM compensator $K_v$ is designed considering the single-stage VCM loop. The optimization problem that will be solved to synthesize $K_v$ is
\begin{equation}\label{eq:opt:sd:single:II}
\begin{aligned}
& {\text{minimize}}
& & \mathrm{\norm{e^{s}}_2^2} \text{  defined in Eq.} (\ref{eq:H2:E:single}) \\
& \text{subject to}
& & \text{Eqs.}~(\ref{eq:Hinf:single}),\\
&&& \text{Eqs.}~(\ref{eq:H2:vU:single})~.
\end{aligned}
\end{equation}
After the VCM compensator $K_v$ has been designed, the MA compensator $K_m$ is subsequently designed considering the dual-stage loop depicted in Fig.~\ref{fig:AppDD:DualSD} and fixing the compensator $K_v$. The optimization problem that will be solved to synthesize $K_m$, while keeping $K_v$ fixed is
\begin{equation}\label{eq:opt:sd:dual:II}
\begin{aligned}
& {\text{minimize}}
& & \mathrm{\norm{e}_2^2} \text{  defined in Eq.} (\ref{eq:H2:E:dual}) \\
& \text{subject to}
& & \text{Eqs.}~(\ref{eq:Hinf:dual:overall}),~(\ref{eq:Hinf:dual:individual:E}),~(\ref{eq:Hinf:dual:individual:U}),\\
&&& \text{Eqs.}~(\ref{eq:H2:vU:dual}),~(\ref{eq:H2:mY:dual}).
\end{aligned}
\end{equation}

The SIMO strategy can be used to obtain the compensators $K_v$ and $K_m$, simultaneously. In this methodology, the dual-stage tracking error in Eq.~(\ref{eq:H2:E:dual}) is minimized, while satisfying $H_\infty$ and $H_2$ constraints for both the VCM single-stage and the dual-stage loops. Therefore, the obtained controller is not necessarily minimizing the single-stage VCM loop tracking error in Eq.~(\ref{eq:H2:E:single}). The optimization problem for the SIMO methodology that will be solved to synthesize the compensator $\bar{K}$ in Eq.~(\ref{eq:AppDD:Kbar_def}) is
\begin{equation}\label{eq:opt:miso:II}
\begin{aligned}
& {\text{minimize}}
& & \mathrm{\norm{e}_2^2} \text{  defined in Eq.} (\ref{eq:H2:E:dual})  \\
& \text{subject to}
& & \text{Eqs.}~(\ref{eq:Hinf:single}),~(\ref{eq:Hinf:dual:overall}),~(\ref{eq:Hinf:dual:individual:E}),~(\ref{eq:Hinf:dual:individual:U}),\\
&&& \text{Eqs.}~(\ref{eq:H2:vU:single}),~(\ref{eq:H2:vU:dual}),~(\ref{eq:H2:mY:dual}).
\end{aligned}
\end{equation}
After obtaining $\bar{K}$, the compensators $K_v$ and $K_m$ are obtained from Eqs.~(\ref{eq:miso:KvKbar}) and (\ref{eq:miso:KmKbar}).

%% file: 5-Example/4-ApplicationDD-DesignSetting.tex
\begin{figure}
	\centering
	\includegraphics[width=8cm]{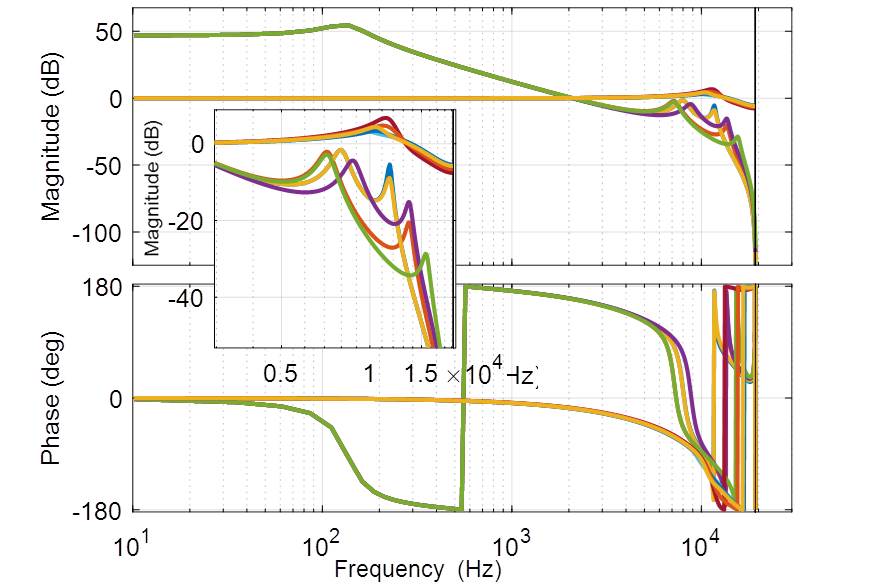}
	\caption{Hard disk drive actuators frequency response measurement data sets used in the design. Five measurements from each actuator are used in the design. $G_v$ and $G_m$ represent VCM and MA actuators' transfer functions. The actuators output units for both the VCM and the MA are 10 nm.}
	\label{fig:SISO:G}
\end{figure}

The data-driven design methodology only uses frequency response measurements of the plant, without requiring models of the actuators. Here, five sets of frequency response data are used to represent dynamics variations for each actuator. These five sets of frequency response data are plotted in Fig. \ref{fig:SISO:G}. If a higher number of frequency response data is required to represent system dynamics, those measurements can be easily included in the design process. A linear frequency grid with 200 points will be used to span the operating frequency region $\Omega$, between 10-19,000 Hz. This frequency grid will be employed throughout this chapter to characterize the open loop and closed loop transfer functions, and synthesize all the compensators.

The frequency response plots shown in Fig.~\ref{fig:SISO:G} were generated using realistic models of both the VCM and the MA. The data-driven design methodology presented in this paper was also used with real frequency response data provided by our HDD industrial research partners. However, this data is considered confidential and proprietary by our partners and cannot be presented here. Therefore, the frequency response data plotted in Fig.~\ref{fig:SISO:G} were produced to mimic real measurements. We emphasize that, although actual transfer functions $G_{vi}(z)$ and $G_{mi}(z)$ were constructed to respectively characterize dynamics variations in the VCM and MA, only the frequency response data generated by these transfer functions, in the form of factorizations $N_{vi}(\omega)$, $M_{vi}(\omega)$ and $N_{mi}(\omega)$, $M_{mi}(\omega)$, were used in the control synthesis algorithms, where $i$ denotes the $i^{th}$ frequency response data set.

\subsubsection{Noise Spectrum}
\begin{figure}
	\centering
	\includegraphics[width=8cm]{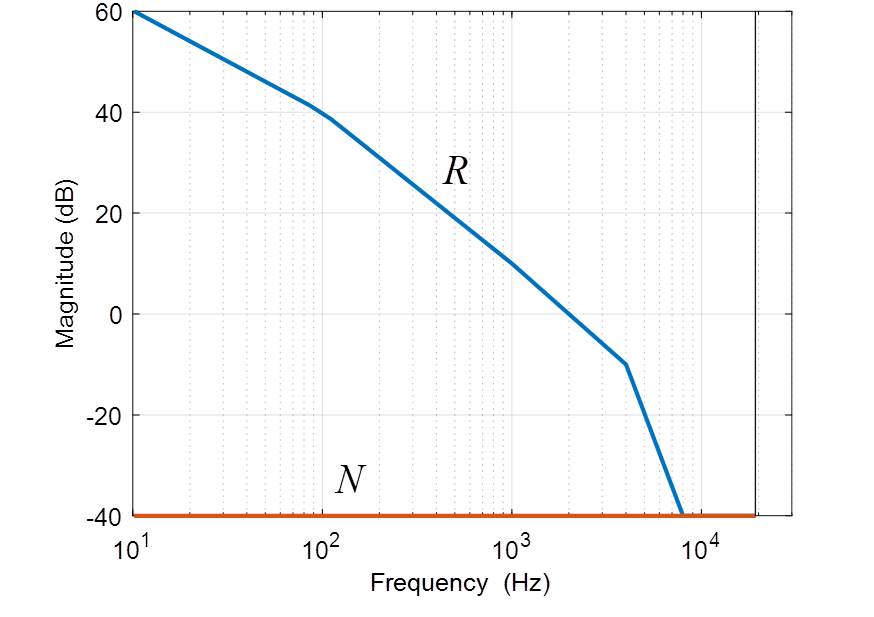}
	\caption{The magnitude Bode plots of the estimated spectrums of external noises applied on the hard disk drive. $R$ and $N$ represent the spectrums for the run-out \textit{r} and the measurement noise \textit{n}. The units for both external noises are 10 nm.}
	\label{fig:SISO:noises}
\end{figure}

The noise spectrums considered for the $H_2$ norm computations in Eqs.~(\ref{eq:H2:E:dual})-(\ref{eq:H2:mY:dual}) are shown in Fig.~\ref{fig:SISO:noises} for the run-out $R$ and measurement noise $N$. The spectrums are denoted with the upper-case letters of the corresponding noises. These spectrums data will be used directly in the design process without any model fitting. According to our HDD industrial research partners, the spectrum for the control input noises can be set to zero, $W_v=0$ and $W_m=0$, since the spectrums given in Fig.~\ref{fig:SISO:noises} are obtained by assuming that the effects of control input noises are lumped into the run-out $R$ and measurement noise $N$ spectrums. Although the spectrum for the control input noises are set to zero, the closed loop transfer functions from control input noises to control inputs and error signals are considered in the design as $H_\infty$ constraints.

\subsubsection{Controller Structure}
The sensitivity decoupling controller structure given in Fig.~\ref{fig:AppDD:DualSD} is used for the dual-stage HDD control design. The estimated model of the MA plant used in the controller structure is considered to be a pure delay
\begin{equation}
\hat{G}_m = \frac{1}{z}.
\end{equation}
Moreover, an integrator of the form $\frac{z}{1-z}$ is incorporated into the VCM controller, in order to eliminate the steady state tracking error due to DC disturbances. 

\subsubsection{Stable Factorizations}
The stable factorizations of a SIMO controller with an integrator included in the first element of the controller was derived in Eqs.~(\ref{eq:MISO:fact:Xi}) and (\ref{eq:MISO:fact:Yi}). In these equations, the value of the parameter $m$ is $2$, since the SIMO controller for the dual-stage HDD has two outputs. The controller order parameter, $n$, will be selected to be $25$.

In the sequential SISO design strategy, first the compensator $K_v$ is obtained using the VCM single-stage loop. Then, the compensator $K_m$ is obtained using the dual-stage loop in Fig.~\ref{fig:AppDD:DualSD} by fixing the compensator $K_v$. The first SISO design step for obtaining the compensator $K_v$ uses the controller stable factorizations in Eqs.~(\ref{eq:MISO:fact:Xi}) and (\ref{eq:MISO:fact:Yi}), which consider an integrator in the controller structure. Since this is a SISO loop, $m=1$ is chosen. The compensator $K_m$ uses stable factorizations given in Eqs.~(\ref{eq:MISO:fact:X}) and (\ref{eq:MISO:fact:Y}) with $m=1$. The compensator order parameters, $n$, for $K_v$ and $K_m$ will be chosen to be $16$ and $20$, respectively. The reasoning behind choosing these values for the compensator order parameters are provided later in this section in the explanation of Fig.~\ref{fig:dual:KSigma}.

The actuators in the HDD are stable actuators. Therefore, the initial stable factorizations of the actuators frequency response measurements are obtained using Eqs. (\ref{eq:MISO:Factorization:Gstable:N}) and (\ref{eq:MISO:Factorization:Gstable:M}). As mentioned in the mixed $H_2/H_\infty$ flowchart in Fig.~\ref{fig:SISO:MixedH2Hinf}, these stable factorizations will be normalized at the end of each iteration using Eq. (\ref{eq:MN:normalize}).

The actuators in the HDD are stable actuators. Therefore, the initial stable factorizations of the actuators frequency response measurements are obtained using Eqs. (\ref{eq:MISO:Factorization:Gstable:N}) and (\ref{eq:MISO:Factorization:Gstable:M}). As mentioned in the mixed $H_2/H_\infty$ flowchart in Fig.~\ref{fig:SISO:MixedH2Hinf}, these stable factorizations will be normalized at the end of each iteration using Eq. (\ref{eq:MN:normalize}).

\subsubsection{Design Scenarios}
Once the stable factorizations for the actuators and controllers are defined, the convex optimization problems mentioned in section~\ref{sec:AppDD:Algorithm} are formulated and solved. The sequential SISO controller design strategy solves the optimization problems formulated in Eqs. (\ref{eq:opt:sd:single:II}) and (\ref{eq:opt:sd:dual:II}) for the compensators $K_v$ and $K_m$, respectively. The SIMO controller $\bar{K}$ in Eq.~(\ref{eq:AppDD:Kbar_def}) is obtained by solving Eq. (\ref{eq:opt:miso:II}). These convex optimization problems are formulated using the YALMIP toolbox\cite{yalmip} in MATLAB software package\cite{MATLAB:2016}. The formulated optimization problems are solved using the MOSEK solver\cite{mosek}. This solver is capable of handling both linear and quadratic programs, where the constraints in the optimization problem are formulated as Linear Matrix Inequalities (LMIs).

The compensators in the sensitivity decoupling structure given in Fig.~\ref{fig:AppDD:DualSD} were designed under four different scenarios each involving a different value of $\eta_{y_m}$, the upper-limit of the MA output stroke average variance. These values are given in Eq.~(\ref{eq:dual:H2:mYnumbers}) and also shown in Table~\ref{table:design:scenarios}. Since the compensators obtained using the sequential SISO design strategy always satisfied these four constraints, only the scenario involving the value of $\eta_{y_m}=44~nm$ will be discussed for the sequential SISO design strategy. Table~\ref{table:design:scenarios} presents the five cases of controller design scenarios that will be evaluated and compared with each other in this section, as well as their respective marker type used in the plots.
\begin{table}[ht]
\caption{Controller design scenarios and their marker type.} 
\centering 
\begin{tabular}{c c c c c } 
\hline\hline\\[-0.8em] 
\text{Index}  & \text{Scenarios} 				& 	\text{Design strategies}	&	\text{$\eta_{y_m}$ ($nm^2$)} 	& \text{Marker type}		\\ [0.5ex] 
\hline \\[-0.5em]
1 & $SIMO_1$  &	SIMO	&	$44^2$ 	& \textcolor{blue}{\large +} \\ [1ex]
2 & $SIMO_2$  &	SIMO	&	$42^2$ 	& \textcolor{magenta}{\large $\times$} \\ [1ex]
3 & $SIMO_3$  &	SIMO	&	$40^2$ 	& \textcolor{green}{\large $\triangleleft$} \\ [1ex]
4 & $SIMO_4$  &	SIMO	&	$38^2$ 	& \textcolor{red}{\large $\triangleright$} \\ [1ex]
5 & $SISO_1$  &	Sequential SISO	&	$44^2$ 	& \textcolor{black}{\large $\circ$} \\ [1ex] 
\hline 
\end{tabular}
\label{table:design:scenarios} 
\end{table}

The design approach proposed in section~\ref{sec:Conrolalg} is an iterative approach. Here, the controllers were designed using $10$ iterations. The $10^{th}$ iteration is used as the stopping criterion for the algorithm in order to have the same number of iterations for all the design scenarios in table~\ref{table:design:scenarios}. However, a more sophisticated stopping criterion, which take into account the rate of optimization objective reduction as a function of the iteration number can be implemented~\cite{boyd2004convex}. As shown in the flowchart given in Fig.~\ref{fig:SISO:MixedH2Hinf}, the first iteration only considers the $H_\infty$ constraints. The second iteration also includes the $H_2$ objective. Finally starting from the third iteration, the full optimization problem including $H_2$ objective, and all $H_2$ and $H_\infty$ constraints are considered. 


%% file: 5-Example/5-ApplicationDD-DesignResult.tex
In this section, the synthesized dual-stage HDD compensators obtained by solving the control objectives in section~\ref{sec:AppDD:Objective} are discussed. First, the design results for the data-driven mixed $H_2/H_\infty$ control methodology are discussed considering all the design scenarios in Table~\ref{table:design:scenarios} using all the plant frequency response data sets in Fig.~\ref{fig:SISO:G}. The plots used to represent these design results utilize the marker types in table~\ref{table:design:scenarios} to distinguish between different scenarios. However, the plots for the same scenario but different frequency response data sets utilize the same marker type and may not be distinguishable from each other at some frequency regions, where the plots are relatively close to each other. Subsequently, an example to describe a limitation of the sequential SISO design strategy as compared to the SIMO design strategy is provided. Last but not the least, the design results for the mixed $H_2/H_\infty$ control problem,
when theorem~\ref{THEOREMMISOHINF} is utilized for imposing $H_\infty$ norm constraints, are compared with the corresponding results when
results in~\cite{karimi2016H2} is used instead of theorem~\ref{THEOREMMISOHINF}. Results in~\cite{karimi2016H2} obtain sufficient conditions for imposing the $H_\infty$ norm constraints, while theorem~\ref{THEOREMMISOHINF} obtains necessary and sufficient conditions.


\subsubsection{Design Scenarios Comparison}\label{sec:Results:Res:scenarios}
The closed loop transfer functions in the final, $10^{th}$ iteration, and their $H_\infty$ constraint limits for the VCM single-stage and dual-stage loops are shown in appendix~B. As shown in the figures, all the closed loop transfer functions satisfy the $H_\infty$ constraints given in Eqs.~(\ref{eq:Hinf:single}), (\ref{eq:Hinf:dual:overall}), (\ref{eq:Hinf:dual:individual:E}) and (\ref{eq:Hinf:dual:individual:U}). The weighting functions used in these equations will shape the closed loop transfer functions. These weighting functions can be numerically shaped, since the data-driven control design methodology is utilized.

\begin{figure}
    \centering
    \begin{subfigure}[b]{8cm}
        \includegraphics[width=8cm]{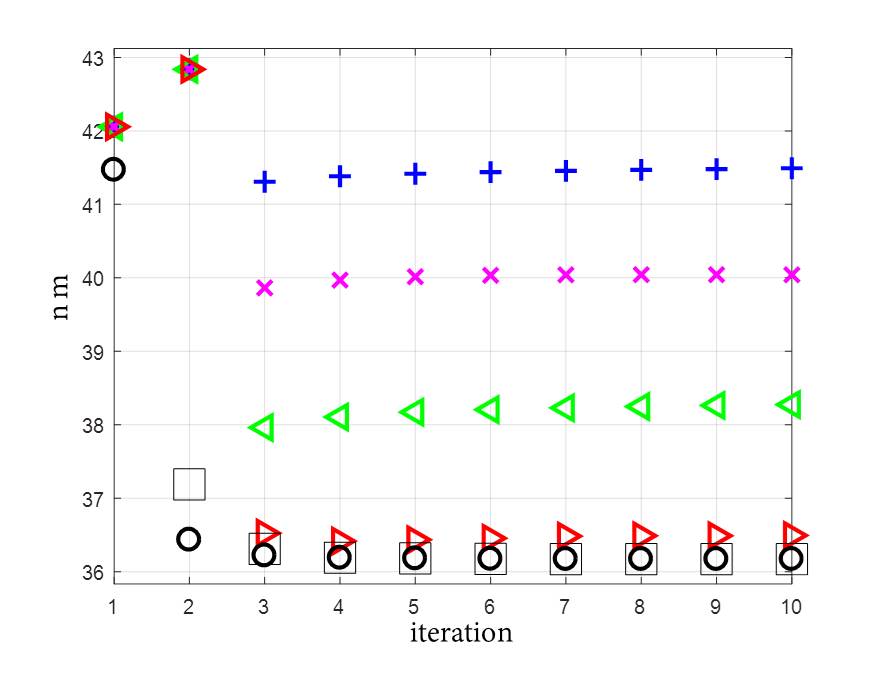}
        \caption{$\sqrt{\frac{1}{l}\sum_{i=1}^{l}\norm{e_i^s}_2}$}
        \label{fig:single:H2:S}
    \end{subfigure}
    \begin{subfigure}[b]{8cm}
        \includegraphics[width=8cm]{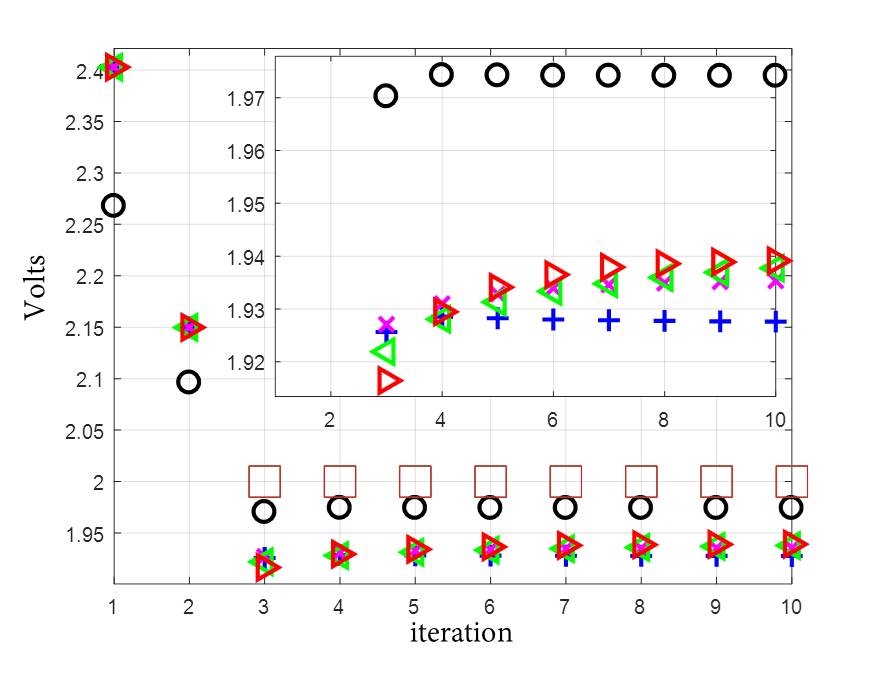}
        \caption{$\sqrt{\frac{1}{l}\sum_{i=1}^{l}\norm{u^s_{vi}}_2}$}
        \label{fig:single:H2:vU}
    \end{subfigure}
    \caption{The $H_2$ norm objective and constraints, for the VCM single-stage over the iterations. The average variances of signals are considered in the optimization problems, however the square roots of the average variances are plotted here. The squares show the upper-bound for the square root of the average variance for the tracking error (a) and the VCM control action (b), where the other markers mentioned in table~\ref{sec:Results:Res:scenarios} show the real values for these variables. The $H_2$ norm objective and constraints are activated starting from the second and third iterations, respectively.}\label{fig:single:H2}
\end{figure}

The $H_2$ norm objective and constraints, for the VCM single-stage loop are shown in Fig.~\ref{fig:single:H2}, considering all the scenarios mentioned in table~\ref{table:design:scenarios}. In the sequential SISO design strategy, the objective is to minimize the average variance of the VCM single-stage tracking error, while designing the VCM controller. However, the objective in the SIMO design strategy is to minimize the average variance of the dual-stage tracking error, without explicitly considering the average variance of the single-stage tracking error as an optimization objective. Therefore, as one can notice in Fig.~\ref{fig:single:H2:S}, the sequential SISO design strategy achieves the smallest average variance for the VCM single-stage tracking error, as compared to the SIMO designs. Moreover, Fig.~\ref{fig:single:H2:vU} shows that the VCM controller designed using the sequential SISO strategy has the highest activity in terms of the VCM control input average variance.

The MA output stroke is restricted by defining an upper-bound on its average variance in Eq.~(\ref{eq:H2:mY:dual}). Reducing the value of this upper-bound will reduce the MA range and consecutively reduce the MA's ability to minimize the tracking error. In this case, the sensitivity decoupling structure suggests that the VCM should compensate for the MA by achieving a smaller average variance of the VCM single-stage tracking error. According to Figs.~\ref{fig:single:H2:S} and \ref{fig:single:H2:vU}, if the MA output stroke is reduced, the VCM single-stage loop achieves a smaller average variance of the tracking error at the cost of having increased average variance of the VCM control input.

\begin{figure}
	\centering
	\includegraphics[width=8cm]{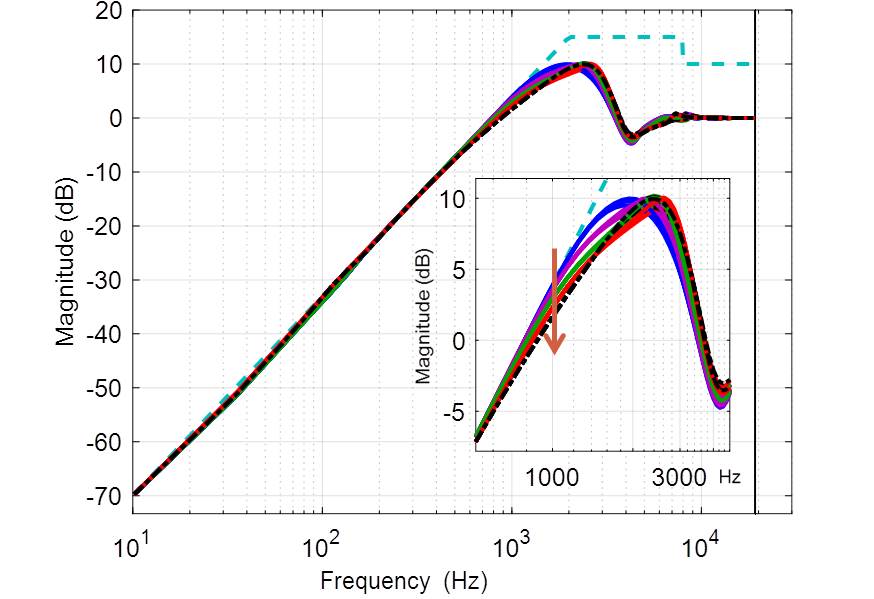}
	\caption{The magnitude Bode plots of the VCM single-stage sensitivity transfer function, $E^s_{r \rightarrow e}$. These plots include 25 closed loop transfer functions for all the 5 design scenarios in table~\ref{table:design:scenarios} using all the 5 frequency response data sets in Fig.~\ref{fig:SISO:G}. The arrow shows the increasing direction for the scenario index numbers in table~\ref{table:design:scenarios}}
	\label{fig:single:Er}
\end{figure}

As shown in Fig.~\ref{fig:single:H2:S}, the controller designed using the sequential SISO design strategy is the most aggressive one in terms of minimizing the average variance of the VCM single-stage tracking error. Also, the SIMO controllers with the more stringent restriction on the MA output stroke will be more aggressive based on this definition. The VCM single-stage sensitivity plots for all these scenarios are shown in Fig.~\ref{fig:single:Er}. In the frequency region between $0-500~Hz$, where the VCM is the dominant actuator, all these scenarios achieve almost the same level of the VCM single-stage tracking error reduction. However in the frequency region between $500-2000~Hz$, where both actuators are active, the more aggressive VCM single-stage controller will improve the VCM single-stage tracking error reduction, in order to compensate for the MA output stroke restriction. 

Since the z-domain transfer functions for all of the actuators frequency response measurements shown in Fig.~\ref{fig:SISO:G} are available for this example, we were able to compute the closed loop poles of the feedback system, for all of the synthesized compensators and for all of the scenarios in table~\ref{table:design:scenarios}, and to verify that all of the compensators designed yield stable feedback systems for all of the plants. The stability margins and bandwidths for all these scenarios are obtained in the frequency domain using the plant frequency response data sets shown in Fig.~\ref{fig:SISO:G}. The worst case stability margins and bandwidths are defined in Eqs.~(\ref{eq:margins:worst:Er})-(\ref{eq:margins:worst:W_PM}) and are shown in table~\ref{table:single:Margins} for all of the scenarios in table~\ref{table:design:scenarios}.
\begin{eqnarray}
\text{worst}(E^s_{r \rightarrow e} \text{peak})&=&\max_i E^s_{(r \rightarrow e)i} \text{peak}\label{eq:margins:worst:Er}\\
\text{worst}(GM)&=&\min_i GM_i\label{eq:margins:worst:GM}\\
\text{worst}(PM)&=&\min_i PM_i\label{eq:margins:worst:PM}\\
\text{worst}(\omega_{GM})&=&\min_i \omega_{GM_i}\label{eq:margins:worst:W_GM}\\
\text{worst}(\omega_{PM})&=&\min_i \omega_{PM_i}\label{eq:margins:worst:W_PM}
\end{eqnarray}
where the subscript $i$ denotes the $i^{th}$ frequency response data sets. The more aggressive controller generally increases the worst case bandwidth, $\text{worst}(\omega_{PM})$, at the cost of having lower stability margins. However, this trend is not completely followed, since the aggressiveness of the controllers is defined in terms of the $H_2$ norm of the closed loop transfer functions, which restricts the integration of the frequency response magnitude over the entire frequency region. Therefore, increasing the restriction on the $H_2$ norm does not always directly affect the bandwidth and stability margins of the open loop transfer function.

\begin{table}[ht]
\caption{Open loop worst case stability margins for the VCM single-stage loop. These margins are obtained by selecting the worst case margins among all frequency response data sets using Eqs.~(\ref{eq:margins:worst:Er})-(\ref{eq:margins:worst:W_PM}).} 
\centering 
\begin{tabular}{c c c c c c} 
\hline\hline\\[-0.8em] 
\multirow{2}{*}{\diagbox[width=8em,trim=lr]{Scenarios}{Worst case}} 				& 	$E^s_{r \rightarrow e}$ peak	&	$GM$ 	& $PM$ 		& 	$\omega_{GM}$ 	& 	$\omega_{PM}$\\ [0.5ex] 
		 			    &	$dB$		& 	$dB$ 	& $degree$ 	& 	$Hz$ 			& 	$Hz$ \\ [0.5ex]
\hline \\[-0.5em]
\text{$SIMO_1$}  &	9.97  & 3.45  &  25.83 	 	& 	2,206 			& 	1,242    \\
\text{$SIMO_2$}  &	10.01  & 3.30  &  28.55 	 	& 	2,319			& 	1,230    \\
\text{$SIMO_3$}  &	10.17  & 3.22  &  31.17 	 	& 	2,427			& 	1,244    \\
\text{$SIMO_4$}  &	10.03 & 3.28   &  32.46	 		& 	2,548			& 	1,276    \\
\text{$SISO_1$}   & 10.10 & 3.28   &  31.45 	 	& 	2,493			&	1,344    \\ [1ex] 
\hline 
\end{tabular}
\label{table:single:Margins} 
\end{table}

\begin{figure}
    \centering
    \begin{subfigure}[b]{8cm}
        \includegraphics[width=8cm]{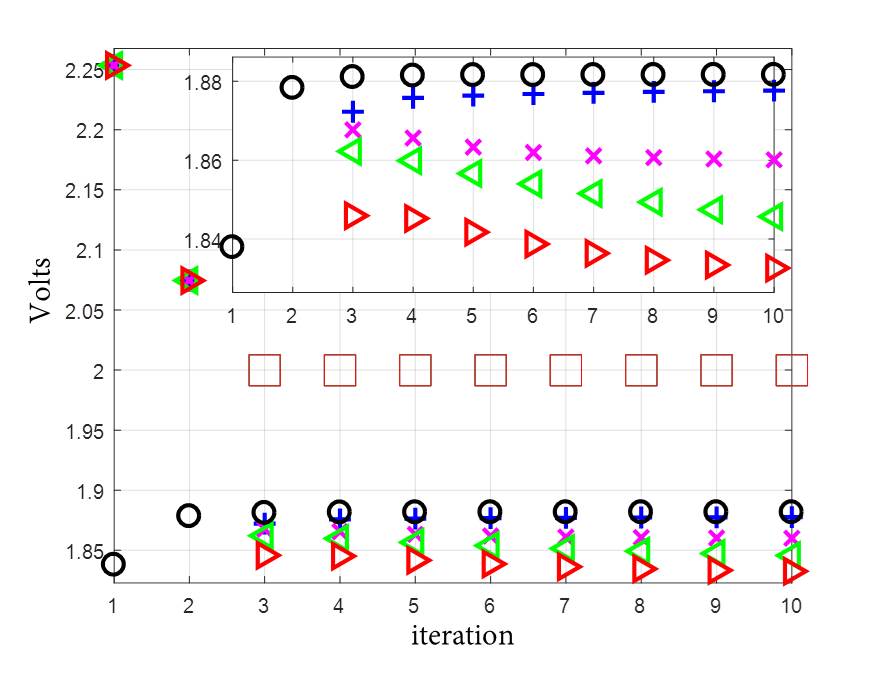}
        \caption{$\sqrt{\frac{1}{l}\sum_{i=1}^{l}\norm{u_{vi}}_2}$}
        \label{fig:dual:H2cst:vU}
    \end{subfigure}
    \begin{subfigure}[b]{8cm}
        \includegraphics[width=8cm]{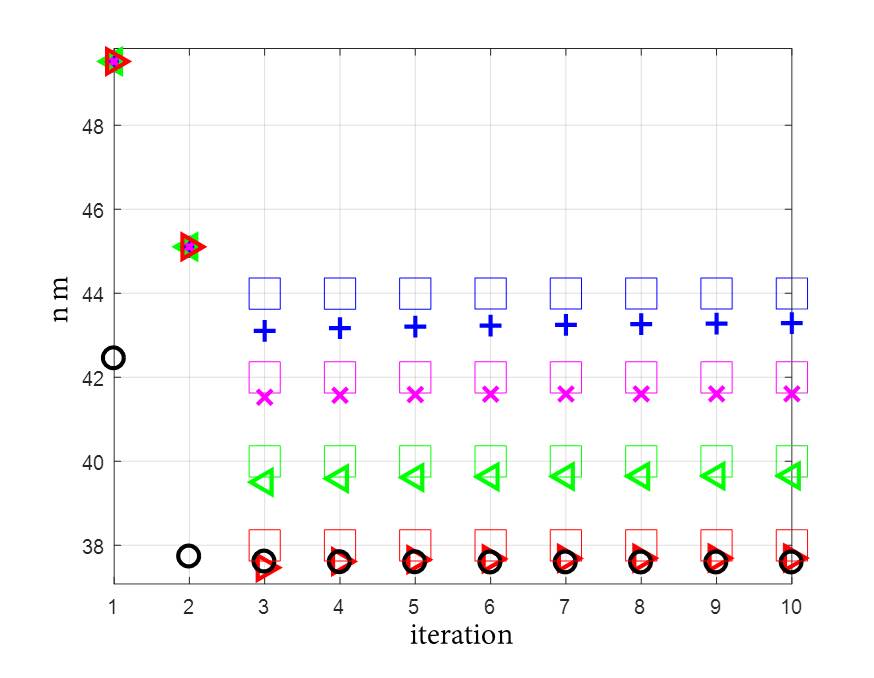}
        \caption{$\sqrt{\frac{1}{l}\sum_{i=1}^{l}\norm{y_{mi}}_2}$}
        \label{fig:dual:H2cst:mY}
    \end{subfigure}
    \caption{The $H_2$ norm constraints for the dual-stage loop over the iterations. The average variances of signals are considered in the optimization problems, however the square roots of the average variances are plotted here. The squares show the upper-bound for the square root of the average variance for the VCM control action (a) and the MA output stroke (b), where the other markers mentioned in table~\ref{sec:Results:Res:scenarios} show the real value for these variables. The $H_2$ norm objective and constraints are activated starting from second and third iterations, respectively.}\label{fig:dual:H2cst}
\end{figure}

The $H_2$ norm constraints for the VCM input and the MA output stroke in the dual-stage HDD are shown in Figs.~\ref{fig:dual:H2cst:vU} and~\ref{fig:dual:H2cst:mY}, respectively. In the sequential SISO design strategy, the VCM is designed to be aggressive. As a result, the average variance of the VCM control input in both the VCM single-stage and dual-stage loops for the sequential SISO design strategy will be higher compared to all the scenarios in the SIMO design strategy, as shown in Figs.~\ref{fig:single:H2:vU} and \ref{fig:dual:H2cst:vU}. As previously mentioned, using a MA with a smaller output stroke will result in a corresponding increase in the average variance of the VCM control input in the VCM single-stage loop, in order to compensate for the MA output stroke reduction. However, the MA with a smaller output stroke decreases the average variance of the VCM control input in the dual-stage loop. This can be justified by the fact that the the smaller MA output stroke requires less VCM movements to compensate for the MA output movements, which are out of phase with the VCM output movements. Fig.~\ref{fig:dual:H2cst:mY} demonstrates that all the controller design scenarios considered in table~\ref{table:design:scenarios} are satisfying their upper-bound constraints for the average variance of the MA output stroke.

\begin{figure}
	\centering
	\includegraphics[width=8cm]{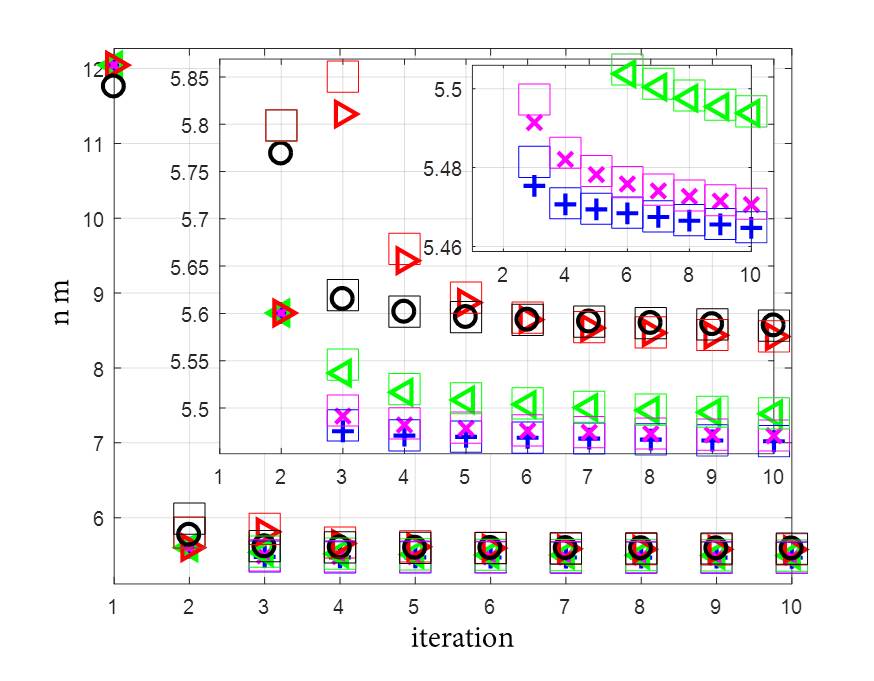}
    \caption{$\sqrt{\frac{1}{l}\sum_{i=1}^{l}\norm{e_i}^2_2}$ for the SIMO feedback system in Fig.~\ref{fig:AppDD:DualSD} as a function of control synthesis iterations: The average variances of signals are considered in the optimization problems, however the square roots of the average variances are plotted here. The $H_2$ norm objective and constraints are activated starting from second and third iterations, respectively.}
    \label{fig:dual:H2obj}
\end{figure}

Fig.~\ref{fig:dual:H2obj} plots the square root of the average variance of the tracking error for the dual-stage feedback system in Fig.~\ref{fig:AppDD:DualSD}. According to Eqs. (\ref{eq:opt:sd:dual:II}) and (\ref{eq:opt:miso:II}), the average variance is considered as the minimization objective in the SIMO design strategy as well as the second step of the sequential SISO design strategy. In the sequential SISO design strategy, the obtained VCM compensator $K_v$, which is synthesized to optimize the average variance of the VCM single-stage tracking error, is kept fixed during the synthesis of the MA compensator $K_m$. This results in a suboptimal overall compensator in terms of minimizing the average variance of the dual-stage tracking error, since the compensator $K_v$ design process in the sequential SISO design strategy does not take into account the dual-stage loop. This is verified in Fig.~\ref{fig:dual:H2obj}, which shows that the sequential SISO design strategy achieves a relatively higher average variance of the dual-stage tracking error compared to all the scenarios in the SIMO design strategy.

The results in Fig.~\ref{fig:dual:H2obj} also suggest that the average variance of the dual-stage tracking error in the SIMO design strategy is a non-linear function of the restriction on the average variance of the MA output stroke. If the average variance upper-bound on the MA output stroke given in Eq. (\ref{eq:dual:H2:mYnumbers}) is 42 nm or greater, the average variance of the dual-stage tracking error will not change significantly. The main reason to this phenomenon is that, according to Fig.~\ref{fig:single:H2:S}, the VCM is able to compensate for this MA output stroke restriction. However, the VCM's ability to compensate for the MA output stroke restriction is limited. If the average variance upper-bound on the MA output stroke is smaller than $42^2~nm$, the decrease in this upper-bound will increase the average variance of the dual-stage tracking error at a higher rate as compared to the case of the MA output stroke with the average variance upper-bound of $42^2~nm$ or greater.

Similar to the VCM single-stage loop, the closed loop poles for all the scenarios in table~\ref{table:design:scenarios} were computed using the z-domain transfer functions of the actuator frequency response measurements. All these closed loop poles were confirmed to be located inside the unit circle. Therefore, all the closed loop transfer functions were stable for all the plants used to generate the frequency response data in the data-driven mixed $H_2/H_\infty$ formulation. Table~\ref{table:dual:Margins} presents the dual-stage open loop stability margins and bandwidths for these design scenarios. These stability margins and bandwidths are reported as the worst case margins and bandwidths using Eqs.~(\ref{eq:margins:worst:Er})-(\ref{eq:margins:worst:W_PM}). In this table, the open loop transfer function is from the track run-out \textit{r} to the position head output \textit{y} in Fig.~\ref{fig:AppDD:DualSD}. As shown in Fig.~\ref{fig:AppDD:DualSD}, the imposition of a more stringent restriction on the MA output stroke, which is achieved by reducing the upper-bound on the average variance of this signal, will result in a lower dual-stage bandwidth. This reduction in the bandwidth will help to improve the stability margins. The dual-stage controller designed using the sequential SISO design strategy will have a relatively lower bandwidth and a higher peak for the sensitivity plot as compared to the $SIMO_4$ case. The higher sensitivity plot peak for the sequential SISO design strategy can be justified by the fact that this strategy will design compensators $k_v$ and $k_m$ in two individual steps and the compensator $K_v$ is designed without considering the dual-stage actuation structure.

\begin{table}[ht]
\caption{Open loop worst case stability margins for the dual-stage loop in Fig.~\ref{fig:AppDD:DualSD}. The open loop transfer function is from the track run-out \textit{r} to the position head output \textit{y}. These margins are obtained by selecting the worst case margins among all frequency response data sets using Eqs.~(\ref{eq:margins:worst:Er})-(\ref{eq:margins:worst:W_PM}).} 
\centering 
\begin{tabular}{c c c c c c} 
\hline\hline\\[-0.8em] 
\multirow{2}{*}{\diagbox[width=8em,trim=lr]{Scenarios}{Worst case}} 				& 	$E_{r \rightarrow e}$ peak	&	$GM$ 	& $PM$ 		& 	$\omega_{GM}$ 	& 	$\omega_{PM}$\\ [0.5ex] 
		 			    &	$dB$		& 	$dB$ 	& $degree$ 	& 	$Hz$ 			& 	$Hz$ \\ [0.5ex]
\hline \\[-0.5em]
\text{$SIMO_1$}  &	9.63  & 8.85  &  19.13 	 & 	12,392 			& 	4,924    \\
\text{$SIMO_2$}  &	9.62  & 8.85  &  19.14 	 & 	12,393 			& 	4,922    \\
\text{$SIMO_3$}  &	9.55  & 8.86  &  19.36 	 &  12,390			& 	4,914    \\
\text{$SIMO_4$}  &	9.43  & 8.87  &  19.74	 &  12,401 			& 	4,884    \\
\text{$SISO_1$}  & 	9.58  & 8.90  &  19.23 	 & 	12,377 			&	4,857    \\ [1ex] 
\hline 
\end{tabular}
\label{table:dual:Margins} 
\end{table}

\begin{figure}
    \centering
    \begin{subfigure}[b]{8cm}
        \includegraphics[width=8cm]{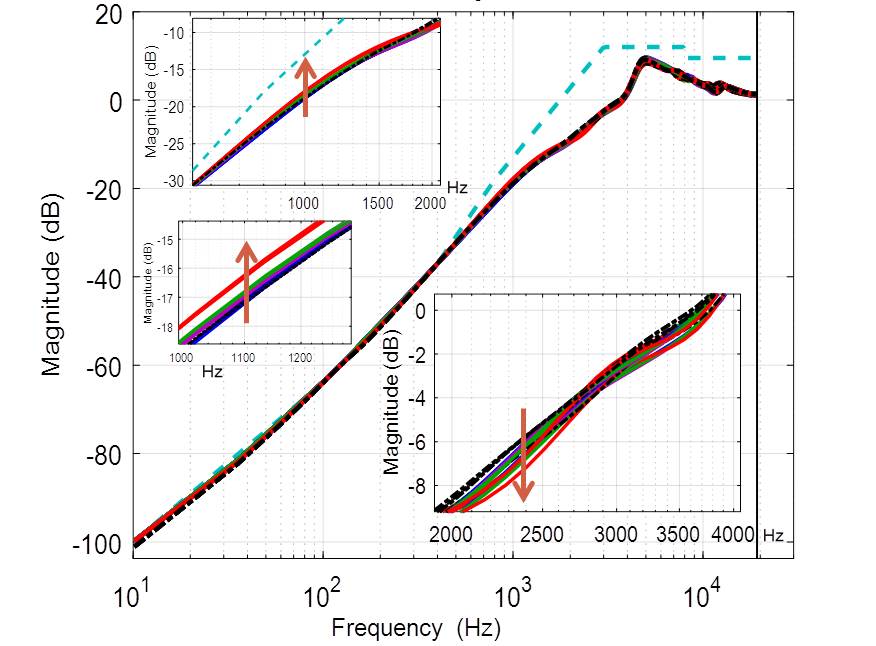}
        \caption{$E_{r \rightarrow e}$}
        \label{fig:dual:Er}
    \end{subfigure}
    \begin{subfigure}[b]{8cm}
        \includegraphics[width=8cm]{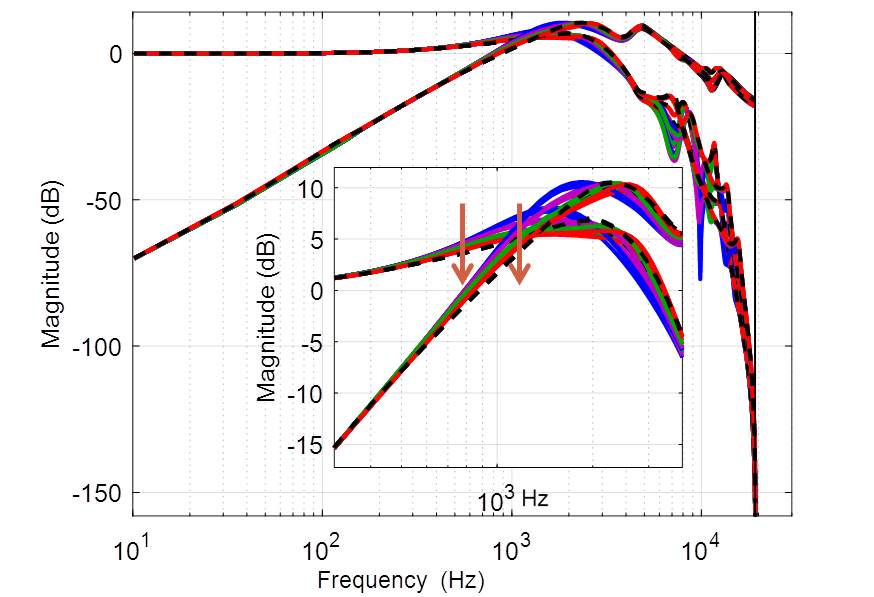}
        \caption{$Y_{r \rightarrow y_v, y_m}$}
        \label{fig:dual:Tr}
    \end{subfigure}
    \caption{The magnitude Bode plots of the closed loop transfer functions for the dual-stage loop. These plots include 25 closed loop transfer functions for all the 5 design scenarios in table~\ref{table:design:scenarios} using all the 5 frequency response data sets in Fig.~\ref{fig:SISO:G}. The arrows show the increasing direction for the scenario index numbers in table~\ref{table:design:scenarios}.}\label{fig:dual:SrTr}
\end{figure}

Fig.~\ref{fig:dual:Tr} plots the frequency responses of the closed loop transfer functions from the track run-out \textit{r} to the actuator outputs $y_v$ and $y_m$. As shown in the figure, the VCM is more active at the low frequency region, while the MA takes over at the mid-frequency region. At the high frequency region, the output of both actuators will reduce, since the system can not be controlled due to the dynamics uncertainty of the actuators.

According to Fig.~\ref{fig:SISO:noises}, the run-out \textit{r} is the dominant external noise and its magnitude has an inverse relationship with frequency. Therefore, a more stringent restriction on the average variance of the MA output stroke defined in Eq. (\ref{eq:H2:mY:dual}) will force the MA output to be more active at higher frequency regions, where the magnitude of run-out is smaller. As shown in Fig.~\ref{fig:dual:Tr}, a controller designed with a more restricted MA output stroke criteria will produce a smaller MA output magnitude in frequencies between $500-2000~Hz$ and a higher MA output magnitude in frequencies between $2000-3000~Hz$. According to Fig.~\ref{fig:dual:Er}, the smaller magnitude of the MA output in lower frequency regions also deteriorates the tracking error rejection in those regions. At higher frequency regions, the larger magnitude of the MA output helps to improve the tracking error rejection.

\begin{figure}
	\centering
	\includegraphics[width=9cm]{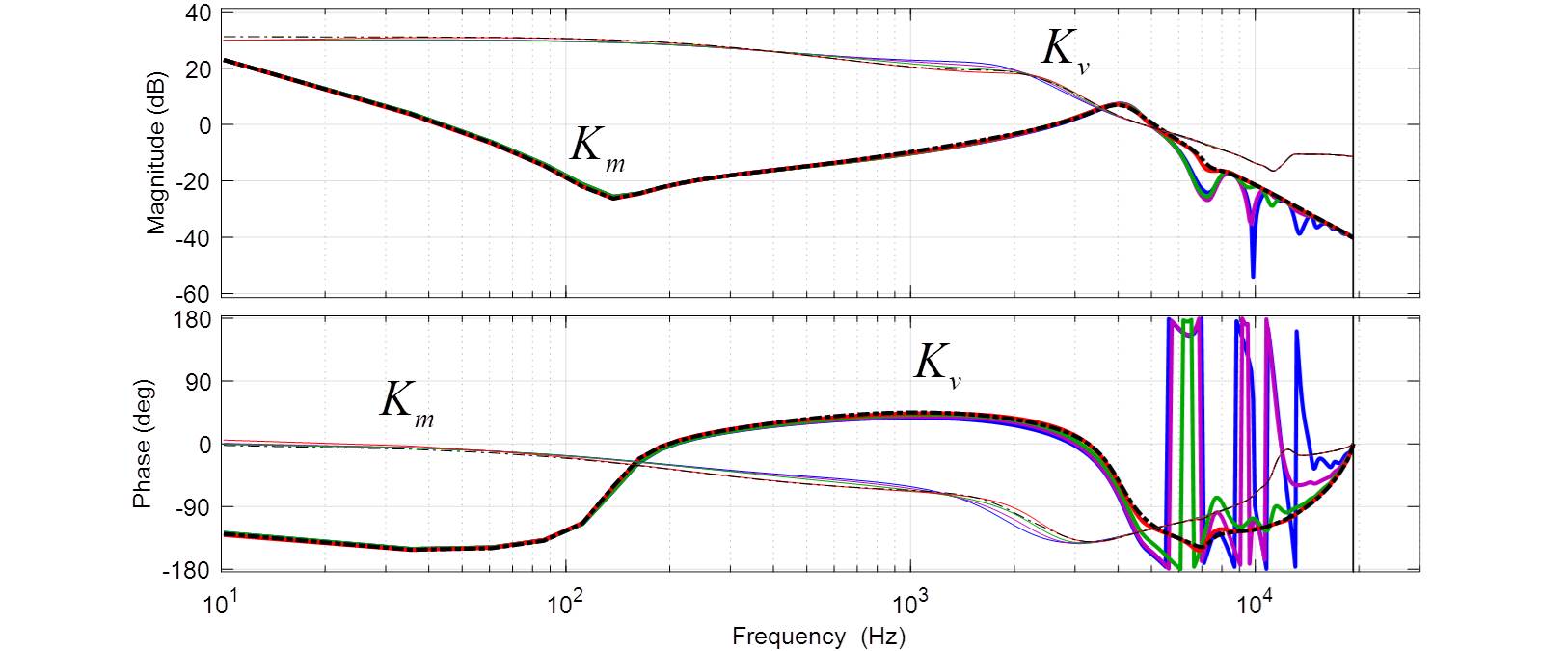}
    \caption{Frequency responses of the compensators $K_v$ and $K_m$ for all the 5 design scenarios in table~\ref{table:design:scenarios} synthesized considering all the 5 plant frequency response data sets plotted in Fig.~\ref{fig:SISO:G}. The frequency responses of the compensator $K_v$ are plotted with thick lines, while the frequency responses of the compensator $K_m$ are plotted with thin lines.}
    \label{fig:dual:K}
\end{figure}

\begin{figure}
    \centering
    \begin{subfigure}[b]{8cm}
        \includegraphics[width=8cm]{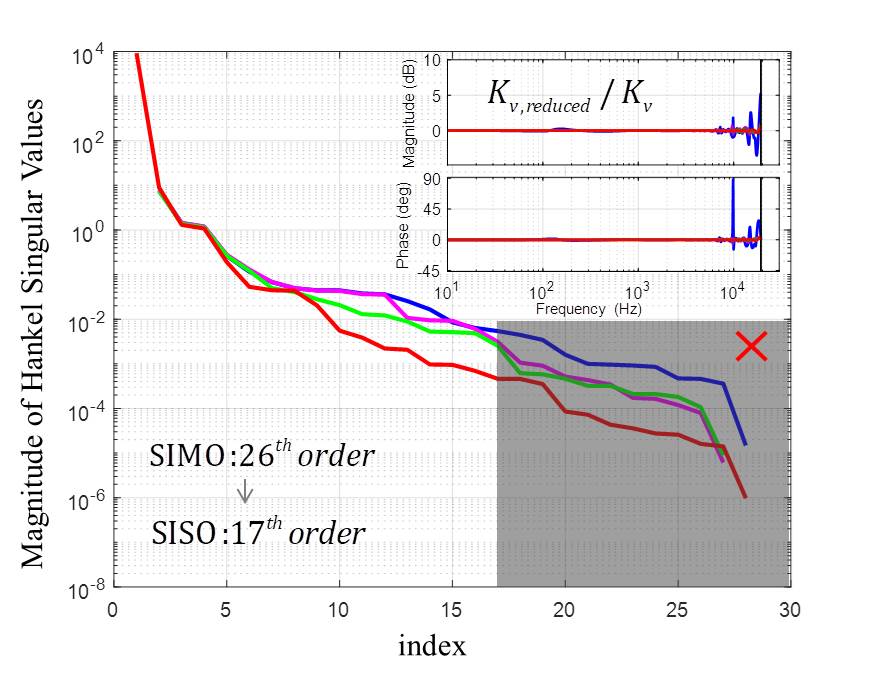}
        \caption{$K_v$}
        \label{fig:dual:KvSigma}
    \end{subfigure}
    \begin{subfigure}[b]{8cm}
        \includegraphics[width=8cm]{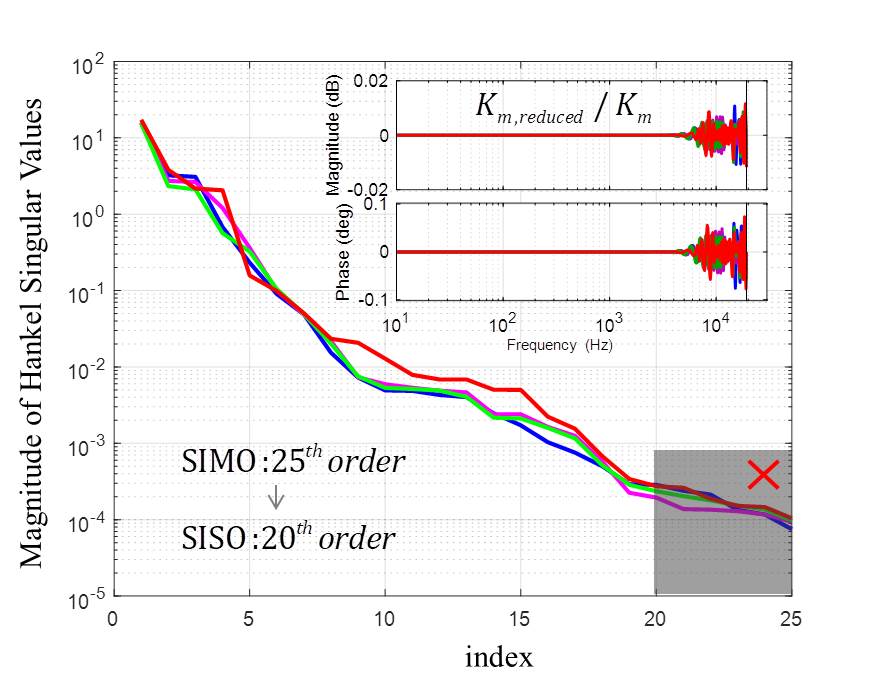}
        \caption{$K_m$}
        \label{fig:dual:KmSigma}
    \end{subfigure}
    \caption{The Hankel singular values for both the $K_v$ and $K_m$ compensators for the design scenarios $SIMO_1$-$SIMO_4$ in table~\ref{table:design:scenarios}. The gray boxes show the singular values which were eliminated in the compensator order reduction. The comparisons between the reduced order and the full order compensators are also provided in these plots.}\label{fig:dual:KSigma}
\end{figure}

The frequency responses of the obtained compensators $K_v$ and $K_m$ are plotted in Fig.~\ref{fig:dual:K} for all the scenarios in table~\ref{table:design:scenarios}. The SIMO design strategy uses the stable factorizations of the controller in Eqs. (\ref{eq:MISO:fact:Xi}) and (\ref{eq:MISO:fact:Yi}) with the controller order parameter $n = 25$ in order to obtain the SIMO controller. After designing the SIMO controller, the compensators $K_v$ and $K_m$ are derived using Eqs. (\ref{eq:miso:KvKbar}) and (\ref{eq:miso:KmKbar}). The Hankel singular values~\cite{hankel2005} of these compensators are plotted in Fig.~\ref{fig:dual:KSigma}. In order to reduce the compensators orders, a few of the Hankel singular values with the smallest magnitudes were eliminated. The gray boxes in Fig.~\ref{fig:dual:KSigma} show the eliminated Hankel singular values. The reduced order compensators $K_v$ and $K_m$ will be $17^{th}$ and $20^{th}$ orders, respectively. The comparisons between the reduced order and the full order compensators are provided in Fig.~\ref{fig:dual:KSigma}. As shown in the figures, the compensator order reduction will not create any significant difference in the low and mid frequency regions of the compensators frequency responses. In high frequency regions, the deviation between the reduced order and the full order VCM compensator $K_v$ will not significantly affect the closed loop transfer functions, since according to Fig.~\ref{fig:SISO:G}, the VCM response has relatively small magnitude at high frequency regions. Moreover, the effect of MA compensator order reduction at high frequency regions is negligible. Therefore, the orders for the compensators $K_v$ and $K_m$ in the sequential SISO design strategy can be reduced to $17^{th}$ and $25^{th}$ orders respectively, which are equal to the orders of the corresponding reduced order compensators in the SIMO design strategy.

\subsubsection{A limitation of the Sequential SISO Design Strategy}\label{sec:Results:Res:Lim}
\begin{figure}
	\centering
	\includegraphics[width=8cm]{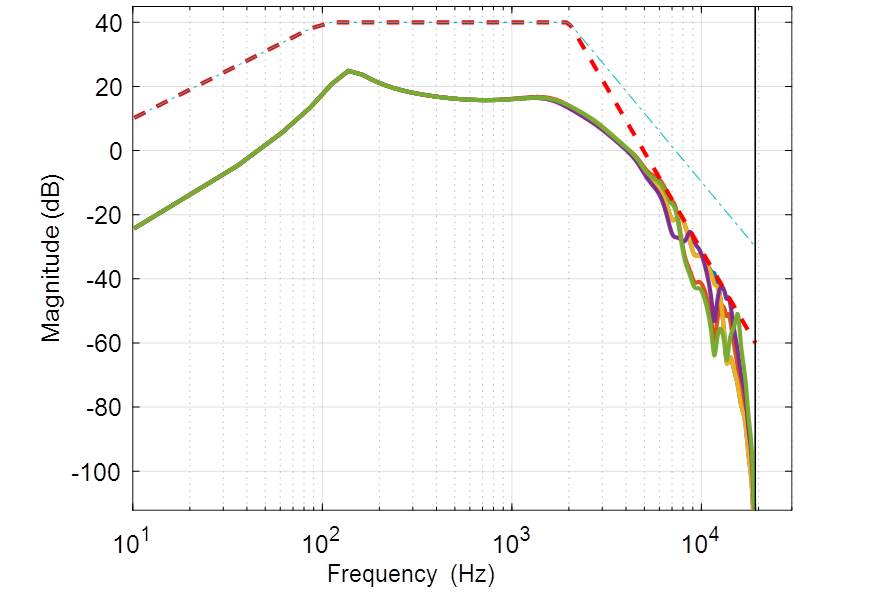}
    \caption{The magnitude Bode plots of the closed loop transfer function from control input disturbance $w_v$ to MA control input $u_m$, $U_{w_v \rightarrow u_m}$. The solid lines show the magnitude Bode plots of the closed loop transfer function $U_{w_v \rightarrow u_m}$ for the design scenario $SIMO_1$ in table~\ref{table:design:scenarios} considering all the 5 plant frequency response data sets in Fig.~\ref{fig:SISO:G}. The red dotted lines represent the more stingest $H_\infty$ constraint limit, while the light blue dotted lines represent the less stingest $H_\infty$ constraint limit used in this section. The SIMO design strategy obtains the controller such that $U_{w_v \rightarrow u_m}$ transfer functions satisfy the more stingest $H_\infty$ constraint limit, while the sequential SISO design strategy cannot find a feasible solution for this $H_\infty$ constraint limit.}
    \label{fig:dual:inf}
\end{figure}

It is worth mentioning that if the SIMO design strategy is successful to find a feasible solution for the optimization problem in Eq. (\ref{eq:opt:miso:II}), there is no guarantee that the sequential SISO design strategy can also find a feasible solution for the optimization problem in Eq. (\ref{eq:opt:sd:dual:II}) by considering the same set of control objectives. As an example, the weighting function used in the $H_\infty$ norm of the closed loop transfer function $U_{w_v \rightarrow u_m}$ was adjusted to force more restrictions at high frequency regions for the scenarios $SISO_1$ and $SIMO_1$ in table~\ref{table:design:scenarios}. In Fig.~\ref{fig:dual:inf}, the less stringent $H_\infty$ constraint limit is shown with light blue dotted lines, while the more stringent $H_\infty$ constraint limit is shown with red dotted lines. The SIMO design strategy is able to synthesize a controller that can satisfy the more stringent constraint. However, the sequential SISO design strategy is not able to find a feasible solution to satisfy this constraint. This result can be partially explained by considering the fact that, in the sequential SISO design strategy the compensator $K_v$ is first synthesized without considering the overall dual-stage loop constraints, and in addition, the compensator $K_m$ is subsequently synthesized without the added flexibility of changing the compensator $K_v$.

\subsubsection{Design Results Comparison Using Two Different Conditions for Imposing $H_\infty$ Constraints}\label{sec:Results:LMIvsIneq}
The results presented in this section, were obtained using theorem~\ref{THEOREMMISOHINF} to prescribe necessary and sufficient convex conditions for imposing the $H_\infty$ constraints considered in the mixed $H_2/H_\infty$ optimization problems given in Eqs. (\ref{eq:opt:sd:single:II})-(\ref{eq:opt:miso:II}). As mentioned in section~\ref{sec:Conrolalg:Hinf}, sufficient convex conditions for imposing the $H_\infty$ constraints were proposed in~\cite{karimi2016H2}. These two types of convex conditions for imposing the $H_\infty$ constraints can be summarized as

\begin{itemize}
\item I) necessary and sufficient convex conditions given in theorem~\ref{THEOREMMISOHINF},
\item II) sufficient convex conditions given in~\cite{karimi2016H2}.
\end{itemize}
The $H_\infty$ sufficient convex conditions in~\cite{karimi2016H2} can be applied to the synthesis of general MIMO compensators, while the $H_\infty$ necessary and sufficient convex conditions in theorem~\ref{THEOREMMISOHINF} is only applied to the synthesis of SIMO compensators.

\begin{figure}
	\centering
	\includegraphics[width=8cm]{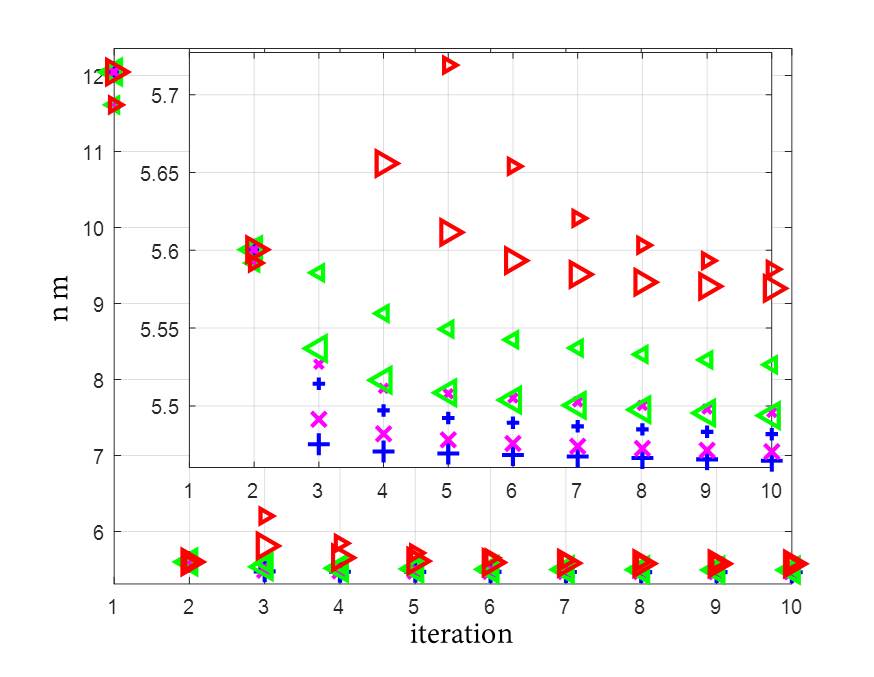}
    \caption{$\sqrt{\frac{1}{l}\sum_{i=1}^{l}\norm{e_i}^2_2}$ for the SIMO feedback system in Fig.~\ref{fig:AppDD:DualSD} as a function of control synthesis iterations, considering the design scenarios $SIMO_1$-$SIMO_4$ in table~\ref{table:design:scenarios}: The average variances of the dual-stage tracking error signal are considered in the optimization problems, however the square roots of the average variances are plotted here. The larger and smaller markers respectively represent conditions I and II as the convex conditions for imposing the $H_\infty$ constraints used in Eq.~(\ref{eq:opt:miso:II}). The $H_2$ norm objective and constraints are considered in the optimization problem starting from the second and third iterations, respectively.}
    \label{fig:HDDAPP:LMIvsIneq}
\end{figure}

In order to compare these two convex conditions for imposing the $H_\infty$ constraints, the SIMO dual-stage controller in Fig.~\ref{fig:AppDD:DualSD} that was synthesizes by solving the mixed $H_2/H_\infty$ control design problem given in Eq.~(\ref{eq:opt:miso:II}), for all the scenarios in table~\ref{table:design:scenarios} using the SIMO design strategy were redesigned using the sufficient convex conditions II instead of using the necessary and sufficient convex condition I, and its performance was compared to the compensator previously designed using condition I. All the designed compensators were able to satisfy the $H_\infty$ and $H_2$ norm constraints. The resulting $H_2$ norm objectives for these two conditions are compared in Fig.~\ref{fig:HDDAPP:LMIvsIneq}, where the circle ($\circ$) and cross ($\times$) marks respectively represent conditions I and II as the convex conditions for imposing the $H_\infty$ constraints. As shown in the figure, a controller designed utilizing the sufficient $H_\infty$ conditions II will produce a larger cost when compared with the cost produced by a controller designed using the necessary and sufficient $H_\infty$ conditions I under the same design scenarios and the same plant data sets. The degraded performance for condition II can be justified by the fact that having only sufficient conditions for imposing the $H_\infty$ constraints introduces conservatism in the optimization problem and leads to an increase in the minimization objective as compared to the necessary and sufficient conditions.

%% file: 6-Conclusion.tex
The frequency based data-driven mixed $H_2/H_\infty$ control design algorithm was studied in order to design feedback loops. The data-driven control design algorithm directly uses the frequency response measurements of the plant in the control design step, rather than fitting an estimated model to those measurements. Therefore, the obtained controller can guarantee the stability and performance level achieved in the design step, if adequate number of frequency response measurements are considered in the design step to represent system dynamics variations.

The data-driven mixed $H_2/H_\infty$ control design problem was converted to a convex local optimization problem, which was solved iteratively. In the proposed algorithm, the $H_2$ and $H_\infty$ norms of the closed loop transfer functions had the flexibility to be considered as the constraints and/or the objective of the optimization problem. The $H_\infty$ norm criteria were used for guaranteeing the closed loop stability and shaping the closed loop transfer functions. The $H_2$ norm criteria were used for constraining or minimizing the variance of signals in the time domain, since $H_2$ norms of transfer functions in the frequency domain are equivalent to the square root of their corresponding signals variances in the time domain.

The necessary and sufficient convex conditions of the $H_\infty$ norm control problem for SISO systems are obtained in~\cite{karimi2016HinfSISO}. These results were extended to MISO systems, where the obtained controller stabilized the given MISO system and satisfied the defined $H_\infty$ constraints. These results were combined with $H_2$ results in~\cite{karimi2016H2} in order to form the mixed $H_2/H_\infty$ control problem. The $H_2$ and $H_\infty$ norms used in this algorithm were defined such that multiple sets of plant measurements could be considered in the design process.

The proposed data-driven mixed $H_2/H_\infty$ control methodology was used to design a track following controller for a dual-stage HDD. The sensitivity decoupling approach was considered as the control structure~\cite{SD_2007}. The controllers in this structure were obtained using either sequential SISO or SIMO data-driven control design strategies. In the sequential SISO strategy, the control problem was decoupled into two SISO problems, and the controller for each actuator was obtained in one individual step. In the SIMO strategy, the complete control block was obtained in one step. It is worth mentioning that the dual-stage controller should be designed such that in the case of MA failure, the single-stage loop remains stable and satisfies predefined performance characteristics. The single-stage and dual-stage stability and performance characteristics were considered together as the constraints and the objectives of the optimization problem.

The dual-stage HDD controller was designed considering the set of five frequency response plant measurement data sets. The closed loop transfer functions for all these data sets were shaped using the weighted $H_\infty$ norm constraints. Since the $H_2$ norms of closed loop transfer functions are directly related to the square roots of variances for the corresponding signals in the time domain, the $H_2$ norm objective and constraints were imposed using the variances of closed loop signals, which were averaged among the set of five frequency response plant measurement data sets. The average variance of the tracking error was considered as the minimization objective, while the VCM control input and the MA output stroke average variances were constrained.

The design results demonstrated that the data driven mixed $H_2/H_\infty$ control design algorithm was successful in satisfying the defined mixed $H_2/H_\infty$ control objectives by designing controllers, which stabilized both the single-stage and the dual-stage loops. Considering the set of five frequency response plant measurement data sets, the SIMO design strategy achieved a smaller average variance of the dual-stage tracking error as well as a higher worst case bandwidth as compared to the sequential SISO design strategy, since it designed the controllers for the VCM and MA simultaneously. Also, the more stringent restriction on the MA output stroke in the SIMO design strategy compromised the worst case dual-stage bandwidth as well as the average variance of the dual-stage tracking error.

%% file: 7-Appendix/7-Appendix_Proof.tex
Theorem~\ref{THEOREMMISOHINF} is stated and proved for SISO systems in \cite{karimi2016HinfSISO}. In this section, the proof presented in \cite{karimi2016HinfSISO} is extended to MISO systems. In order to prove theorem~\ref{THEOREMMISOHINF}, Lemma~\ref{lemma:miso:ineq} is first proved~\cite{rantzer1994convex}.

All transfer functions are considered in the discrete time frequency domain. However to make the notations simple, the frequency dependence arguments $(e^{j\omega})$ and $(\omega)$ will not be written in the transfer functions.


\begin{lemma}\label{lemma:miso:ineq}
Assume the transfer function $W_{U_{w \rightarrow u}} U_{w \rightarrow u}$ used in theorem~\ref{THEOREMMISOHINF} is bounded and analytic in the right half plane, where $U_{w \rightarrow u}$ is written in terms of plant and controller stable factorizations in Eq.~(\ref{eq:MISO:UwDefinition}). The $H_\infty$ norm defined in Eq.~(\ref{eq:MISO:theorem:Hinf}) is satisfied if and only if the following inequality holds over the entire frequency region, $\Omega$,
\begin{equation}\label{eq:MISO:lemma:Ineq}
\forall \omega \in \Omega:~\gamma^{-1} \bar{\sigma}( W_{U_{\omega}}X\tilde{N}F )<Re( \tilde{N}XF + \tilde{M}YF )
\end{equation}
where $F\in \mathbb{RH}_\infty$ is a stable proper rational scalar transfer function and $\bar{\sigma(M)}$ denotes the maximum singular value of the matrix $M$.
\end{lemma}

\subsection{Proof of Lemma~\ref{lemma:miso:ineq}}
\begin{proof}
Considering the definition of $H_\infty$ norm in Eq.~(\ref{eq:intro:MISO:Hinf_def}), the weighted $H_\infty$ norm criterion given in Eq.~(\ref{eq:MISO:theorem:Hinf}) can be written as
\begin{equation}
\forall \omega \in \Omega:~\bar{\sigma}( W_{U_{w \rightarrow u}}X\tilde{N}( \tilde{N}X + \tilde{M}Y )^{-1} ) < \gamma.
\end{equation}
According to the dimensions of plant and controller stable factorizations given in Eqs.~(\ref{eq:MISO:G_parameterization}) and (\ref{eq:MISO:K_parameterization}), the term $\tilde{N}X + \tilde{M}Y$ will be a scalar. Therefore, the above equation can be simplified as follows
\begin{equation}\label{eq:MISO:lemma:Ineq_abs}
\forall \omega \in \Omega:~\gamma^{-1}\bar{\sigma}( W_{U_{w \rightarrow u}}X\tilde{N} ) < \abs{ \tilde{N}X + \tilde{M}Y }.
\end{equation}
where $\abs{G}=\abs{G(\omega)}$ denotes the magnitude of $G(\omega)$.

At each given frequency $\omega \in \Omega$, a disk centered at $z_0 = \tilde{N}X + \tilde{M}Y$ with radius $r=\gamma^{-1}\bar{\sigma}( W_{U_{w \rightarrow u}}X\tilde{N} )$ will be considered. This disk is shown in Fig.~\ref{fig:MISO:LemmaProof},
\begin{figure}
	\centering
	\includegraphics[width=8cm]{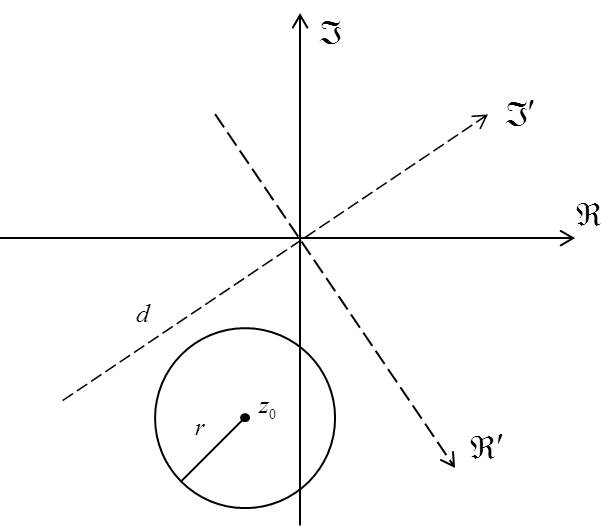}
	\caption{$\mathbb{R}$ and $\mathbb{I}$ represent the real and imaginary axes, respectively. The prime notation is used to represent the rotated axes.}
	\label{fig:MISO:LemmaProof}
\end{figure}
where all the points inside this disk will be represented as
\begin{equation}\label{eq:MISO:lemma:disk}
\abs{z-z_0} < r
\end{equation}

By combining Eqs.~(\ref{eq:MISO:lemma:Ineq_abs}) and (\ref{eq:MISO:lemma:disk}), it can be concluded that the origin $z = (0, 0)$ is not located inside this disk. Therefore, as one can see in Fig.~\ref{fig:MISO:LemmaProof}, there exists a line $d$ passing through the origin that does not intersect the disk. A unitary rotation matrix $f(j\omega)$ will be considered which makes line $d$ the new imaginary axis. The center and radius of the disk in the new rotated system will be given by
\begin{eqnarray}
z^f_0 = \tilde{N}Xf + \tilde{M}Yf\label{eq:MISO:lemma:z0_rotated},\\
r^f = \gamma^{-1}\bar{\sigma}( W_{U_{\omega}}X\tilde{N}f).\label{eq:MISO:lemma:r_rotated}
\end{eqnarray}
Since line $d$ is the new imaginary axis and is not intersecting with the disk, the real component of the disk center must be always greater than the disk radius
\begin{equation}\label{eq:MISO:lemma:r_l_Rez0}
r^f < Re( z^f_0 ).
\end{equation}
Plugging in Eqs.~(\ref{eq:MISO:lemma:z0_rotated}) and (\ref{eq:MISO:lemma:r_rotated}) into Eq.~(\ref{eq:MISO:lemma:r_l_Rez0}) will result in the following inequality
\begin{equation}\label{eq:MISO:lemma:Ineqfjw}
\forall \omega \in \Omega:~\gamma^{-1}\bar{\sigma}( W_{U_{w \rightarrow u}}X\tilde{N}f )<Re( \tilde{N}Xf + \tilde{M}Yf ).
\end{equation}
According to \cite{rantzer1994convex}, $f$ can be approximated by the frequency response data of a stable proper rational scalar transfer function $F$, if and only if
\begin{equation}
V = \frac{1}{ \abs{\tilde{N}X + \tilde{M}Y}-\gamma^{-1}\bar{\sigma}(W_{U_{w \rightarrow u}}X\tilde{N}) }
\end{equation}
is analytic in the right half plane for all $\bar{\gamma}>\gamma$. In order to prove that $V$ is analytic in the right half plane, we first assume $\bar{\gamma} \to \infty$. Therefore, $V = \frac{1}{ \abs{\tilde{N}.X + \tilde{M}.Y} }$ is stable and analytic according to stability of $U_{w \rightarrow u}$. If $\bar{\gamma}$ is reduced from infinity to $\gamma$, the poles of $V$ will move continuously, and according to Eq. (\ref{eq:MISO:lemma:Ineq_abs}), it can be shown that
\begin{equation}
\forall \omega \in \Omega: V^{-1}(j\omega)\neq 0.
\end{equation}
Therefore, the function $V$ will remain analytic in the right half plane for all $\bar{\gamma}>\gamma$. As a result, the rotation matrix $f(j\omega)$ can be estimated by a stable proper rational scalar transfer function $F$, which will prove the inequality given in Eq.~(\ref{eq:MISO:lemma:Ineq}).
\end{proof}

\subsection{Proof of Theorem~\ref{THEOREMMISOHINF}}

\begin{proof}
The equivalence of statements $I$ and $II$ presented in theorem~\ref{THEOREMMISOHINF} will be proved here.
\begin{itemize}
\item ($I \Rightarrow II$) Assume the stabilizing controller is factorized as $K=X_0Y_0^{-1}$. According to Lemma~\ref{lemma:miso:ineq}, the $H_\infty$ norm criterion given in Eq.~(\ref{eq:MISO:theorem:Hinf}) can be written as
\begin{equation}\label{eq:MISO:theorem:proofItoII}
\forall \omega \in \Omega:  \gamma^{-1}\bar{\sigma}( W_{U_{w \rightarrow u}}X_0\tilde{N}F )<Re( \tilde{N}X_0F + \tilde{M}Y_0F ).
\end{equation}
Since $F$ is a stable proper rational scalar transfer function, it can be merged inside the controller factorizations such that $X=X_0F$ and $Y=Y_0F$. This merge will not change the controller and $K=XY^{-1}=X_0Y_0^{-1}$. Therefore, Eq.~(\ref{eq:MISO:theorem:proofItoII}) will result in Eq.~(\ref{eq:MISO:theorem:II}) given in statement $II$ of the theorem.

\item ($II \Rightarrow I$) The real part of any complex number is always less than its magnitude
\begin{equation}
Re(\ \tilde{N}X + \tilde{M}Y\ ) < \abs{\ \tilde{N}X + \tilde{M}Y\ }.
\end{equation}
Therefore, Eq.~(\ref{eq:MISO:theorem:II}) will result in the following equation
\begin{equation}\label{eq:MISO:theorem:proof:inf_l_abs}
\forall \omega \in \Omega :~ \gamma^{-1} \bar{\sigma}(W_{U_{\omega}}X\tilde{N}) < \abs{\ \tilde{N}X + \tilde{M}Y\ }.
\end{equation}
Since $\tilde{N}X + \tilde{M}Y$ is a scalar term, Eq.~(\ref{eq:MISO:theorem:proof:inf_l_abs}) can be written as
\begin{equation}\label{eq:MISO:theorem:proof:Hinf_l_gamma}
\forall \omega \in \Omega :~ \bar{\sigma}(W_{U_{w \rightarrow u}}X\tilde{N}(\tilde{N}X + \tilde{M}Y)^{-1}) < \gamma
\end{equation}
According to the definition of the weighted $H_\infty$ norm given in Eq. (\ref{eq:intro:MISO:Hinf_def}), the above equation is basically equivalent to the $H_\infty$ criterion given in Eq.~(\ref{eq:MISO:theorem:Hinf}).

Now, we have to show that the controller $K$ stabilizes the plant $G$. The stability is analyzed using the Nyquist stability theorem~\cite{robustbook1996}. According to Eq. (\ref{eq:MISO:cl:MN}), the closed loop transfer functions can be written as follows
\begin{equation}\label{eq:MISO:cl:MN_D}
\begin{bmatrix}
E_{r \rightarrow e}            & E_{n \rightarrow e}               & E_{w \rightarrow e}      \\
U_{r \rightarrow u}     		 & U_{n \rightarrow u} 		       & U_{w \rightarrow u}      \\
Y_{r \rightarrow y}     		 & Y_{n \rightarrow y} 		       & Y_{w \rightarrow y}
\end{bmatrix}
=\frac{1}{D}
\begin{bmatrix}
\tilde{M}Y      & -X\tilde{N}        & -\tilde{N}Y \\
\tilde{M}X      &  \tilde{M}X   	   & -\tilde{N}X \\
\tilde{N}X		 &  \tilde{N}X		   &  \tilde{N}Y
\end{bmatrix},
\end{equation}
where 
\begin{equation}\label{eq:MISO:D}
D = \tilde{N}X + \tilde{M}Y.
\end{equation}

Since all the elements of the numerator matrix in Eq. (\ref{eq:MISO:cl:MN_D}) are stable factorizations of the plant and controller, they will have stable poles. Therefore, the closed loop systems in Eq. (\ref{eq:MISO:cl:MN_D}) will be stable if all the zeros of $D$ are stable.

In order to prove that the zeros of $D$ are stable, the Nyquist plot for $D$ will be considered. Since $D$ is a linear function of stable factorizations, it will not have any unstable poles. Moreover, Eq.~(\ref{eq:MISO:theorem:II}) will result in
\begin{equation}
Re(D(\omega))>0,
\end{equation}
which means that $D$ will not encircle around the origin. As a result, the Nyquist stability theorem will conclude that $D$ will not have any unstable zeros. Therefore, all the closed loop systems in Eq. (\ref{eq:MISO:cl:MN_D}) are stable and the controller $K$ stabilizes the plant $G$.
\end{itemize}
\end{proof}

%% file: 7-Appendix/7-Appendix_Hinf.tex

The closed loop transfer functions considered in the $H_\infty$ norm constraints given in Eqs. (\ref{eq:Hinf:single}), (\ref{eq:Hinf:dual:overall}), (\ref{eq:Hinf:dual:individual:E}) and (\ref{eq:Hinf:dual:individual:U}) are plotted in this section. These plots include the closed loop transfer functions, as well as the weighting functions used to shape those transfer functions. The closed loop transfer functions are obtained for all the design scenarios in table~\ref{table:design:scenarios} using all the frequency response data sets in Fig.~\ref{fig:SISO:G}. These plots utilize the color code in table~\ref{table:design:scenarios} to distinguish between different scenarios. However, the plots for the same scenario but different frequency response data sets utilize the same marker type and may not be distinguishable from each other at some frequency regions, where the plots are relatively close to each other. These plots include 25 closed loop transfer functions for all the 5 design scenarios in table~\ref{table:design:scenarios} using all the 5 frequency response data sets in Fig.~\ref{fig:SISO:G}.

\begin{figure}
    \centering
    \begin{subfigure}[b]{7cm}
        \includegraphics[width=7cm]{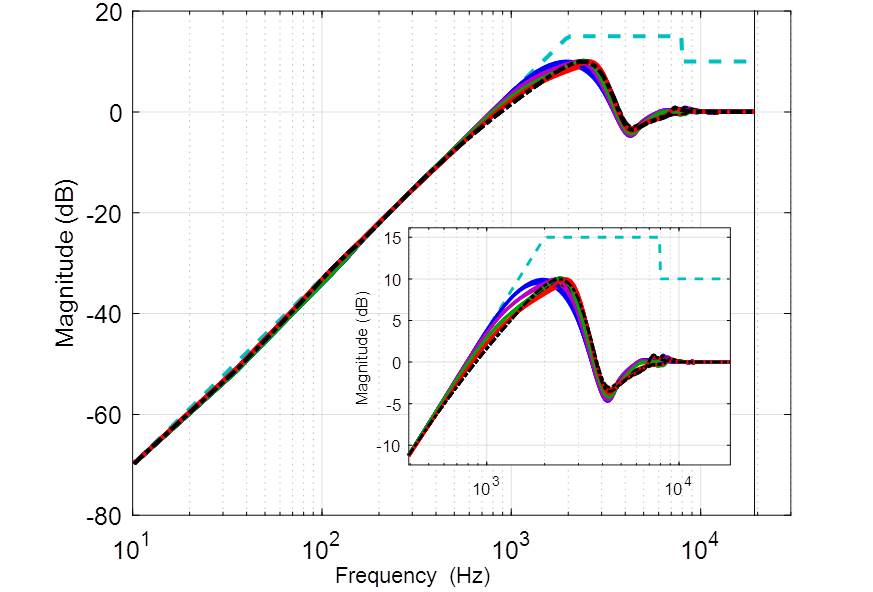}
        \caption{$E^s_{r \rightarrow e}$, $\abs{W_{E_{r \rightarrow e}}}^{-1}$}
        \label{fig:single:Sr}
    \end{subfigure}
    ~ 
    \begin{subfigure}[b]{7cm}
        \includegraphics[width=7cm]{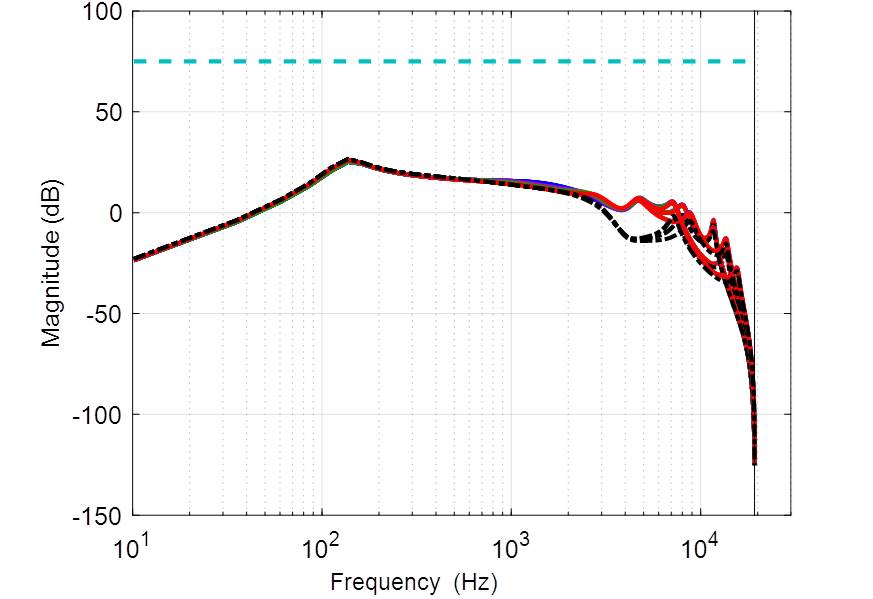}
        \caption{$E^s_{w_v \rightarrow e}$, $\abs{W_{E^s_{w_v \rightarrow e}}}^{-1}$}
        \label{fig:single:Sw}
    \end{subfigure}
    ~ 
    \begin{subfigure}[b]{7cm}
        \includegraphics[width=7cm]{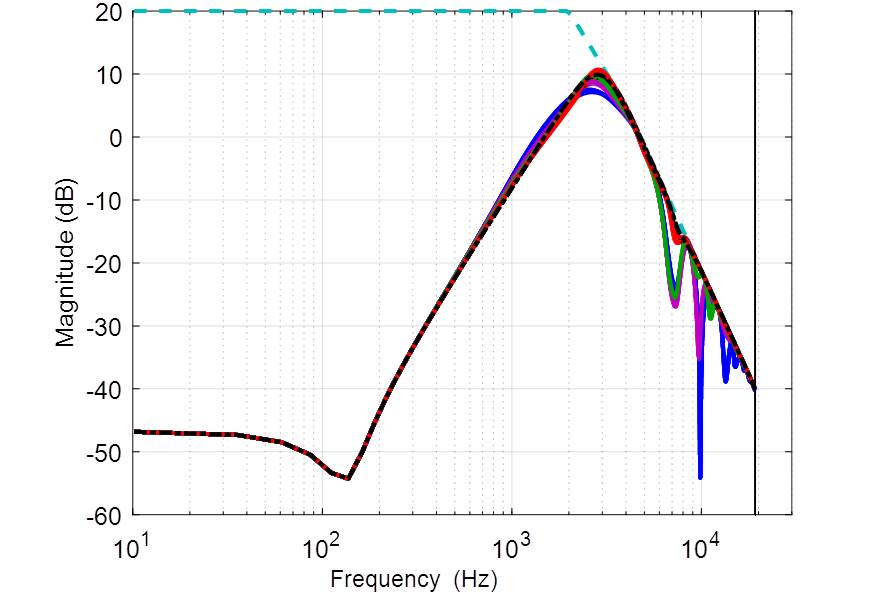}
        \caption{$U^s_{r \rightarrow u_v}$, $\abs{W_{U^s_{r \rightarrow u_v}}}^{-1}$}
        \label{fig:single:Ur}
    \end{subfigure}
    ~ 
    \begin{subfigure}[b]{7cm}
        \includegraphics[width=7cm]{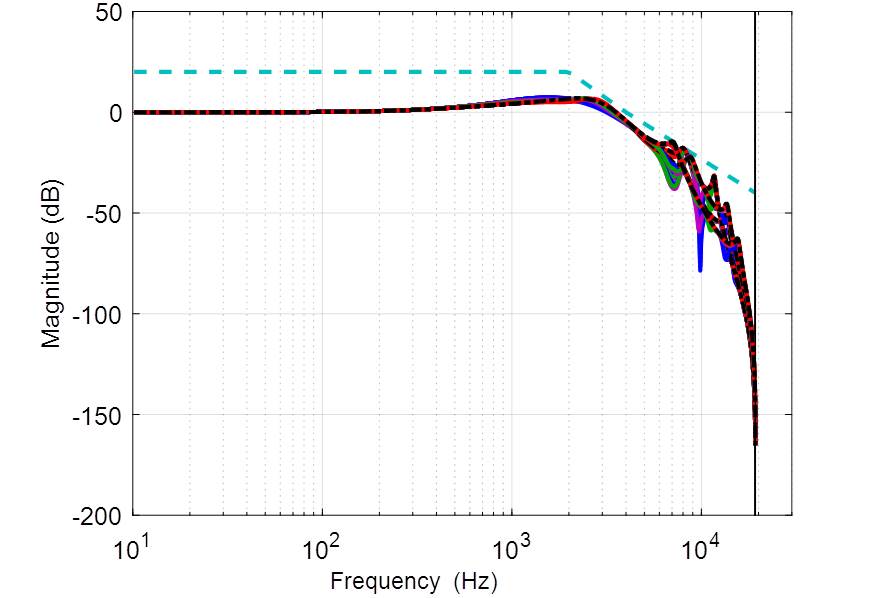}
        \caption{$U^s_{w_v \rightarrow u_v}$, $\abs{W_{U^s_{w_v \rightarrow u_v}}}^{-1}$}
        \label{fig:single:Uw}
    \end{subfigure}
    \caption{The magnitude Bode plots of the single-stage closed loop transfer functions ($H$) and the inverse of the $H_\infty$ weighting functions magnitudes ($\abs{W_H}^{-1}$). The $\abs{W_H}^{-1}$ functions are shown with the light blue dotted lines. The $H_\infty$ norm criteria are provided in Eq. (\ref{eq:Hinf:single}).}\label{fig:single:Hinf}
\end{figure}

\begin{figure}
    \centering
    \begin{subfigure}[b]{7cm}
        \includegraphics[width=7cm]{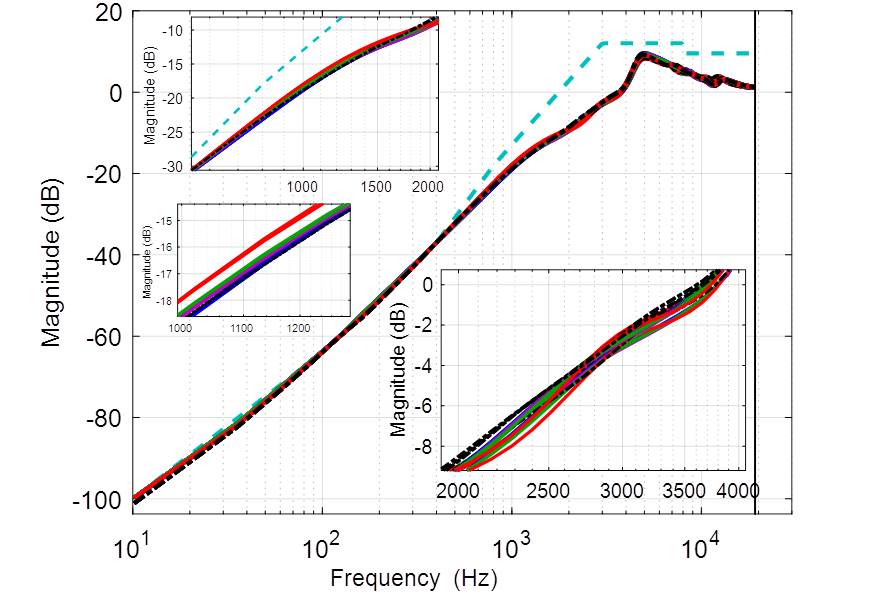}
        \caption[.]{$\bar{\sigma}(E_{r \rightarrow e})$, $\underline{\sigma}^{-1}(W_{E_{r \rightarrow e}})$}
        \label{fig:dual:Sr_limit}
    \end{subfigure}
    \begin{subfigure}[b]{7cm}
        \includegraphics[width=7cm]{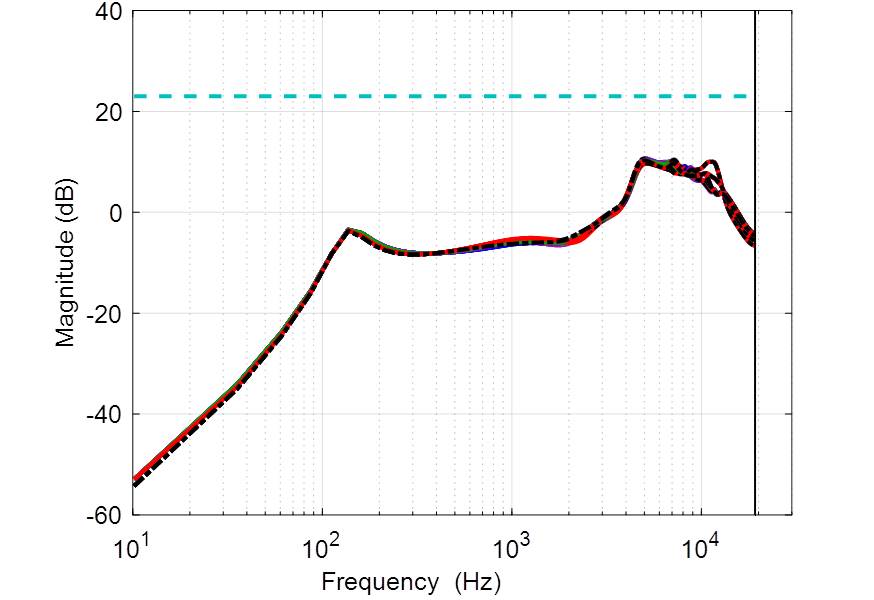}
        \caption[.]{$\bar{\sigma}(E_{w \rightarrow e})$, $\underline{\sigma}^{-1}(W_{E_{w \rightarrow e}})$}
        \label{fig:dual:Sw_limit}
    \end{subfigure}
    \begin{subfigure}[b]{7cm}
        \includegraphics[width=7cm]{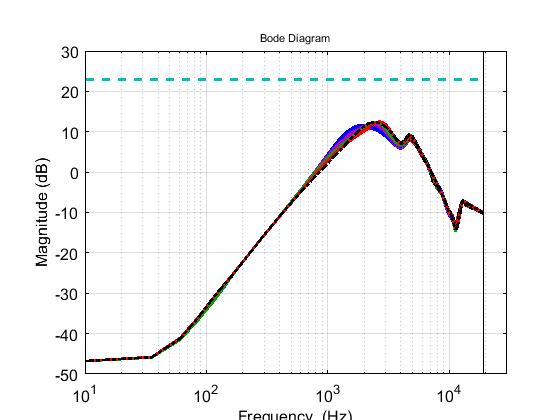}
        \caption[.]{$\bar{\sigma}(U_{r \rightarrow u})$, $\underline{\sigma}^{-1}(W_{U_{r \rightarrow u}})$}
        \label{fig:dual:Ur_limit}
    \end{subfigure}
    \begin{subfigure}[b]{7cm}
        \includegraphics[width=7cm]{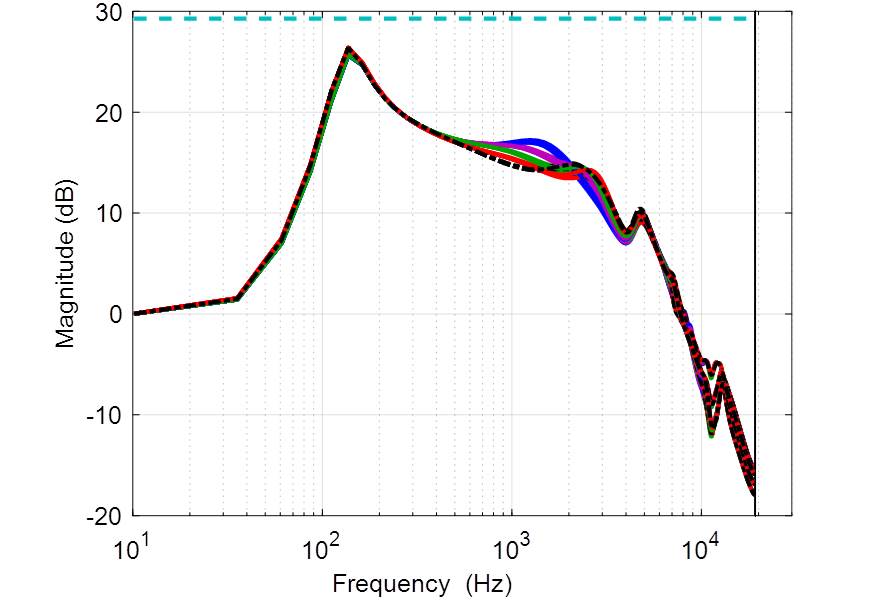}
        \caption[.]{$\bar{\sigma}(U_{w \rightarrow u})$, $\underline{\sigma}^{-1}(W_{U_{w \rightarrow u}})$}
        \label{fig:dual:Uw_limit}
    \end{subfigure}
    \caption[.]{The maximum singular values of the dual-stage closed loop transfer functions $H$ shown as $\bar{\sigma}(H)$ and the inverse of the minimum singular values of the $H_\infty$ weighting functions ($\underline{\sigma}^{-1}(W_H)$). The $\underline{\sigma}^{-1}(W_H)$ functions are shown with the light blue dotted lines. The $H_\infty$ norm criteria are provided in Eq. (\ref{eq:Hinf:dual:overall}).}\label{fig:dual:Hinf_limit}
\end{figure}

\begin{figure}
    \centering
    \begin{subfigure}[b]{7cm}
        \includegraphics[width=7cm]{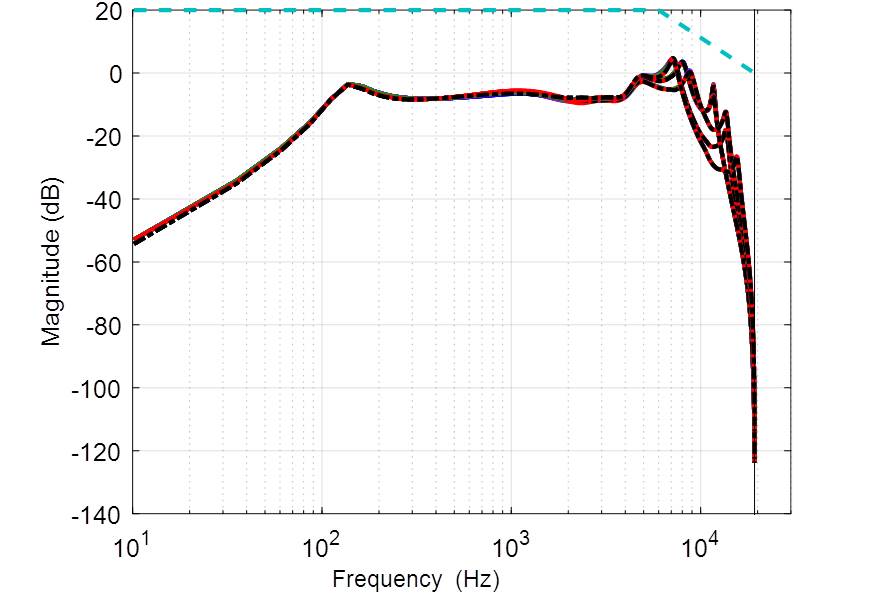}
        \caption[.]{$E_{w_v \rightarrow e}$, $\abs{W_{E_{w_v \rightarrow e}}}^{-1}$}
        \label{fig:indv:Srv}
    \end{subfigure}
    \begin{subfigure}[b]{7cm}
        \includegraphics[width=7cm]{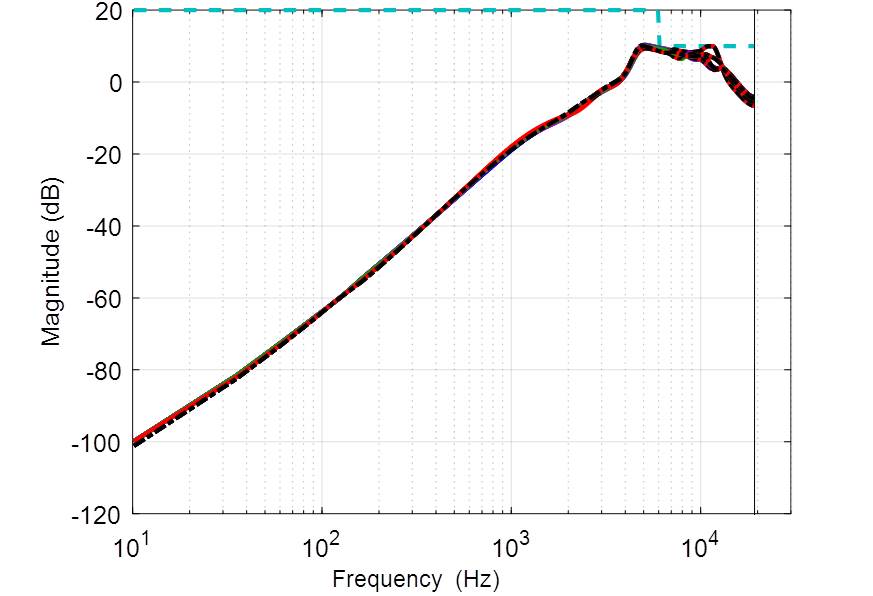}
        \caption[.]{$E_{w_m \rightarrow e}$, $\abs{W_{E_{w_m \rightarrow e}}}^{-1}$}
        \label{fig:indv:Srm}
    \end{subfigure}
    \caption{The magnitude Bode plots of the individual SISO closed loop transfer functions to the tracking error in the dual-stage HDD ($H$) and the inverse of their $H_\infty$ weighting functions magnitudes ($\abs{W_H}^{-1}$). The $\abs{W_H}^{-1}$ functions are shown with the light blue dotted lines. The $H_\infty$ norm criteria are provided in Eq. (\ref{eq:Hinf:dual:individual:E}).}\label{fig:indv:Sr}
\end{figure}

\begin{figure}
    \centering
    \begin{subfigure}{7cm}
        \includegraphics[width=7cm]{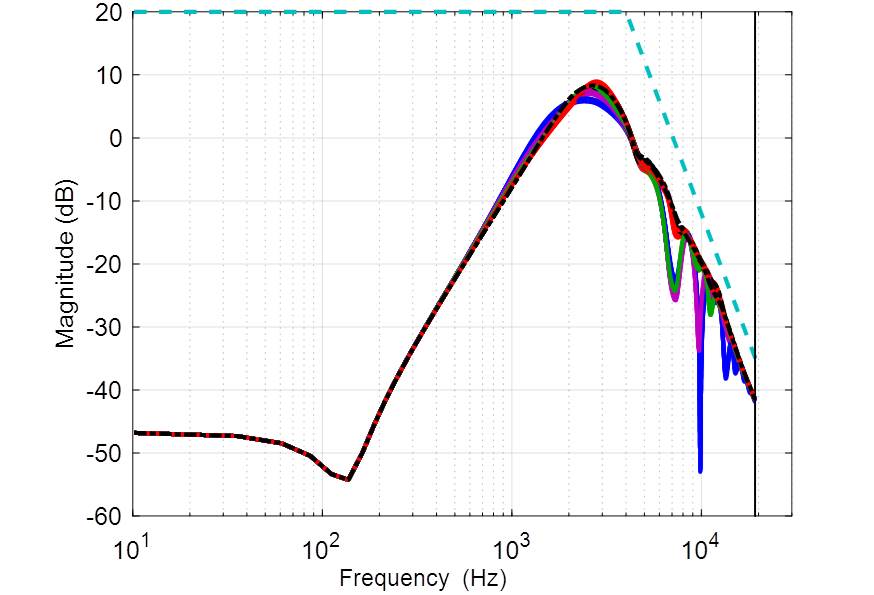}
        \caption[.]{$U_{r \rightarrow u_v}$, $\abs{W_{U_{r \rightarrow u_v}}}^{-1}$}
        \label{fig:indv:vUr}
    \end{subfigure}
    \begin{subfigure}{7cm}
        \includegraphics[width=7cm]{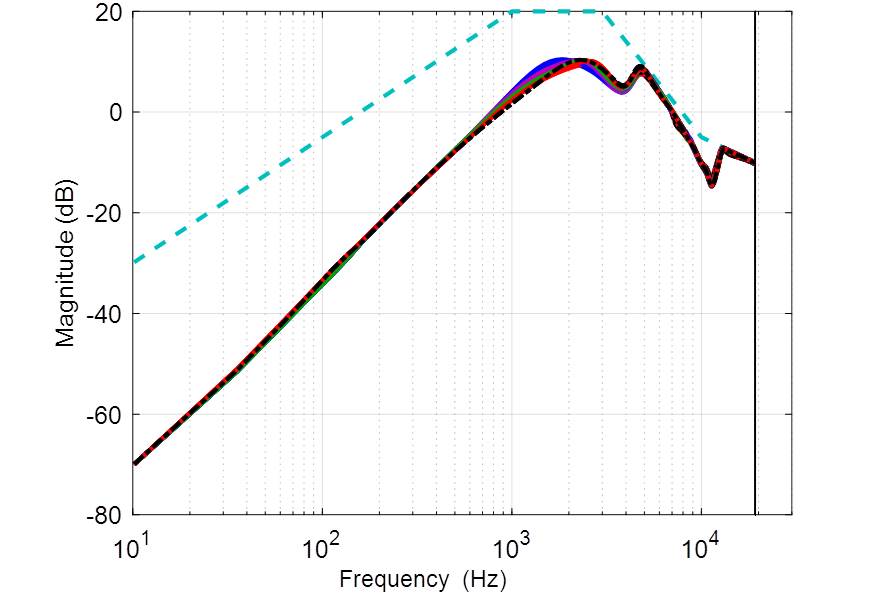}
        \caption[.]{$U_{r \rightarrow u_m}$, $\abs{W_{U_{r \rightarrow u_m}}}^{-1}$}
        \label{fig:indv:mUr}
    \end{subfigure}
        \caption{The magnitude Bode plots of the individual SISO closed loop transfer functions from track run-out $r$ to the control inputs in the dual-stage HDD ($H$) and the inverse of their $H_\infty$ weighting functions magnitudes ($\abs{W_H}^{-1}$). The $\abs{W_H}^{-1}$ functions are shown with the light blue dotted lines. The $H_\infty$ norm criteria are provided in Eq. (\ref{eq:Hinf:dual:individual:U}).}\label{fig:indv:Ur}
\end{figure}

In the case of SISO transfer functions, the inverse of the weighting functions magnitude shapes the magnitude of the closed loop transfer functions. The inverse of weighting functions for the single-stage loop as well as individual SISO loops in the dual-stage loop are shown in Figs.~\ref{fig:single:Hinf}, \ref{fig:dual:Sr_limit}, \ref{fig:indv:Sr} and \ref{fig:indv:Ur} with light blue dotted lines. The closed loop transfer functions are also shown with their corresponding color code mentioned in table~\ref{table:design:scenarios}.

\begin{figure}
    \centering
    \begin{subfigure}[b]{7cm}
        \includegraphics[width=7cm]{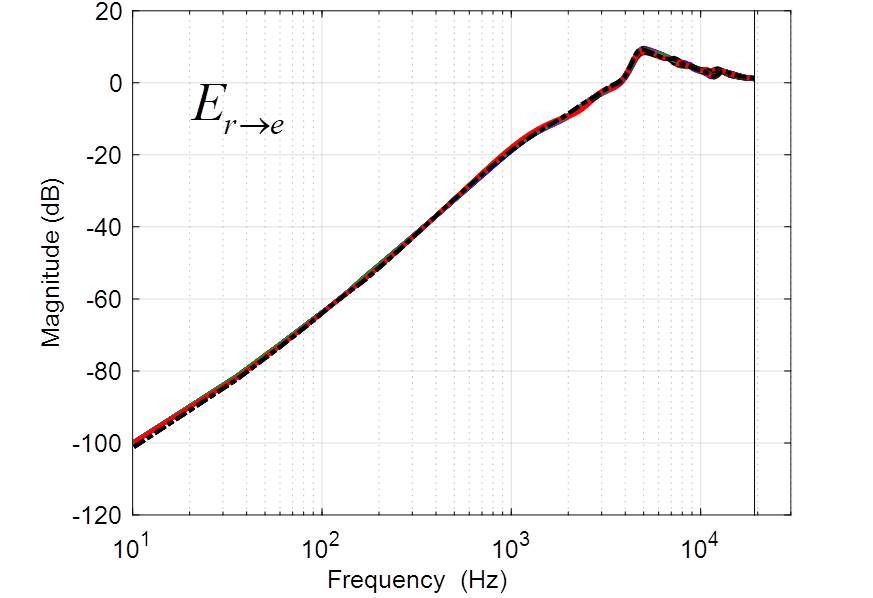}
        \caption{$E_{r \rightarrow e}$}
        \label{fig:dual:Sr}
    \end{subfigure}
    ~ 
    \begin{subfigure}[b]{7cm}
        \includegraphics[width=7cm]{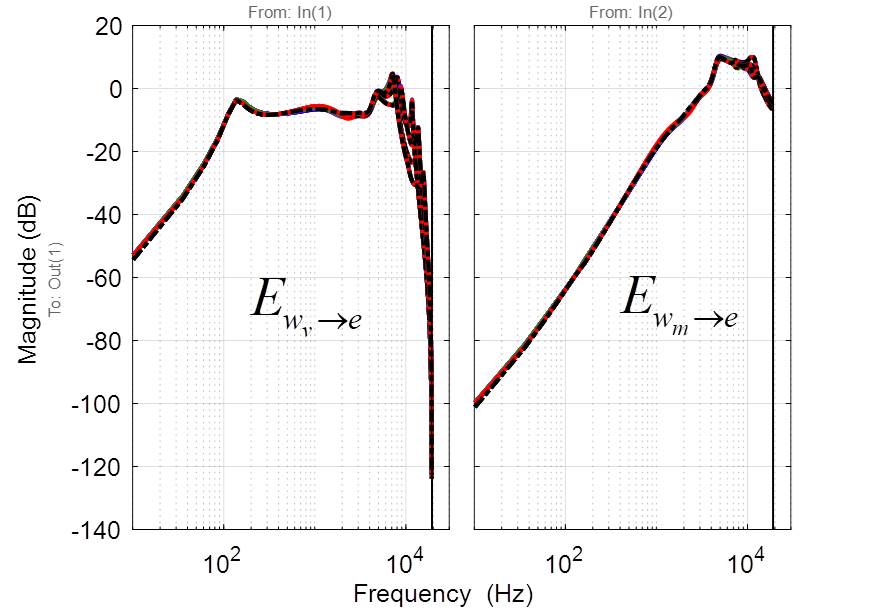}
        \caption[.]{$E_{w \rightarrow e}$, $w=\matrixonetwo{w_v}{w_m}^T$}
        \label{fig:dual:Sw}
    \end{subfigure}
    ~ 
    \begin{subfigure}[b]{7cm}
        \includegraphics[width=7cm]{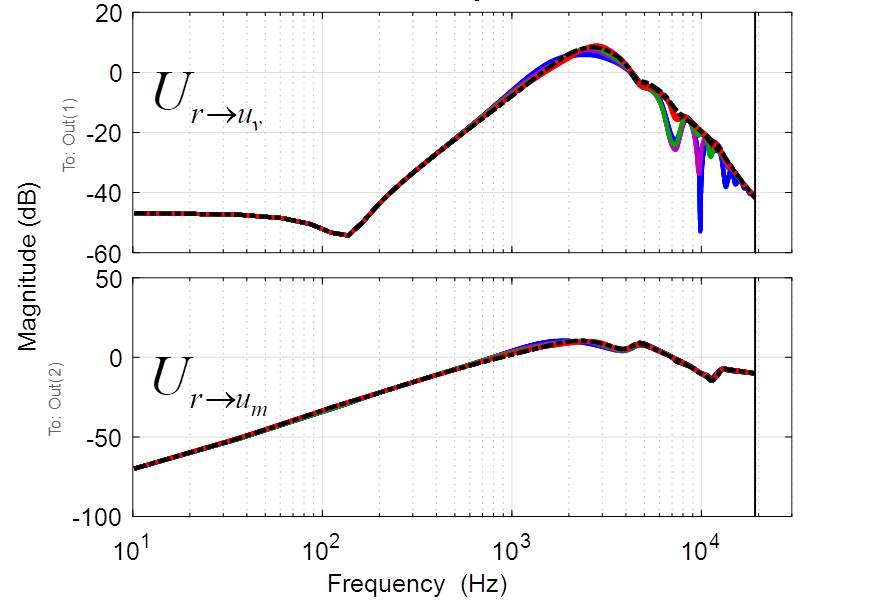}
        \caption[.]{$U_{r \rightarrow u}$, $u=\matrixonetwo{u_v}{u_m}^T$}
        \label{fig:dual:Ur}
    \end{subfigure}
    ~ 
    \begin{subfigure}[b]{7cm}
        \includegraphics[width=7cm]{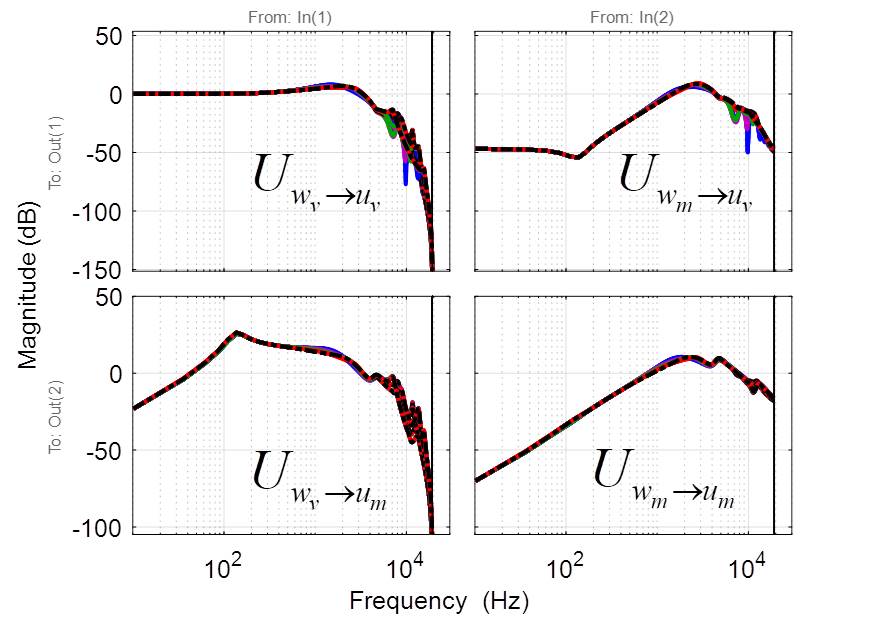}
        \caption[.]{$U_{w \rightarrow u}$, $w=\matrixonetwo{w_v}{w_m}^T$, $u=\matrixonetwo{u_v}{u_m}^T$}
        \label{fig:dual:Uw}
    \end{subfigure}
    \caption{The magnitude Bode plots of the dual-stage closed loop transfer functions.}\label{fig:dual:Hinf}
\end{figure}

The transfer functions for the dual-stage loops are plotted in Fig.~\ref{fig:dual:Hinf}. $E_{r \rightarrow e}$ is a SISO and $E_{w \rightarrow e}$, $U_{r \rightarrow u}$, $U_{w \rightarrow u}$ are MIMO transfer functions. The weighting functions used to shape the MIMO transfer functions in Eq.~(\ref{eq:Hinf:dual:overall}) are as follows
\begin{eqnarray}
\forall \omega \in \Omega:
W_{E_{w \rightarrow e}}(\omega)=
\begin{bmatrix}
0.10 \\
0.10 
\end{bmatrix},\\
\forall \omega \in \Omega:
W_{E_{r \rightarrow u}}(\omega)=
\begin{bmatrix}
0.10 & 0.00\\
0.00 & 0.10
\end{bmatrix},\\
\forall \omega \in \Omega:
W_{E_{w \rightarrow u}}(\omega)=
\begin{bmatrix}
0.10 & 0.10\\
0.04 & 0.10
\end{bmatrix}.
\end{eqnarray}

In the case of MIMO transfer functions, the weighting functions in the MIMO transfer functions shape the maximum singular values of the closed loop transfer functions by limiting them to be smaller than the inverse of minimum singular values of the weighting transfer functions across the entire frequency region. Figs.~\ref{fig:dual:Sw_limit}-\ref{fig:dual:Uw_limit} plot the maximum singular values of closed loop transfer functions with the color code mentioned in table~\ref{table:design:scenarios} and the inverse of the minimum singular values of the weighting functions with light blue dotted lines.

All the values corresponding to SISO and MIMO closed loop transfer functions plotted in this section are smaller than their upper-bounds shown with light blue dotted lines. Therefore, all these closed loop transfer functions satisfy the $H_\infty$ constraints defined in Eqs.~(\ref{eq:Hinf:single}), (\ref{eq:Hinf:dual:overall}), (\ref{eq:Hinf:dual:individual:E}) and (\ref{eq:Hinf:dual:individual:U}).


\begin{figure}
    \centering    
    \begin{subfigure}{7cm}
        \includegraphics[width=7cm]{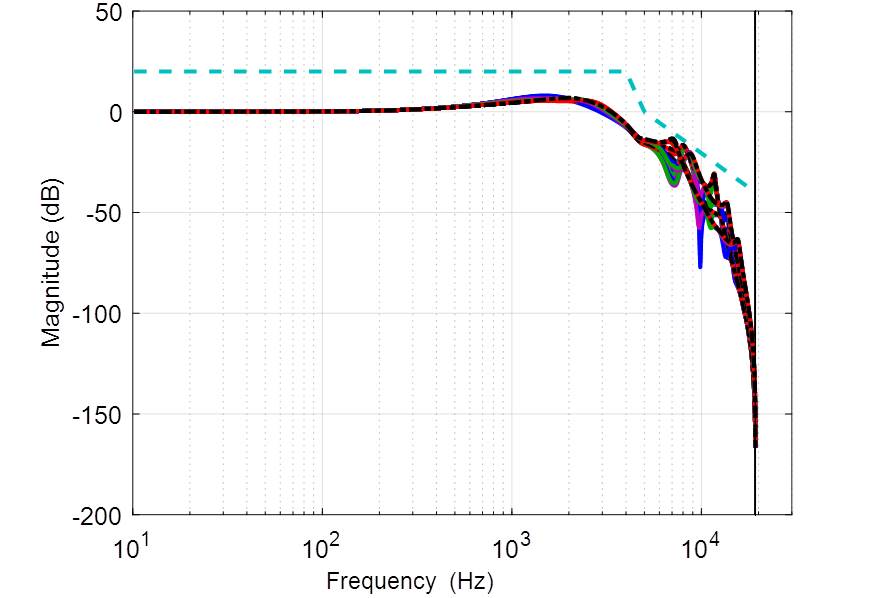}
        \caption[.]{$U_{w_v \rightarrow u_v}$, $\abs{W_{U_{w_v \rightarrow u_v}}}^{-1}$}
        \label{fig:indv:vUrv}
    \end{subfigure}
    \begin{subfigure}{7cm}
        \includegraphics[width=7cm]{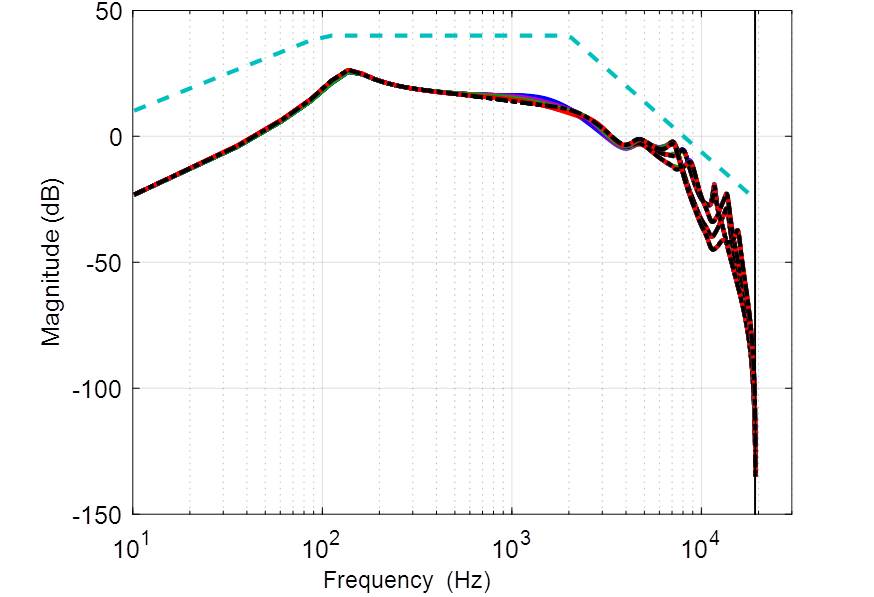}
        \caption[.]{$U_{w_v \rightarrow u_m}$, $\abs{W_{U_{w_v \rightarrow u_m}}}^{-1}$}
        \label{fig:indv:mUrv}
    \end{subfigure}
    \begin{subfigure}{7cm}
        \includegraphics[width=7cm]{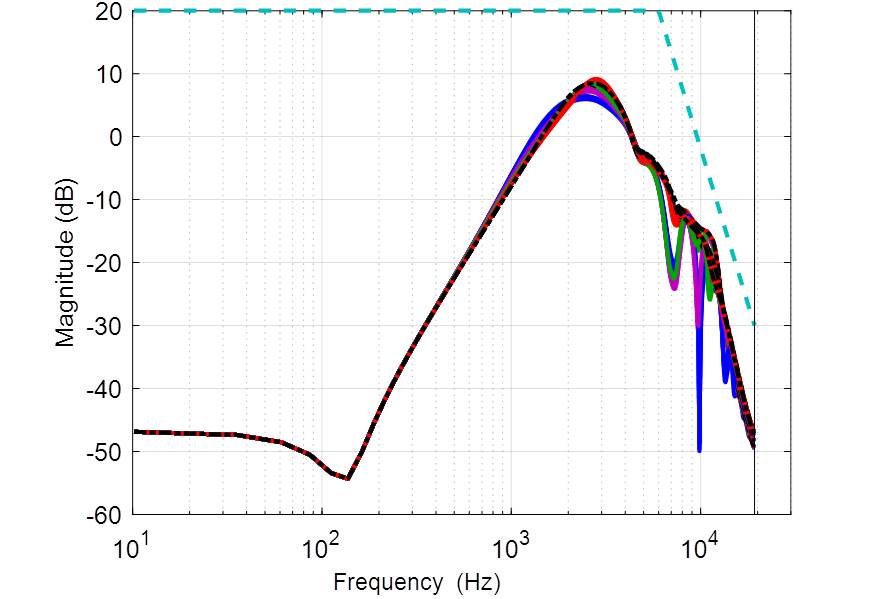}
        \caption[.]{$U_{w_m \rightarrow u_v}$, $\abs{W_{U_{w_m \rightarrow u_v}}}^{-1}$}
        \label{fig:indv:vUrm}
    \end{subfigure}
    \begin{subfigure}{7cm}
        \includegraphics[width=7cm]{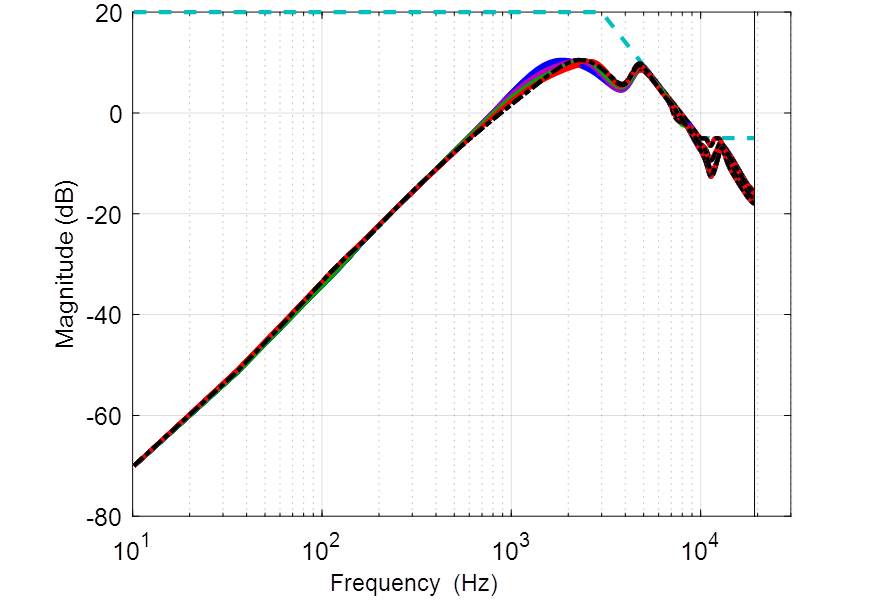}
        \caption[.]{$U_{w_m \rightarrow u_m}$, $\abs{W_{U_{w_m \rightarrow u_m}}}^{-1}$}
        \label{fig:indv:mUrm}
    \end{subfigure}
    \caption{The magnitude Bode plots of the individual SISO closed loop transfer functions from control input disturbances to the control inputs in the dual-stage HDD ($H$) and the inverse of their $H_\infty$ weighting functions magnitudes ($\abs{W_H}^{-1}$). The $\abs{W_H}^{-1}$ functions are shown with the light blue dotted lines. The $H_\infty$ norm criteria are provided in Eq. (\ref{eq:Hinf:dual:individual:U}).}\label{fig:indv:Uw}
\end{figure}


%% file: main.bbl
\providecommand{\bysame}{\leavevmode\hbox to3em{\hrulefill}\thinspace}
\providecommand{\MR}{\relax\ifhmode\unskip\space\fi MR }
\providecommand{\MRhref}[2]{%
  \href{http://www.ams.org/mathscinet-getitem?mr=#1}{#2}
}
\providecommand{\href}[2]{#2}
\begin{thebibliography}{10}

\bibitem{abramovitch2002brief}
Daniel Abramovitch and Gene Franklin, \emph{A brief history of disk drive
  control}, IEEE Control Systems \textbf{22} (2002), no.~3, 28--42.

\bibitem{aggarwal1997micro}
Sanjay~K Aggarwal, David~A Horsley, Roberto Horowitz, and Albert~P Pisano,
  \emph{Micro-actuators for high density disk drives}, American Control
  Conference, 1997. Proceedings of the 1997, vol.~6, IEEE, 1997,
  pp.~3979--3984.

\bibitem{al2006hard}
Abdullah Al~Mamun, GuoXiao Guo, and Chao Bi, \emph{Hard disk drive:
  mechatronics and control}, vol.~23, CRC press, 2006.

\bibitem{al2003dual}
Abdullah Al~Mamun, Iven Mareels, TH~Lee, and Arthur Tay, \emph{Dual stage
  actuator control in hard disk drive-a review}, Industrial Electronics
  Society, 2003. IECON'03. The 29th Annual Conference of the IEEE, vol.~3,
  IEEE, 2003, pp.~2132--2137.

\bibitem{hankel2005}
Athanasios~C Antoulas, \emph{Approximation of large-scale dynamical systems},
  SIAM, 2005.

\bibitem{mosek}
MOSEK ApS, \emph{The mosek optimization toolbox for matlab manual. version 7.1
  (revision 28).}, 2015.

\bibitem{atsumi2003vibration}
Takenori Atsumi, Toshihiro Arisaka, Toshihiko Shimizu, and Takashi Yamaguchi,
  \emph{Vibration servo control design for mechanical resonant modes of a
  hard-disk-drive actuator}, JSME International Journal Series C Mechanical
  Systems, Machine Elements and Manufacturing \textbf{46} (2003), no.~3,
  819--827.

\bibitem{OmidPhDThesis}
Omid Bagherieh, \emph{Estimation, identification and data-driven control design
  for hard disk drives}, University of California, Berkeley, 2017.

\bibitem{datadriven_book_H2}
Alexandre~Sanfelice Bazanella, Luc{\'\i}ola Campestrini, and Diego Eckhard,
  \emph{Data-driven controller design: the h2 approach}, Springer Science \&
  Business Media, 2011.

\bibitem{boyd2004convex}
Stephen Boyd and Lieven Vandenberghe, \emph{Convex optimization}, Cambridge
  university press, 2004.

\bibitem{ding2000design}
Jiagen Ding, M~Tomizukas, and H~Numasato, \emph{Design and robustness analysis
  of dual stage servo system}, American Control Conference, 2000. Proceedings
  of the 2000, vol.~4, IEEE, 2000, pp.~2605--2609.

\bibitem{galdos2010h}
Gorka Galdos, Alireza Karimi, and Roland Longchamp, \emph{H-infinity controller
  design for spectral mimo models by convex optimization}, Journal of Process
  Control \textbf{20} (2010), no.~10, 1175--1182.

\bibitem{goodwin1992quantifying}
Graham~C Goodwin, Michel Gevers, and Brett Ninness, \emph{Quantifying the error
  in estimated transfer functions with application to model order selection},
  IEEE Transactions on Automatic Control \textbf{37} (1992), no.~7, 913--928.

\bibitem{hernandez1999dual}
Daniel Hernandez, Sung-Su Park, Roberto Horowitz, and Andrew~K Packard,
  \emph{Dual-stage track-following servo design for hard disk drives}, American
  Control Conference, 1999. Proceedings of the 1999, vol.~6, IEEE, 1999,
  pp.~4116--4121.

\bibitem{herrmann2004hdd}
Guido Herrmann and Guoxiao Guo, \emph{Hdd dual-stage servo-controller design
  using a $\mu$-analysis tool}, Control engineering practice \textbf{12}
  (2004), no.~3, 241--251.

\bibitem{SD_2007}
Roberto Horowitz, Yunfeng Li, Kenn Oldham, Stanley Kon, and Xinghui Huang,
  \emph{Dual-stage servo systems and vibration compensation in computer hard
  disk drives}, Control Engineering Practice \textbf{15} (2007), no.~3,
  291--305.

\bibitem{hou2013model}
Zhong-Sheng Hou and Zhuo Wang, \emph{From model-based control to data-driven
  control: survey, classification and perspective}, Information Sciences
  \textbf{235} (2013), 3--35.

\bibitem{huang2001active}
Fu-Ying Huang, Tetsuo Semba, Wayne Imaino, and Francis Lee, \emph{Active
  damping in hdd actuator}, IEEE Transactions on Magnetics \textbf{37} (2001),
  no.~2, 847--849.

\bibitem{karimi2016H2}
Alireza Karimi and Christoph Kammer, \emph{A data-driven approach to robust
  control of multivariable systems by convex optimization}, arXiv preprint
  arXiv:1610.08776 (2016).

\bibitem{karimi2016HinfSISO}
Alireza Karimi, Achille Nicoletti, and Yuanming Zhu, \emph{Robust h-infinity
  controller design using frequency-domain data via convex optimization},
  International Journal of Robust and Nonlinear Control (2016).

\bibitem{HDDvsSSD2011}
Vamsee Kasavajhala, \emph{Solid state drive vs. hard disk drive price and
  performance study}, Proc. Dell Tech. White Paper (2011), 8--9.

\bibitem{khargonekar1991mixed}
Pramod~P Khargonekar and Mario~A Rotea, \emph{Mixed h-2/h-infinity control: a
  convex optimization approach}, IEEE Transactions on Automatic Control
  \textbf{36} (1991), no.~7, 824--837.

\bibitem{kobayashi2001track}
Masahito Kobayashi and Roberto Horowitz, \emph{Track seek control for hard disk
  dual-stage servo systems}, IEEE Transactions on Magnetics \textbf{37} (2001),
  no.~2, 949--954.

\bibitem{ljung1999system}
Lennart Ljung, \emph{System identification}, Wiley Online Library, 1999.

\bibitem{yalmip}
Johan Lofberg, \emph{Yalmip: A toolbox for modeling and optimization in
  matlab}, Computer Aided Control Systems Design, 2004 IEEE International
  Symposium on, IEEE, 2004, pp.~284--289.

\bibitem{MATLAB:2016}
MATLAB, \emph{version 9.1.0 (r2016b)}, The MathWorks Inc., Natick,
  Massachusetts, 2016.

\bibitem{pintelon2012system}
Rik Pintelon and Johan Schoukens, \emph{System identification: a frequency
  domain approach}, John Wiley \& Sons, 2012.

\bibitem{rantzer1994convex}
Anders Rantzer and A~Megretski, \emph{A convex parameterization of robustly
  stabilizing controllers}, IEEE Transactions on Automatic Control \textbf{39}
  (1994), no.~9, 1802--1808.

\bibitem{richardson1982parameter}
Mark~H Richardson and David~L Formenti, \emph{Parameter estimation from
  frequency response measurements using rational fraction polynomials},
  Proceedings of the International Modal Analysis Conference, 1982,
  pp.~167--182.

\bibitem{StatsDataCenters}
Cisco Systems, \emph{Amount of data actually stored in data centers worldwide
  from 2015 to 2020 (in exabytes). statista - the statistics portal, statista},
  jun 2017.

\bibitem{tay1989indirect}
TT~Tay, JB~Moore, and R~Horowitz, \emph{Indirect adaptive techniques for fixed
  controller performance enhancement}, International Journal of Control
  \textbf{50} (1989), no.~5, 1941--1959.

\bibitem{yi1999two}
Li~Yi and Masayoshi Tomizuka, \emph{Two-degree-of-freedom control with robust
  feedback control for hard disk servo systems}, IEEE/ASME transactions on
  mechatronics \textbf{4} (1999), no.~1, 17--24.

\bibitem{convexconcave2003}
Alan~L Yuille and Anand Rangarajan, \emph{The concave-convex procedure}, Neural
  computation \textbf{15} (2003), no.~4, 915--936.

\bibitem{robustbook1996}
Kemin Zhou, John~Comstock Doyle, Keith Glover, et~al., \emph{Robust and optimal
  control}, vol.~40, Prentice hall New Jersey, 1996.

\end{thebibliography}
